\documentclass[letterpaper,dvipsnames]{article}
\pdfoutput=1 

\usepackage{jheppub} 
\usepackage{hyperref}
\usepackage{tikz}
\usepackage[normalem]{ulem}
\usepackage{amsmath}
\usepackage[T1]{fontenc} 
\usepackage{braket}
\usepackage{subcaption}
\usepackage{soul}
\usepackage{array}
\usepackage[dvipsnames]{xcolor}
\usepackage{cancel}
\usepackage{orcidlink}
\usepackage{slashed}
\usepackage{comment}
\excludecomment{supercomment}
\usepackage[many]{tcolorbox}    	
\usepackage{dsfont}
\usepackage{stmaryrd} 
\usepackage{bm}
\usepackage{mathtools}
\usepackage{mathrsfs}

\newcommand{\be}{\begin{eqnarray}}
\newcommand{\ee}{\end{eqnarray}}

\definecolor{colorPSgold}{rgb}{0.996,0.745,0.063}
\definecolor{colorPSblue}{rgb}{0.000,0.322,0.624}

\DeclareMathAlphabet{\mathbbb}{U}{bbold}{m}{n} 

\definecolor{colorRTD}{rgb}{.2,.2,.7}

\definecolor{colorRTD2}{rgb}{.2,.3,.7}

\newtcolorbox{boxA}{
    fontupper = \bf,
    boxrule = 1.5pt,
    colframe = black 
}

\newtcolorbox{boxE}{
    enhanced, 
    boxrule = 0pt, 
    borderline = {0.75pt}{0pt}{main}, 
    borderline = {0.75pt}{2pt}{sub} 
}

\usepackage[compat=1.1.0]{tikz-feynman}

\author[a]{Paolo Bilisco\orcidlink{https://orcid.org/0009-0004-7515-1353},}
\affiliation[a]{Institut de Physique Th\'eorique, Universit\'e Paris-Saclay, CNRS, CEA, F-91191 Gif-sur-Yvette, France}

\author[a,b]{Raffaele Tito D'Agnolo\orcidlink{https://orcid.org/0000-0002-8271-6840},}
\affiliation[b]{School of Natural Sciences, Institute for Advanced Study, Princeton, NJ 08540, USA}

\author[c,d]{Arthur Platschorre\orcidlink{https://orcid.org/0000-0003-4106-6287}}
\affiliation[c]{Rudolf Peierls Centre for Theoretical Physics, 
University of Oxford, Parks Road, Oxford OX1 3PU, United Kingdom}
\affiliation[d]{Department of Physics, LEPP, Cornell University, Ithaca, NY 14853, United States}

\author[e]{and Pablo~Sesma\orcidlink{https://orcid.org/0000-0002-9035-3322}}
\affiliation[e]{Institut de Física d’Altes Energies (IFAE) and The Barcelona Institute of Science
and Technology (BIST), Campus UAB, 08193 Bellaterra, Barcelona, Spain}

\emailAdd{paolo.bilisco@ipht.fr}
\emailAdd{raffaele-tito.dagnolo@ipht.fr}
\emailAdd{acp256@cornell.edu}
\emailAdd{psesma@ifae.es}

\title{Deconfinement on $\mathbb{Z}_{F}$ and Axion Domain Walls}

\abstract{We investigate the nontrivial dynamics on $\mathbb{Z}_{F}$ domain walls and compute their tension in quantum chromodynamics (QCD) compactified on a circle with $N$ colors and $F < N$ massless fundamental flavors. In the process, we obtain insights on the $\eta'$ meson's potential in a regime where it is heavy and comparable in mass to other QCD resonances, as in Nature. We employ a center-stabilized compactification of QCD on a small circle of size $L$ with color-flavor twisted boundary conditions for the quarks. We demonstrate that probe particles in color representations of $N$-ality $N/\mathrm{gcd}(N,F)$ are deconfined on the $\mathbb{Z}_{F}$ domain walls whenever $N$ and $F$ are not coprime. This is reminiscent of deconfinement on axion domain walls in pure Yang--Mills. Finally, in the double scaling limit $N\rightarrow \infty$ keeping $NL$ and $F$ fixed, the infrared spectrum spans an emergent dimension. We establish that the $\mathbb{Z}_{F}$ domain walls are dual to electrostatic charge distributions in this emergent dimension. This duality allows us to compute their tension analytically and find a different tension scaling at large-$N$ between $\mathbb{Z}_{F}$ domain walls in QCD and axion domain walls in pure Yang--Mills, in center-stabilized compactifications. }

\begin{document} 

\maketitle

\section{Introduction}
The nature of the confining vacuum of quantum chromodynamics (QCD) has been the subject of intense experimental and theoretical study for over half a century. At a perturbative level, it is known that the infrared phase of QCD depends strongly on the number of colors $N$ and the number of active fundamental flavors $F$. When the number of flavors $F$ exceeds the ratio $\frac{11}{2}N$, the theory is infrared free, whilst for $F$ just below $\frac{11}{2}N$, there exists a conformal window in which the theory flows to a nontrivial fixed point. Below the conformal window the QCD vacuum is strongly coupled and confining in the infrared. In the confining vacuum the $U(F)_L \times U(F)_R$ chiral symmetry of the (massless) fundamental flavors is non-linearly realized, resulting in $F^2-1$ pseudo-Nambu--Goldstone bosons: the pions. 

In this paper we study the tension of  domain walls that interpolate between the different confining QCD vacua, as well as the nontrivial dynamics living on them. We refer to these as $\mathbb{Z}_F$ domain walls, since the vacua they connect are related by a transformation in $\mathbb{Z}_F\subset U(1)_A$. 
We place a particular emphasis on the scenario in which $N$ and $F$ are not coprime, i.e. $\mathrm{gcd}(N,F)>1$, where the  domain walls are expected to have a rich internal structure, but we study also the regime $\mathrm{gcd}(N,F)=1$.

To this end, we consider a calculable  model of QCD, compactified on a circle. In this model, both the tension and domain wall profile of the domain walls that interpolate between the confining vacua are calculable. At large $N$, we find an interesting duality, that allows us to compute the properties of the domain walls in our model from an emergent electrodynamics in an additional spatial dimension.

The results obtained in this paper are relevant to the study of the $\eta'$ meson, the would-be Nambu--Goldstone of $U(1)_A$. Due to the chiral anomaly~\cite{tHooft:1986ooh}, the $\eta'$ acquires a mass of the order of the confinement scale, and therefore serves as an essential probe for the non-perturbative dynamics of QCD. The calculable model we consider allows us to obtain insights on the $\eta'$ potential, whose structure is still object of speculation in the literature~\cite{Csaki:2023yas, Kitano:2025ezk}. The $\mathbb{Z}_F$ domain walls, which are the main focus of this work, can be regarded as $\eta'$ domain walls in various limits of QCD, such as large-$N$ and softly-broken supersymmetric QCD, since the $\eta'$ remains part of the infrared spectrum. Before moving on to introducing the compactified model of QCD we use, we briefly review the literature on the $\eta'$ meson in these limits of QCD.  

At large $N$ and finite $F$~\cite{tHooft:1973alw,Witten:1979kh,Witten:1980sp,DiVecchia:1980yfw}, the low energy QCD spectrum consists of massless pions and the $\eta'$ meson, whose squared mass is of $\mathcal{O}(1/N)$. The leading terms in the large-$N$ expansion of $\eta'$ potential are analytically calculable~\cite{Witten:1980sp,DiVecchia:1980yfw} and depend on the physical ratio $\eta'/N$. In order to be compatible with the transformation\footnote{Here and in the following we normalize the $\eta'$ meson to its decay constant and treat it as an angular variable with periodicity $\eta' \equiv \eta'+2\pi F$. The transformation $\eta' \rightarrow \eta'+2\pi$ is a $\mathbb{Z}_{F} \subset U(1)_{\mathrm{axial}}$ symmetry of massless QCD.} $\eta' \rightarrow \eta'+2\pi$, the potential has $N$ branches with cusps at $\eta' =\pi, \ 3\pi, \ \mathrm{etc}$. At these cusps, the derivatives of the potential become singular and the effective theory  breaks down. It is plausible\footnote{Further insights about the cusp can be obtained in 
gauge-gravity duals at large $\lambda = N g^{2}$ 't Hooft coupling~\cite{Witten:1998zw,Aharony:1999ti,Maldacena:2000yy}.} that this breakdown in the low energy theory occurs because of the rapid rearrangement of heavy degrees of freedom near the cusp not appropriately captured by the low energy effective theory that describes the $\eta'$ and  pions.

At finite $N$ and $F$, a similar  breakdown occurs in the infrared description of another QCD(-like) theory, softly-broken supersymmetric QCD \cite{Draper:2018mpj}, with supersymmetry breaking from anomaly-mediation~\cite{Randall:1998uk,Giudice:1998xp,Arkani-Hamed:1998dti,Murayama:2021xfj,Csaki:2022cyg,Csaki:2023yas,Csaki:2024lvk}. The infrared spectrum of supersymmetric QCD consists of massless pions and a massless $\eta'$ boson. In the presence of soft supersymmetry-breaking gaugino masses (or more generally a source of $R$-symmetry breaking), the light $\eta'$ meson obtains a potential that depends on the ratio $\eta'/(N-F)$ whenever $|N-F|\geq 1$~\cite{Csaki:2023yas}. This potential has $N-F$ branches and becomes singular at $\eta' =\pi, \ 3\pi, \ \mathrm{etc}$ where  derivatives of the potential diverge. When $N=F$, the $\eta'$ meson becomes heavy and its potential incalculable~\cite{Kitano:2025ezk}.

It is thought that the singularities in these limits of QCD(-like) theories occurs because the $\eta'$ meson is parametrically lighter than the other heavy QCD resonances. This large gap between the $\eta'$ meson and the other QCD resonances generates a cuspy (i.e. singular) effective theory, which prevents a careful study of $\eta'\rightarrow \eta'+2\pi$ domains walls (i.e. $\mathbb{Z}_F$ domain walls) that cross the singularity. Indeed, while the infrared spectrum of QCD is analytically tractable in both large-$N$ and softly-broken supersymmetric QCD scenarios, the tension of $\eta'\rightarrow \eta'+2\pi$ domain walls that cross the singularity is incalculable, and can drop below the standard BPS bound~\cite{Kogan:1997dt}. The missing energy density that raises the tension to or above the BPS bound is attributed to the rearrangement of  heavy fields at the cusp. Both the BPS bound and $D$-brane estimates suggest a tension $T$ scaling as $T \sim N\Lambda_{\mathrm{QCD}}^{3}$~\cite{Dvali:1996xe,Kovner:1997ca,Kogan:1997dt,Witten:1997ep,Witten:1998uka}.

Whenever $\mathrm{gcd}(N,F)>1$, powerful anomaly-matching constraints~\cite{tHooft:1979rat,Gaiotto:2014kfa, Gaiotto:2017yup} tell us that an effective $\eta'$-only theory must break down between $\eta'$ and $\eta'+2\pi$, as the $\eta'$-only theory fails to saturate the ultraviolet anomaly inflow of the mixed ’t Hooft anomaly between the baryon-color-flavor bundles and the $\mathbb{Z}_{2F}$ axial symmetry~\cite{Cordova:2019jnf,Cordova:2019uob, Anber:2019nze, Tanizaki:2017qhf}. The anomaly inflow can be matched by nontrivial degrees of freedom living on the $\mathbb{Z}_{F}$ domain walls, such as massless charged degrees of freedom or a TQFT. In softly-broken supersymmetry, it has been suggested that the anomaly inflow on $\mathbb{Z}_{F}$ walls can be matched by an additional $U(1)_{F-N}$ Chern--Simons theory~\cite{Kitano:2025ezk}. Taken together, these considerations show that the breakdown in the effective field theory of both large-$N$ QCD and softly-broken sQCD obscures both the tension and the rich internal structure of $\mathbb{Z}_{F}$ domain walls.

In order to access this rich internal structure and calculate the tension of the $\mathbb{Z}_{F}$ domain walls, we employ a center-stabilized compactification of QCD on a small circle of size $L \ll (N\Lambda_{\rm QCD})^{-1}$~\cite{Unsal:2008ch, Poppitz:2021cxe}, as reviewed in Section \ref{sec:twistedQCD}. To realize an infrared spectrum with chiral symmetry breaking~\cite{Unsal:2008eg}, the $F$ fundamental fermions are taken to have $U(F)_{V}$ color-flavor twisted boundary conditions around the compact circle. The twisted boundary conditions explicitly break part of the $U(F)_{V} \times U(F)_{A}$ flavor symmetries of the massless quarks, and hence the resulting (center-stabilized) twisted QCD theory does not lie in the same universality class as untwisted QCD. 

Importantly, as shown in Appendix~\ref{app:cartan-cfc-anomalies}, the reduction of the flavor symmetries in twisted QCD does not remove the 't Hooft anomaly between the baryon-color-flavor bundles and the $\mathbb{Z}_{2F}$ axial symmetry~\cite{Anber:2019nze,Tanizaki:2017qhf}. This gives a solid basis for the comparison with $4d$ QCD, as it implies that the $\mathbb{Z}_{F}$ domain walls in twisted QCD should have a similarly rich internal structure whenever $\mathrm{gcd}(N,F)>1$, whilst the semiclassical control makes the wall profiles and tensions accessible to explicit calculations.   

In this model of QCD, the infrared spectrum is analytically tractable~\cite{Cherman:2016hcd} and singularities are absent. One can explicitly demonstrate the confinement of color charges~\cite{Unsal:2008ch} and the spontaneous breaking of chiral symmetry, resulting in a calculable Chiral Lagrangian~\cite{Cherman:2016hcd,Cherman:2017dwt}. The resulting infrared spectrum has been well-studied, including its symmetries, phases and anomalies \cite{Tanizaki:2017qhf,Cherman:2017dwt,Anber:2023urd}. Its glueball and massive meson spectrum are in Ref.~\cite{Aitken:2017ayq}.

In Section \ref{sec:deconf}, we investigate the dependence of the vacuum and domain walls of twisted QCD on the number of colors $N$ and the number of massless fundamental fermion flavors $F$. We first demonstrate that the infrared spectrum of twisted QCD decomposes into massive and massless modes. The massless modes are identified as a set of massless pions, the Nambu--Goldstone bosons of the spontaneously broken chiral symmetry of the fundamental fermions in the QCD vacuum. We find that, collectively, the set of massive modes plays the same role in twisted QCD as the single $\eta'$ meson plays in large-$N$ QCD or in softly-broken sQCD, as explained in Section \ref{sec:centerstabQCDmon}.

Twisted QCD admits $\mathbb{Z}_{F}$ domain walls that interpolate between vacua related by the $\mathbb{Z}_{F}$ axial symmetry. As the infrared spectrum of twisted QCD is without singularities, both the tension and the effective theory on these domain walls are analytically calculable and are uniquely set by the color-flavor twisted boundary conditions of the quark. Whenever $N$ and $F$ are not coprime, we show that probe particles in color representation of $N$-ality $N/\mathrm{gcd}(N,F)$ under the $SU(N)$ gauge group are deconfined on the $\mathbb{Z}_{F}$ domain walls.

It is interesting to compare this prediction to the common lore that suggests that an $\eta'$ meson and its $\mathbb{Z}_{F}$ domain walls in four-dimensional QCD 
act similar to a dynamical $\theta_{\mathrm{YM}}$ angle and its domain walls in four-dimensional pure Yang--Mills. In center-stabilized pure gluodynamics, the vacuum structure as a function of the $\theta_{\mathrm{YM}}$-angle is analytically calculable~\cite{Bhoonah:2014gpa,Aitken:2018mbb,Anber:2017rch,Aitken:2018kky}.  Physical quantities depend on the ratio $\theta_{\mathrm{YM}}/N$ and the theory remains $2\pi$ periodic in $\theta_{\mathrm{YM}}$ due to the rearrangement of heavy  dual photons near the CP-preserving angle $\theta_{\mathrm{YM}} =\pi, \ 3\pi, \ \mathrm{etc}$. In concordance with anomaly arguments~\cite{Cox:2019aji,Anber:2018xek, Poppitz:2020tto}, one can prove that this rearrangement of dual photons near $\theta_{\mathrm{YM}} =\pi$ results in the deconfinement of probe color charges of $N$-ality $1$ on axion domain walls $\theta_{\mathrm{YM}} \rightarrow \theta_{\mathrm{YM}} + 2\pi$~\cite{Poppitz:2012nz,Chen:2020syd, Anber:2020xfk}. A key ingredient to demonstrate this deconfinement is the $\mathbb{Z}_{N}$ center symmetry of pure Yang--Mills. In twisted QCD, a key ingredient to demonstrate the deconfinement on $\mathbb{Z}_{F}$ domain walls is a remnant  color-flavor-center symmetry $\mathbb{Z}_{\mathrm{gcd}(N,F)} \subset \mathbb{Z}_{N}$~\cite{Cherman:2017tey, Kanazawa:2019tnf} whenever $\mathrm{gcd}(N,F)>1$.

The tension and profile of the  $\mathbb{Z}_{F}$ domain walls is numerically calculable for any $F$ and $N >F$. In Section~\ref{sec:emerge}, we show that one can analytically solve for the $\mathbb{Z}_{F}$ domain wall profile and tension in the limit $N\rightarrow \infty$, keeping $NL$ and $F$ fixed. In this double-scaling limit, the infrared spectrum of twisted QCD spans an emergent dimension~\cite{Cherman:2016jtu} with an emergent Abelian gauge field. We utilize this fact to reduce solving for domain wall profiles at large $N$ to  electrostatic problems in one higher dimension. The tension of the domain wall maps to the energy of the electrostatic charge distribution in the emergent dimension. In pure center-stabilized gluodynamics, we find that the tension of the $\theta_{\mathrm{YM}}$ domain walls scales as $T \sim \pi N\ln{(N)} \Lambda_{\mathrm{YM}}^{3}$. In  twisted QCD, we find that the tension of the $\mathbb{Z}_{F}$ domain walls scales as $T \sim N \Lambda_{\mathrm{QCD}}^{3}$. 
We show that the disparity between the tensions is due to the massless pion directions, which allow the $\mathbb{Z}_{F}$ domain wall tension to relax. Remarkably, this result suggests that the common lore connecting $\eta'$ and $\theta_{\mathrm{YM}}$ has its limits. 

In Section~\ref{sec:real} we make contact between the results obtained in twisted QCD and QCD in the real world, and comment on potential implications for the smoothness of the $\eta'$ potential. We conclude in Section~\ref{sec:conclusion} and discuss potential future avenues of research.

\section{Twisted QCD }
\label{sec:twistedQCD}

The idea to study Yang--Mills theory and QCD on compact dimensions smaller than $\Lambda_{\mathrm{QCD}}^{-1}$ is an old one (see for instance Refs.~\cite{Bjorken:1979hv,vanBaal:2000zc,Luscher:1982ma}). Throughout this section, we provide a review of the compactification of QCD on a small circle with a center-stabilizing potential, largely overlapping with earlier works on the topic~\cite{Unsal:2008ch,Aitken:2017ayq}. Most of the concepts and tools reviewed in this section are applied in Sections \ref{sec:deconf} and \ref{sec:emerge}.

Here we review how the infrared spectrum of QCD on a small circle looks in the confined phase in the presence of $n_{f}$ massive adjoint Weyl fermions. The infrared dynamics is shown to be dominated by monopole-instantons. We  adopt $U(F)_{V}$  twisted boundary conditions for the fundamental quarks around the compact direction, which determine the fermion zero mode distribution over the monopoles. Following Ref.~\cite{Cherman:2016hcd}, we find that this results in spontaneously broken axial symmetries. The resulting Chiral Lagrangian, including the massive states of the infrared spectrum, is studied in Section \ref{sec:deconf}.

\subsection{Center-Stabilized Pure Yang--Mills Theory}\label{sec:Center-Stabilized pure Yang--Mills theory}

In this section, we review the main features of the standard center-stabilized construction of pure gluodynamics, $SU(N)$ Yang--Mills theory, described by the action
\begin{equation}
    S = -\frac{1}{2g^{2}} \int d^{4}x \ \mathrm{Tr}\, G_{\mu \nu} G^{\mu \nu}\,, \qquad \mathrm{with} \qquad G_{\mu \nu} = \partial_{\mu}A_{\nu} - \partial_{\nu} A_{\mu} -i [A_{\mu},A_{\nu}] \, .
    \label{eq:ym}
\end{equation}
We normalize the generators in the fundamental representation such that $\mathrm{Tr}\,(T^{a}T^{b})=\delta^{ab}/2$.
In order to remain in the perturbative regime of the strong coupling constant $g$, we place the theory on $\mathbb{R}^3\times S^1$, where the compact circle $S^1$ has circumference $L \ll \Lambda_{\mathrm{QCD}}^{-1}$. The compact coordinate is identified as $x_3\sim x_3+L$, while the non-compact coordinates are $(x_0,x_1,x_2)\in \mathbb{R}^3$. The four-dimensional gauge fields decompose as $A_{\mu}=(A_i,A_3)$, with $i=0,1,2$, where $A_i$ are the three-dimensional gauge fields and $A_3$ behaves locally as an adjoint scalar. After fixing the $x_3$-dependent gauge redundancy, $\partial_3 A_3=0$~\cite{Poppitz:2021cxe}, the physical information contained in the zero mode of $A_3$ is captured by the holonomy around the circle, which acts as an adjoint Higgs field:
\begin{equation}
    \Omega(x)=\mathcal{P}\exp\left(i\int_{S^1}A_3 \right)\, .
\end{equation}
The infrared dynamics of the three-dimensional gauge theory is set by the expectation value of the gauge-invariant trace of the Polyakov loop $\mathrm{Tr}\,\Omega$ taken in the fundamental representation. 

The pure $SU(N)$ Yang--Mills theory has a $\mathbb{Z}_{N}$ $1$-form symmetry~\cite{Gaiotto:2014kfa} (see Refs.~\cite{Brennan:2023mmt,Gomes:2023ahz,Bhardwaj:2023kri} for a review). The $1$-form symmetry acts on closed Wilson loops and transforms them by a $\mathbb{Z}_{N}$ phase. On $\mathbb{R}^3\times S^1$, the $1$-form symmetry decomposes into a $\mathbb{Z}_{N}$ $1$-form symmetry acting on the Wilson loops in $\mathbb{R}^{3}$ and an ordinary $\mathbb{Z}_{N}$ $0$-form center symmetry that acts on the Polyakov loop $\Omega$ as
\begin{equation}
    \mathrm{Tr}\,\Omega \rightarrow e^{2\pi i n/N} \mathrm{Tr}\,\Omega\, ,\qquad n\in\mathbb{Z}_N\,.
    \label{eq:center symmetry action on Omega}
\end{equation}
The Polyakov loop is the order parameter for the $\mathbb{Z}_{N}$ $0$-form center symmetry, which we will henceforth simply refer to as the $\mathbb{Z}_{N}$ center symmetry. 

 In compactified Yang--Mills, integrating out the non-zero Kaluza--Klein modes of the gauge field and ghosts at $1$-loop generates the Gross--Pisarski--Yaffe (GPY) contribution to the effective potential of the holonomy $\Omega(x)$, reading~\cite{Gross:1980br,Unsal:2008ch,Poppitz:2021cxe}
\begin{equation}
  V_{\rm GPY}^{(3d)}(\Omega)=-\frac{2}{\pi^2L^3}\sum_{n=1}^{\infty}\frac{1}{n^4}\left(\left|\mathrm{Tr}\,(\Omega^n)\right|^2-1 \right)\,.
    \label{eq:GPY potential}
\end{equation}
This potential is invariant under the action of the center symmetry in Eq.~\eqref{eq:center symmetry action on Omega}. The negative sign favors a minimum at $\Omega= z\,\mathbf{1}_N$, with $z\in\mathbb{Z}_N$, for which $|\mathrm{Tr}\, \Omega|=N$. Hence, in pure gluodynamics at small circle sizes, the perturbative GPY potential spontaneously breaks the $\mathbb{Z}_{N}$ center symmetry. The long-distance theory below the compactification scale is a three-dimensional non-Abelian $SU(N)$ gauge theory and Wilson loops in the non-compact directions exhibit an area law.

The spontaneously broken $\mathbb{Z}_{N}$ center symmetry at small circle sizes for pure gluodynamics stands in sharp contrast to four dimensional Yang--Mills, where the full $\mathbb{Z}_{N}$ $1$-form center symmetry group is unbroken and all closed Wilson lines exhibit an area. This phase transition can be avoided by introducing $n_{f}$ periodic adjoint Weyl fermions, which stabilize the potential~\eqref{eq:GPY potential} around a center-symmetry-preserving minimum. Inspired by supersymmetry, adding a single massless adjoint Weyl fermion (gaugino) with periodic boundary conditions completely cancels the bosonic GPY potential in Eq.~\eqref{eq:GPY potential}. More generally, with $n_f$ massless periodic adjoint Weyl fermions, the one-loop holonomy potential becomes
\begin{equation}
    V_{1\rm -loop}^{(3d)}(\Omega)=\frac{2(n_f-1)}{\pi^2 L^3}\sum_{n=1}^{\infty}\frac{1}{n^4}\left(\left|\mathrm{Tr}\left(\Omega^n\right)\right|^2-1\right)\,.
\end{equation}

For $n_f=1$, this potential vanishes, as in $\mathcal N=1$ super-Yang--Mills. For $n_f>1$, its sign is opposite to that of pure Yang--Mills, and the minimum has the desired center-symmetry~\cite{Unsal:2008ch,Myers:2007vc,Myers:2009df,Hosotani:1983xw}.
 A convenient representative of the center-symmetric minimum, chosen in the fundamental Weyl chamber\footnote{See Refs.~\cite{Anber:2015wha,DiFrancesco:1997nk} for a refresher on Weyl chambers and weights. The fundamental Weyl chamber for $SU(N)$ is fixed by demanding that the phases in Eq.~\eqref{eq:center symmetric holonomy} are in increasing order within the $[0,2\pi)$ domain. } where each subsequent phase is increasing,  is~\cite{Aitken:2017ayq}
\begin{equation}
    \Omega_{\mathrm{cs}}=e^{-\pi i (N-1)/N} \mathrm{diag} \left(1, e^{2\pi i/N}, e^{4\pi i/N},\ldots, e^{2\pi i (N-1)/N} \right)\,,
    \label{eq:center symmetric holonomy}
\end{equation}
where the overall phase of $\Omega_{\mathrm{cs}}$ ensures that $\Omega_{\mathrm{cs}}$ is an element of $SU(N)$.

Importantly, note that if $n_f$ is sufficiently large, the adjoint fermions may be given masses while still preserving the center-symmetric minimum~\cite{Lai:2022ybv,Unsal:2010qh,Anber:2014lba,Misumi:2014raa}, with asymptotic freedom requiring $n_f\leq 5$~\cite{Lai:2022ybv}. We therefore work with a sufficient number of periodic adjoint Weyl fermions, with masses chosen in the center-stabilizing regime $m_{\rm adj}\lesssim m_W=2\pi/(NL)$, where $m_W$ is the mass of the lightest off-diagonal gauge bosons in the center-symmetric background.

In this regime, the adjoint fermions are sufficiently light to stabilize the center-symmetric holonomy, while remaining parametrically heavy compared to the non-perturbative infrared scale. They therefore decouple from the low-energy dynamics that we study below. Our next task is to describe the non-perturbative low-lying modes of the energy spectrum sketched in Fig.~\ref{fig:energyspec}.

\subsection{Monopole-Instantons \& Confinement}\label{sec:magsym}

\begin{figure}
    \centering

\begin{tikzpicture}[
    >=stealth,
    font=\small,
    scale=1,
    every node/.style={align=center},
    level/.style={thick},
    state/.style={draw, rounded corners=1pt, minimum width=3cm,
                  minimum height=0.42cm, fill=white},
    fermion/.style={state, dashed},
    gauge/.style={state},
    scaleLine/.style={densely dashed, gray!70, thick}
]


\def\xYM{2.0}
\def\yL{6.0}
\def\yNL{4.0}
\def\yQCD{2.5}
\def\ymphi{0.5}
\def\ympion{-1}
\def\xF{6.2}

\draw[->, thick,font=\normalsize] (-0.22,\ympion) -- (-0.2,7.2)
    node[above] {$E$};

\foreach \y in {\yL,\yNL,\yQCD,\ymphi,\ympion} {
   \draw[thick] (-0.2,\y) -- ++(-0.12,0);
};

\node[font=\bfseries] at (\xYM,7.0) {Pure Yang--Mills};
\node[font=\bfseries] at (\xF,7.0) {Twisted QCD};


\node[anchor=east, font=\small] at (-0.3,\yQCD)
    {$\Lambda_{\rm QCD}$};

\node[anchor=east, font=\small] at (-0.3,\yNL)
    {$2\pi/(NL)$};

\node[anchor=east, font=\small] at (-0.3,\yL)
    {$2\pi/L$};

\node[anchor=east, font=\small] at (-0.3,\ymphi)
    {$m_{\phi}$};

\node[anchor=east, font=\small] at (-0.3,\ympion)
    {$0$};


\node[state] at (\xYM,\yNL+0.25) {$SU(N) \rightarrow U(1)^{N-1}$};

\node[state] at (\xYM,\yL) {KK modes};

\node[state] at (\xYM,\yNL-0.25) {$n_{f}$ Weyl Adjoints};

\node[state] at (\xYM,\ymphi) {Dual photons};


\node[state] at (\xF,\yNL+0.25) {$SU(N) \rightarrow U(1)^{N-1}$};
\node[state] at (\xF,\yNL-0.25) {$n_{f}$ Weyl Adjoints};

\node[state] at (\xF,\yL) {KK modes};

\node[state] at (\xF,\yNL-0.75) {$F$ Quarks};

  \draw [decorate, decoration={brace, amplitude=10pt}] 
    (\xF+1.65,\ymphi+0.5) -- (\xF+1.65,\ymphi-0.5)
       node [midway, right=9pt] {$F$};
    
\node[state] at (\xF,\ymphi+0.5) {Single-Flavor QCD};
\node[state] at (\xF,\ymphi) {Single-Flavor QCD};
\node[state] at (\xF,\ymphi-0.5) {Single-Flavor QCD};

\node[state] at (\xF,\ympion) {Pions};




\draw[scaleLine] (0.0,1.75) -- (8.0,1.75);
\node[font=\scriptsize, text=gray] at (\xYM,1.5)
    {non-perturbative};
\node[font=\scriptsize, text=gray] at (\xF,1.5)
    {non-perturbative};

\end{tikzpicture}

    \caption{Energy Spectrum of center-stabilized Yang--Mills and flavor-twisted QCD on a small circle of size $L$. The $SU(N)$ gauge group of Yang--Mills is spontaneously broken to the Cartan subgroup $U(1)^{N-1}$, resulting in $N(N-1)$
    massive off-diagonal “W-bosons” with masses $ \sim m_{W}$.  Center-stabilization is provided by $n_{f} >1$ adjoint Weyl fermions with masses $\lesssim m_{W}$. The non-perturbative infrared spectrum consists of $N-1$ dual photons of mass $m_{\phi}$, see Eq.~\eqref{eq:massmodes}. Center-stabilized twisted QCD has $F$ additional fundamental quarks at scales of order $m_{W}$. The infrared spectrum contains the massive modes of $F$ versions of single-flavor QCD and $F-1$ massless pions, which generically have mixed gauge identifications (see Section \ref{eq:lownenergychiralmixedgauge}).}
    \label{fig:energyspec}
\end{figure}
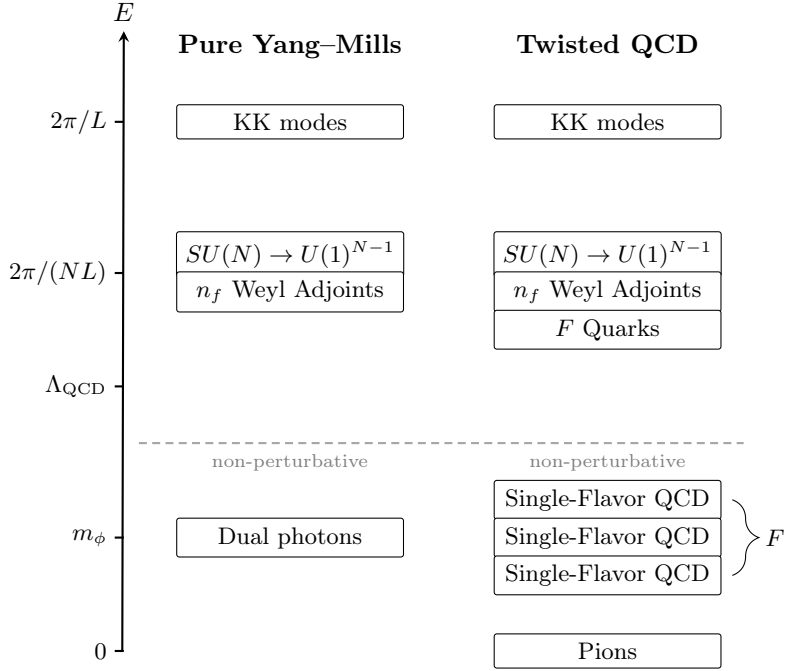

In this section, we review the infrared spectrum of center-stabilized pure gluodynamics with an emphasis on the non-perturbative  dynamics that generates it and on the symmetries of the low-energy theory. We first study the generation of a mass gap due to the formation of a semiclassical gas of monopoles.  

The vacuum expectation value of the adjoint-valued Polyakov loop in Eq.~\eqref{eq:center symmetric holonomy}  spontaneously breaks the $SU(N)$ gauge theory down to its Cartan torus $U(1)^{N-1}$. The long-distance theory abelianizes and contains $N-1$ photons. Rather than choosing an explicit $(N-1)$-dimensional basis of $\mathfrak{su}(N)$, it is convenient to embed its Cartan subalgebra into a diagonal Cartan subalgebra of $\mathfrak{u}(N)$, and use $N$ constrained components\footnote{Following Ref.~\cite{Unsal:2008ch}, we use a redundant diagonal basis of $\mathfrak{u}(N)$, with generators normalized as $\mathrm{Tr}\,(H^aH^b)=\delta^{ab}$. This differs from the conventional normalization $\mathrm{Tr}\,(T^aT^b)=\delta^{ab}/2$ of the $\mathfrak{su}(N)$ generators used in Eq.~\eqref{eq:ym}.}:
\begin{equation}
    A_i=\sum_{a=1}^N A_i^{a}H^{a}\,,\qquad H^{a}_{rs}=\delta_{ra}\delta_{sa}\,,\qquad \sum_{a=1}^{N}A^{a}_i=0\, .
     \label{eq:ndimbasis}
\end{equation}

The matrices $H^{a}$ are the diagonal generators of $\mathfrak{u}(N)$: they are not individually traceless, and their traceless linear combinations span the Cartan subalgebra of $\mathfrak{su}(N)$. The constraint in Eq.~\eqref{eq:ndimbasis} removes the direction proportional to $\sum_{a=1}^NH^{a}=\mathbf{1}_N$, leaving precisely $N-1$ independent photons. Equivalently, if the identity component is temporarily retained as part of the redundant $\mathfrak{u}(N)$ notation, it can be viewed as an overall $U(1)$ mode that decouples from the $SU(N)$ dynamics.

The off-diagonal charged gauge bosons of $SU(N)$ obtain masses of the order of $m_{W} =2\pi/(NL)$, and the gauge coupling $g^{2}$ stops running at the scale $m_{W}$. In order to be in a perturbative regime, we therefore have to demand that the compact radius $L$ satisfies
\begin{equation}
   \eta=  N L \Lambda_{\mathrm{QCD}} \ll 1 \,.
   \label{eq:etasmall}
\end{equation}

The long distance theory inherits the unbroken $\mathbb{Z}_{N}$ center symmetry in Eq.~\eqref{eq:center symmetry action on Omega}. 
The action of the center symmetry permutes the eigenvalues of the holonomy in Eq.~\eqref{eq:center symmetric holonomy} and takes it outside the fundamental Weyl chamber. In order to return to the fundamental Weyl chamber, we have to perform the inverse permutation by a gauge transformation $P \subset SU(N)$. The gauge transformation permutes the three-dimensional gauge bosons. The effect of the $\mathbb{Z}_{N}$ center symmetry on the infrared spectrum is therefore to permute the Cartan photons,
\begin{equation}
    A^{a}_{i} \rightarrow A^{a+1}_{i} \,.
    \label{eq:znsym}
\end{equation}

At the semiclassical level, the infrared dynamics of center-stabilized Yang--Mills is dictated by the existence of $N$ dynamical magnetic monopoles, which appear as instantons in three dimensions. The first $N-1$ are the ’t Hooft--Polyakov monopoles~\cite{Polyakov:1974ek,tHooft:1974kcl}, whose magnetic charges are classified by the spontaneous breaking of the $SU(N)$ gauge theory down to its Cartan subgroup $U(1)^{N-1}$:
\begin{equation}
    \pi_2\big(SU(N)/U(1)^{N-1}\big)\cong \pi_1\big( U(1)^{N-1}\big)\cong \mathbb{Z}^{N-1}\, .
\end{equation}

The magnetic charges of the monopoles lie in the co-root lattice of $SU(N)$, which coincides with its root lattice. A convenient basis for the  ’t Hooft--Polyakov monopoles is formed by the $N-1$ $U(1)$ subsets of $SU(N)$ generated by the diagonal Cartan matrices $H^{a} - H^{a+1}$. In this basis, magnetic monopoles carry unit magnetic charge under the gauge theory associated with $A^{a}_{i}$ and negative unit magnetic charge under the gauge theory associated with $A^{a+1}_{i}$.

The magnetic monopoles naturally couple to the magnetic duals of the Cartan photons. In light of this, we choose to describe the infrared theory in terms of the  $N$ constrained magnetic dual variables, the dual photons $\sigma_{a}$, defined as
\begin{equation}
    \frac{g^{2}}{2\pi L}\partial_{i}\sigma_{a} = \epsilon_{ijk}  \partial_{j}A^{a}_{k}\,, \qquad  \sum_{a=1}^{N} \sigma_{a} = 0 \,.
    \label{eq:dualphotons}
\end{equation}

The action of the dual photons can be obtained from Eq.~\eqref{eq:dualphotons}, starting from the original action of the $SU(N)$ Yang--Mills field in Eq.~\eqref{eq:ym}, and has the form\footnote{Here $\vec F\equiv(F^{1},\ldots, F^{N})$ denotes the collection of Abelian field-strength $F^{a}=dA^{a}$. The tracelessness constraint in Eq.~\eqref{eq:ndimbasis} implies $\sum_{a=1}^{N}F^{a}=0$.}
\begin{equation}
    S \supset -\frac{L}{4g^2}\int d^3x\,\sum_{a=1}^N F_{ij}^{a}F_{ij}^{a}=\int d^{3}x \   \frac{g^{2}}{8 \pi^{2}L}\ \sum_{a=1}^N\partial^{\mu} \sigma_{a}   \partial_{\mu} \sigma_{a} \,.
    \label{eq:dualphotonkin}
\end{equation}
The dual action inherits all the symmetries of the original theory. The $\mathbb{Z}_{N}$ center symmetry (Eq.~\eqref{eq:znsym}) acts on the dual variables by a permutation,
\begin{equation}
    \sigma_{a} \rightarrow \sigma_{a+1} \,.
    \label{eq:dualphotonperm}
\end{equation}

The coupling of the dual photons to the $N-1$ magnetic monopoles can be represented by the insertion of a monopole vertex of the form~\cite{Unsal:2008ch}
\begin{equation}
    \mathcal{M}_k(x)\sim e^{-S_0}e^{i\vec{\alpha}_{k} \cdot \vec{\sigma}(x)}\,,   \qquad   S_0=\frac{8\pi^2}{g^2 N}\,, \qquad k=1,\ldots, N-1\, \,,
    \label{eq:magneticshift}
\end{equation}
where $\vec{\alpha}_k$ is the simple root of $\mathfrak{su}(N)$ such that $\vec{\alpha}_{k} \cdot \vec{\sigma} = \sigma_{k} - \sigma_{k+1}$. The latter provides the correct coupling for a monopole of unit magnetic charge with respect to dual photon $\sigma_{j}$ and negative unit magnetic charge with respect to $\sigma_{j+1}$. Here $S_{0}$ is the semiclassical action cost of the ’t Hooft--Polyakov monopole. 

Beyond the $N-1$ ’t Hooft--Polyakov monopoles we have just described, the compactified $SU(N)$ Yang--Mills theory on $\mathbb{R}^{3}\times S^{1}$ gives one additional fundamental Kaluza--Klein monopole~\cite{Lee:1997vp, Lee:1998bb,Davies:1999uw,Csaki:2017cqm}. Its magnetic charge is the affine co-root, which coincides with the affine root for $SU(N)$. The existence of this $N$-th monopole can be derived from the $\mathbb{Z}_{N}$ permutation symmetry of the dual photons: in order for the partition function to be invariant under the $\mathbb{Z}_{N}$ permutation of the dual photons (Eq.~\eqref{eq:dualphotonperm}), an additional Kaluza--Klein monopole vertex must exist with the coupling
\begin{equation}
    \mathcal{M}_N(x)\sim e^{-S_0}e^{i \vec{\alpha}_{N} \cdot \vec{\sigma}(x)} \qquad \mathrm{and} \qquad \vec{\alpha}_{N} \cdot \vec{\sigma} = \sigma_{N}-\sigma_{1}\,.
    \label{eq:KKmon}
\end{equation}

The field profiles of the monopoles consist of both a magnetic field associated with the three-dimensional gauge fields and a winding of the Higgs-field, the adjoint-valued Polyakov loop $\Omega(x)$, at spatial infinity. A winding in the Polyakov loop, the “fourth” gauge boson, is associated with an electric field in the four-dimensional $SU(N)$ gauge theory. The combined magnetic and electric fields make the magnetic monopoles instantons in the four-dimensional $SU(N)$ theory. A careful calculation reveals that the magnetic monopoles carry fractional $1/N$ four-dimensional instanton charge~\cite{Kraan:1998sn}, as can also be seen from the monopole action in Eq.~\eqref{eq:magneticshift}. The ordinary four-dimensional integer-valued instantons therefore consist of a chain of the $N$ distinct magnetic monopoles.

The existence of the dynamical magnetic monopoles explicitly breaks a set of emergent $U(1)^{N-1}_{\mathrm{mag}}$ $0$-form magnetic symmetries of the infrared spectrum, that are not present in the ultraviolet theory. The $U(1)^{N-1}_{\mathrm{mag}}$ magnetic symmetries measure the magnetic charges of probe particles and are associated with the conserved currents $\vec j_m=\vec F/(2\pi)$, whose conservation follows perturbatively from $d\vec F=0$. The magnetic symmetries act on the dual photons by a linear shift with zero trace:
\begin{equation}
    \sigma_{a} \rightarrow \sigma_{a} + c_{a}\,, \qquad \mathrm{with} \qquad \sum_{a=1}^{N} c_{a} = 0\mod 2\pi\,.
    \label{eq:sigma-shifts}
\end{equation}
The semiclassical gas of monopoles and anti-monopoles generates a potential for the dual photons~\cite{Unsal:2008ch} 
\begin{equation}
    V_{\rm mon}(\vec \sigma)=-\zeta \sum_{a=1}^N\cos\left(\sigma_{a} - \sigma_{a+1}\right) \,,\qquad \zeta = A \lambda^{-2} m_{W}^{3} e^{- S_{0}} \sim\frac{\Lambda_{\rm QCD}^3}{\lambda^2}(NL\Lambda_{\rm QCD})^{(2-2n_f)/3}  \,,    \label{eq:potentialym}
\end{equation}
where $A$ is an order $1$ number that depends on the regularization scheme, and we defined the ’t Hooft coupling $\lambda = N g^{2}$. The sum also makes use of the periodic definition $\sigma_{N+1} \equiv \sigma_{1} $. 

The monopole gas explicitly breaks the emergent $U(1)^{N-1}_{\mathrm{mag}}$ magnetic shift symmetries, and the dual photons acquire a mass. This is the standard Polyakov mechanism \cite{Polyakov:1976fu} for confinement in $(2+1)$-dimensional gauge theories with compact gauge groups. The dual photon mass screens magnetic fields and confines electric fields into flux tubes. The mass of the dual photon can be obtained by diagonalizing the potential~\eqref{eq:potentialym} and kinetic term~\eqref{eq:dualphotonkin} around the minima, resulting in $N-1$ excitations of masses \cite{Unsal:2008ch}
\begin{equation}
   m_{k}^{2} = m_{\phi}^{2}  \sin^2{\left(\frac{\pi k}{N} \right)}\,, \qquad   m_{\phi}^2 = C\Lambda_{\mathrm{QCD}}^{2} \  \lambda^{-3} \ \eta^{\gamma} \sim\frac{\Lambda_{\rm QCD}^2}{\lambda^3}(NL\Lambda_{\rm QCD})^{(5-2n_f)/3}\,,
   \label{eq:massmodes}
\end{equation}
where $k=1,\dots, N-1$.
Here, $C$ is a dimensionless coefficient of $\mathcal{O}(1)$, while $\gamma = b_{0}/N-2 =\left( 5-2n_{f} \right)/3$ depends on the running of the gauge coupling and on the number of adjoint fermions needed for the center-stabilization mechanism. The definition of $\eta \ll 1$ is in Eq.~\eqref{eq:etasmall}. Interestingly, the infrared spectrum of center-stabilized Yang--Mills in the double-scaling limit $N\rightarrow \infty$ and $\eta$ fixed becomes gapless at large $N$~\cite{Cherman:2016jtu}. The non-perturbative, massive dual photon modes complete the energy spectrum of center-stabilized pure Yang--Mills in Fig.~\ref{fig:energyspec}.  

The $U(1)^{N-1}$ gauge theory of dual photons also has a set of gauge identifications inherited from the $SU(N)$ gauge theory. These gauge identifications are most conveniently described in the monopole coordinates $\vec{\alpha}_{i} \cdot \vec\sigma $, which satisfy the constraint $\sum_{i=1}^{N} \vec{\alpha}_{i} \cdot \vec\sigma =0$. In this basis, a gauge identification corresponds to shifting a single monopole coordinate $\vec{\alpha}_{i} \cdot \vec\sigma$ by $2\pi$ and another single monopole coordinate $\vec{\alpha}_{j \neq i} \cdot \vec\sigma$ by $-2\pi$,
\begin{equation}
    \vec{\alpha}_{i} \cdot \vec\sigma \rightarrow \vec{\alpha}_{i} \cdot \vec\sigma + 2\pi \qquad \mathrm{and} \qquad \vec{\alpha}_{j \neq i} \cdot \vec\sigma \rightarrow \vec{\alpha}_{j \neq i} \cdot \vec\sigma - 2\pi\,.
    \label{eq:gaugetransmagmon}
\end{equation}
It can be seen from Eq.~\eqref{eq:magneticshift} that this gauge identification is compatible with the form of the monopole vertices and the dual photon potential in Eq.~\eqref{eq:potentialym}. In Sections \ref{sec:deconfzfdw} and \ref{sec:deconfnewzf}, we use these gauge transformations to identify the color charge of the deconfined particle on the $\mathbb{Z}_{F}$ domain wall.

\subsection{Color-Flavor Twisted Boundary Conditions}
\label{sec:cfctwist}

In Section~\ref{sec:Center-Stabilized pure Yang--Mills theory}, we discussed the infrared dynamics of center-stabilized pure gluodynamics. A key ingredient was the $\mathbb{Z}_{N}$ center symmetry. In sharp contrast to pure Yang--Mills, QCD compactified on a small circle generically admits no nontrivial order parameters for the center symmetry like the Polyakov loop. This is because the fundamental fermions screen color charge and explicitly break the center symmetry.  In this section, we show that the infrared dynamics of $3d$ QCD depends strongly on the boundary conditions of the fundamental fermions around the compact direction, as previously discussed in Refs.~\cite{Cherman:2016hcd, Cherman:2017tey}. Whenever $\mathrm{gcd}(N,F)>1$, a remnant $\mathbb{Z}_{\mathrm{gcd}(N,F)}$ color-flavor-center symmetry plays the role of the original $\mathbb{Z}_{N}$ center symmetry and provides new order parameters. In Section \ref{sec:deconfnewzf}, we use this  symmetry to show that probe color charges are deconfined on $\mathbb{Z}_{F}$ domain walls. 

Quantum chromodynamics consists of an $SU(N)$ Yang--Mills theory and $F$ fundamental fermions. We take the fundamental fermions $\psi_j$ to be massless and the high-energy Lagrangian to be
\begin{equation}
    S = \int d^{4}x \left[  - \  \frac{1}{2g^{2}} \mathrm{Tr}\,G_{\mu \nu}G^{\mu \nu} + \sum_{j=1}^Fi\bar{\psi}_{j} \slashed D \psi_{j} \right]\,.
    \label{eq:QCD}
\end{equation}

The addition of $F$ fundamental fermions modifies the global symmetry structure of the theory. In $4d$ for zero quark masses, the fermions have a faithfully acting symmetry group
\begin{equation}
    G_F=\frac{U(F)_L\times U(F)_R}{\mathds{Z}_N}\,,
    \label{eq:flavorsym}
\end{equation}
where $U(F)_L$ and $U(F)_{R}$  transform left- and right-handed fermions, respectively. The $\mathbb{Z}_{N}$ quotient is generated by the element $(e^{2\pi i/N} \mathds{1}_F,\,e^{2\pi i/N} \mathds{1}_F)$. This quotient specifies that part of the symmetry group is actually a gauge identification in the $SU(N)$ gauge theory.

The left and right-handed symmetries can be decomposed into $U(F)_{V}$ vector and $U(F)_{A}$ axial symmetries. The  $U(1)_{A}$ diagonal subgroup of $U(F)_{A}$ is explicitly broken by the chiral anomaly and plays an important role in the dynamics of the infrared spectrum.\footnote{For readers not used to lower dimensions, it is useful to recall that a $4d$ Dirac fermion $\psi=(q_{\alpha},q^{c\dagger\dot\alpha})$ gives two independent two-component fermions in $3d$, $q$ and $q^c$, at each Kaluza--Klein level. The $U(1)_A$ and $U(1)_V$ transformations continue to act in $3d$ as in $4d$: $U(1)_A$ acts as $q\to e^{i\alpha}q$, $q^c\to e^{i\alpha}q^c$, while $U(1)_V$ acts as $q\to e^{i\beta}q$, $q^c\to e^{-i\beta}q^c$. Although irreducible $3d$ spinors have no chirality and the inherited singlet axial rotation acts as a common phase rotation of $q$ and $q^c$, $U(1)_A$ remains anomalous. Its explicit breaking by the $4d$ anomaly is reproduced in the semiclassical theory by the monopole-instanton vertices. When background sources are included, the inherited anomalies can also be encoded in holonomy-dependent Chern--Simons and mixed topological terms.}

As in the pure Yang--Mills case, we work at small circle size and stabilize the gauge holonomy near the center-symmetric value $\Omega_{\rm cs}$ by introducing a sufficient number $n_f$ of periodic adjoint Weyl fermions, with appropriate masses.
The gauge group is abelianized, $SU(N)\to U(1)^{N-1}$, and the long-distance dynamics is described by the Cartan photons, or equivalently by the $N-1$ dual photons $\vec\sigma$ in Eq.~\eqref{eq:dualphotons}. The new ingredient relative to the pure-glue theory is the choice of boundary conditions around the compact direction for the fermions. A choice of boundary conditions can be described by the flavor matrix $U$ as 
\begin{equation}
    \psi(x,x_3+L)=U\, \psi(x,x_3)\,, \qquad U\in U(F)_V\,.
    \label{eq:flavor-twisted-bc}
\end{equation}
Different choices of $U$ define different compactified theories. Constant vector flavor rotations act on $U$ by conjugation, so only its conjugacy class is physical. We choose a convenient flavor basis in which
\begin{equation}
    U=\mathrm{diag}\left(e^{i\kappa_1},\ldots, e^{i\kappa_F} \right)\, .
    \label{eq:generalflavor}
\end{equation}
For generic distinct phases $\kappa_j$, the twisted boundary conditions explicitly break the non-Abelian flavor symmetries of Eq.~\eqref{eq:flavorsym} down to their Cartan subgroup
\begin{equation}
    U(1)^{F-1}_{\cal V} \times U(1)^{F-1}_{\cal A} \times U(1)_{V} \times U(1)_{A}\,,
    \label{eq:u1s}
\end{equation}
where we have separated the overall $U(1)_{V}$ phase of the vector symmetry and the overall $U(1)_{A}$ axial phase. The curly $\cal V$ and $\cal A$ represent the orthogonal Cartan $U(1)$'s. Importantly, the fact that generic boundary conditions break the full vector and axial symmetries of Eq.~\eqref{eq:flavorsym} down to the diagonal subgroups means that we are only able to describe the Chiral Lagrangian involving the pions associated with purely diagonal axial transformations. 

In the special case where the number of colors $N$ and the number of fundamental flavors $F$ are not coprime, that is $\mathrm{gcd}(N,F) >1$, special flavor twisted boundary conditions for the fermions around the compact direction can result in a color-flavor-center (CFC) $\mathbb{Z}_{\mathrm{gcd}(N,F)}$ symmetry~\cite{Briceno:2013hya,Liu:2016yij,Cherman:2016vpt,Cherman:2016hcd, Cherman:2017tey}. The CFC symmetry is the remnant of the $\mathbb{Z}_{N}$ center symmetry of the pure-glue theory in the presence of fundamental matter and flavor-twisted boundary conditions~\cite{Cherman:2017tey}. 

To recover the CFC in the case $\mathrm{gcd}(N,F) >1$, observe that a center transformation by $z\in \mathbb{Z}_N$ acts as a flux through the compact circle and changes the boundary condition of the fundamental fermions as
\begin{equation}
    \psi(x,x_3+L)=U\, \psi(x,x_3)\longmapsto \psi(x,x_3+L)=z\, U\, \psi(x,x_3)\, .
    \label{eq:colorcenterferm}
\end{equation}

For ordinary periodic boundary conditions, this is not a symmetry: the fundamental quarks explicitly break the Yang--Mills center symmetry by screening the color charge.  However, for special flavor-twisted boundary conditions, some center actions can be compensated by flavor permutations. Consider the following flavor-center symmetric boundary conditions around the compact direction:
\begin{equation}
    U_{\rm F}= e^{i \theta_{V}}e^{-\pi  i(F-1)/F}\mathrm{diag}\left(1,\nu,\nu^2,\ldots, \nu^{F-1}\right)\,, \qquad \nu=e^{2\pi i/F}\,.
    \label{eq:flavor-center-twist}
\end{equation}
Here $\theta_{V}$ is a background holonomy angle for the vector $U(1)_V$ symmetry, normalized so that each quark has unit charge. This common phase shifts the quark Kaluza--Klein momenta without affecting the CFC symmetry. We choose it so that the eigenvalues of the flavor holonomy $U_{\rm F}$ remain sufficiently separated from those of the center-symmetric color holonomy $\Omega_{\rm cs}$ in Eq.~\eqref{eq:center symmetric holonomy}, ensuring that the perturbative quark modes remain heavy compared with the infrared scales in the dual-photon description used below~\cite{Cherman:2016hcd,Aitken:2017ayq}.\footnote{For massless four-dimensional quarks, the tree-level three-dimensional masses for the Kaluza--Klein modes are~\cite{Cherman:2016hcd,Aitken:2017ayq}
\[
M_{a,j,n}=\frac{|2\pi n +\kappa_j -\vartheta_a|}{L}\,,\qquad \kappa_j=\theta_V-\frac{\pi(F-1)}{F}+\frac{2\pi (j-1)}{F}\,,\qquad j=1,\ldots, F
\]
where $a=1,\ldots, N$, $j=1,\ldots, F$, $n\in \mathbb{ Z}$, and $e^{i\vartheta_a}$ are the eigenvalues of $\Omega_{\rm cs}$. A mode becomes massless when $\kappa_j=\vartheta_a\mod 2\pi$. These perturbative Kaluze--Klein modes should not be confused with the fermion zero modes localized on monopole-instantons.
} 

The action of a color-center transformation in Eq.~\eqref{eq:colorcenterferm} by the color-center element $ z_d=e^{2\pi i/d}=e^{2\pi i\, (N/d)/N}\in \mathbb{Z}_N$ with $d=\gcd(N,F)$ changes the boundary-condition matrix as~\cite{Cherman:2017tey}
\begin{equation}
    U_{\rm F}\longmapsto z_d\,U_{\rm F}=e^{i \theta_{V}}e^{-\pi  i(F-1)/F}\mathrm{diag}\left(\nu^{F/d},\nu^{F/d+1},\ldots, \nu^{F/d+F-1} \right)\, .
    \label{eq:flavorcentersymmetric}
\end{equation}

This is the same set of eigenvalues as the original $U_{\rm F}$, just cyclically permuted by the integer $F/d$. Therefore the color-center transformation can be undone by a flavor permutation. In the abelianized variables of Eq.~\eqref{eq:dualphotons}, the generator of this combined symmetry acts by a simultaneous cyclic permutation of color labels of the dual photons and the flavor labels of the fundamental fermions,
\begin{equation}
    \sigma_{a}\longmapsto \sigma_{a+N/d}\,, \qquad \psi_j\longmapsto \psi_{j+F/d}\, ,
    \label{eq:CFC-action}
\end{equation}
with indices understood modulo $N$ and $F$, respectively. The Polyakov loop $\Omega$ and gauge invariant fermion bilinears are order parameters of the CFC symmetry and can be used to characterize the distinct phases of the infrared theory. In the heavy-quark limit, $m\gg \Lambda_{\rm QCD}$, the quarks decouple and the CFC symmetry reduces to the ordinary $\mathbb{Z}_{N}$ Yang--Mills center symmetry. 

The low energy spectrum and its symmetries strongly depend on the choice of the flavor twist matrix $U$. In what follows, we  work with generic, non-zero flavor twists for  $F$ and $N$ whenever $\mathrm{gcd}(N,F)=1$, but with a CFC-respecting flavor twist whenever $F$ and $N$ are not coprime. The resulting CFC symmetry whenever $\mathrm{gcd}(N,F)>1$ will be responsible for the deconfinement on the $\mathbb{Z}_{F}$ domain wall, as discussed in Section~\ref{sec:deconfnewzf}.

\subsection{Center-Stabilized QCD}
\label{sec:centerstabQCDmon}
In this section, we combine the results of the previous sections to discuss the infrared physics of center-stabilized QCD in the presence of $F$ fundamental fermions. The corresponding energy spectrum is presented on the right column of Fig.~\ref{fig:energyspec}. The infrared dynamics is naturally described by the $N$ constrained dual photons defined in Eq.~\eqref{eq:dualphotons}. For generic flavor-twisted boundary conditions, the fermions (except their zero modes) acquire masses of order of the compactification radius $\sim 1/(NL)$  due to the Polyakov loop and the flavor boundary conditions, and are absent from the light perturbative spectrum on a small circle, as shown in Fig.~\ref{fig:energyspec}. However, the fermions can become light near the $N$ monopole cores, and their zero modes play an important role in the non-perturbative breaking of the $U(1)_{A}$ axial symmetry.

A four-dimensional instanton carries $2F$ fundamental fermion zero modes, as dictated by the Atiyah--Singer index theorem~\cite{Atiyah:1968mp}. On $\mathbb{R}^3\times S^1$, the instanton decomposes into its $N$ fundamental monopole constituents, and the Callias--Nye--Singer index theorem determines how the fermion zero modes are distributed among them~\cite{Nye:2000eg,Poppitz:2008hr,Cherman:2016hcd}. The distribution depends crucially on the fermion boundary conditions around the circle.

For trivial boundary conditions, the zero-mode distribution is flavor-blind and all $2F$ fundamental zero modes are localized on the same monopole. Choosing labels so that this monopole is $\mathcal{M}_1$, the corresponding monopole operators take the schematic form
\begin{equation}
    \mathcal{M}_1\sim e^{-S_0}e^{i \vec{\alpha}_{1} \cdot \vec{\sigma}}\det_{a,b}\big(\bar\psi_{L,a}\psi_{R,b}\big)\,, \qquad \mathcal{M}_k\sim e^{-S_0}e^{i\vec{\alpha}_{k} \cdot \vec{\sigma}}\,,\quad k=2,\ldots,N\, .
    \label{eq:njllike}
\end{equation}

Importantly, in the small-circle regime, the small coupling $e^{-S_0}$ in the would-be Nambu--Jona-Lasinio-like Lagrangian~\cite{Nambu:1961tp,Nambu:1961fr} induced by the monopole vertices of Eq.~\eqref{eq:njllike} does not result in chiral symmetry breaking: the semiclassical dynamics leaves the non-Abelian $SU(F)_A$ axial symmetry unbroken~\cite{Unsal:2008eg}. This is in sharp contrast to ordinary QCD.

Flavor-twisted boundary conditions change this conclusion by explicitly breaking the off-diagonal part of the $SU(F)_{A}$ axial symmetry, allowing the fermion zero modes to disperse over the monopoles and generating a fermionic condensate, as we show in Section~\ref{sec:condensate}. This results in a low-energy dynamics qualitatively similar to that of ordinary QCD. 

\begin{figure}[t]
    \centering
    \includegraphics[width=0.9\linewidth]{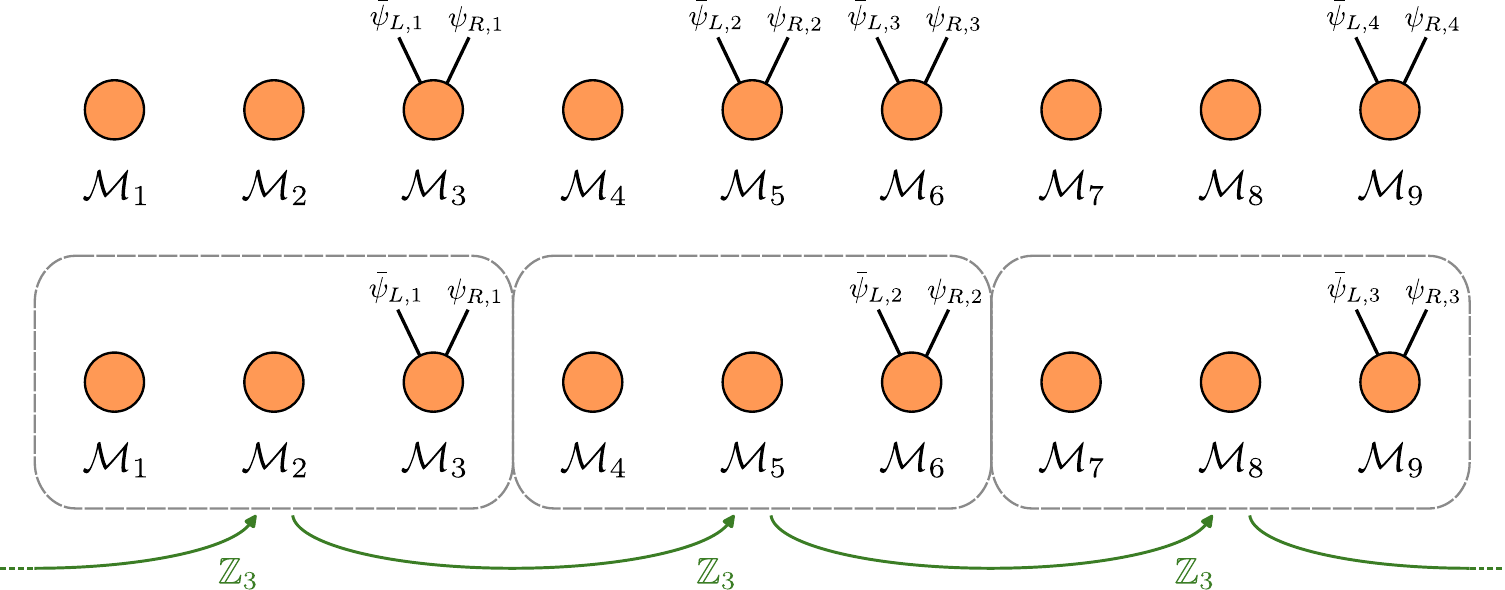}
    \caption{Distribution of fermion zero modes for $N=9$, $F=4$ (top) and $F=3$ (bottom). When $\mathrm{gcd}(N,F)=1$ (top), the distribution does not show any symmetry pattern. Conversely, when $\mathrm{gcd}(N,F)\ne1$ (bottom), the monopole lattice has a CFC-symmetric pattern and consists of $\mathrm{gcd}(N,F)$ unit cells.  Under the action of the $\mathbb{Z}_{\mathrm{gcd}(N,F)}$ CFC symmetry, the unit cells are permuted. }
    \label{fig:fermion-disttrib}
\end{figure}

For generic flavor-twisted boundary conditions and assuming $F< N$, $F$ monopoles carry one chiral pair of fundamental zero modes each, while the remaining $N-F$ monopoles are purely bosonic. We denote the set of monopole labels that carry fundamental zero modes by $\mathcal{F}=\{n(j)~|~j=1,\ldots, F\}$, where $j$ labels a zero mode pair and $n(j)$ is the label of the monopole that the $j$-th pair sits on. We denote the set of monopoles without fermion zero modes as the complement $\mathcal{B}$. For generic flavor-twists with $F< N$, these sets have sizes $|\mathcal{F}|=F$ and $|\mathcal{B}|=N-F$. The monopole vertices are then 
\begin{equation}
    \mathcal{M}_{n(j)}\sim e^{-S_0}e^{i\vec{\alpha}_{n(j)} \cdot \vec{\sigma}}\left(\bar\psi_{L,j}\psi_{R,j}\right)\,, \quad n(j) \in \mathcal{F} \,,\qquad \mathcal{M}_{b}\sim e^{-S_0}e^{i \vec{\alpha}_{b} \cdot \vec{\sigma}}\,, \quad b \in \mathcal{B}\,.
    \label{eq:monopolefermionvertex}
\end{equation}
The distribution according to the Callias--Nye--Singer index theorem for the case $N=9$ and $F=4$ with arbitrarily chosen flavor phases $\kappa_{j}$ is drawn in Fig.~\ref{fig:fermion-disttrib} (top).

Whenever $\mathrm{gcd}(N,F) >1$, the distribution of fermion zero modes over the monopoles is CFC symmetric~\cite{Cherman:2016hcd} due to the flavor-center-symmetric twist of Eq.~\eqref{eq:flavorcentersymmetric}. In this case, the fermion zero mode distribution is best described by first dividing the lattice of $N$ monopoles into $\mathrm{gcd}(N,F)$ unit cells  each containing $N/\mathrm{gcd}(N,F)$ monopoles, see Fig.~\ref{fig:fermion-disttrib} (bottom) for the case $N=9$ and $F=3$, which has $3$ unit cells with $3$ monopoles per cell.

Each unit cell contains $F/\mathrm{gcd}(N,F)$ fermion zero mode pairs. Within a unit cell, the distribution of the fermion zero modes over the monopoles depends on the  flavor-center-symmetric twist. However, crucially, for an arbitrary CFC-symmetric twist, the fermion zero mode pattern is repeated across all the unit cells. Reproducing the unit cell $\mathrm{gcd}(N,F)$ times yields the whole monopole lattice, as can be seen in Fig.~\ref{fig:fermion-disttrib} (bottom) for the case $N=9$ and $F=3$.

This repetition of the fermion zero modes across unit cells is consistent with the color-flavor-center $\mathbb{Z}_{\mathrm{gcd}(N,F)}$ symmetry (Eq.~\eqref{eq:CFC-action}). Under this permutation symmetry, the unit cells including the fermion zero modes each move by one unit cell. This can be seen in Fig.~\ref{fig:fermion-disttrib} (bottom) which has a $\mathbb{Z}_{3}$ permutation symmetry that moves the monopoles and fermion zero modes by one unit cell.

We now study the continuous symmetries of twisted QCD by repeating the discussion on the magnetic symmetries in Section~\ref{sec:magsym} in the presence of fermion zero modes. As in pure gluodynamics, the monopole gas explicitly breaks the perturbative $U(1)^{N-1}_{\mathrm{mag}}$ magnetic shift symmetries of the dual photons. The 
$U(1)^{F-1}_{\mathcal{A}} \subset SU(F)_{\mathcal{A}}$ traceless axial symmetries, that remain after the explicit breaking by the flavor-twist, act on fermion bilinears as
\begin{equation}
    \bar\psi_{L,j}\psi_{R,j}\longmapsto e^{i\epsilon_j} \bar\psi_{L,j}\psi_{R,j}\,,\qquad\sum_{j=1}^F\epsilon_j=0\,.
    \label{eq:axial transformation}
\end{equation}
The monopole vertices $\mathcal{M}_{n(j)}$ carrying fermion zero modes remain invariant under the $U(1)^{F-1}_{\mathcal{A}}$ axial symmetry if the fermion transformation is accompanied by the dual-photon shift
\begin{equation}
    \vec{\alpha}_{n(j)} \cdot \vec{\sigma}\longmapsto \vec{\alpha}_{n(j)} \cdot \vec{\sigma}-\epsilon_j\,, \qquad j\in 1,\dots,F\,.
    \label{eq:pionshift}
\end{equation}
As the transformation is already traceless, it preserves the mathematical identity $\sum_{i=1}^{N} \vec\alpha_{i} \cdot \vec\sigma =0$, and the infrared spectrum has a $U(1)^{F-1}_{\mathcal{A}}$ set of axial shift symmetries. By the gauge transformations of the magnetic monopole coordinates in Eq.~\eqref{eq:gaugetransmagmon}, we see that $U(1)^{F-1}_{\mathcal{A}}$  transformations involving only $2\pi$ valued shifts, $\epsilon_i/2\pi \in  \mathbb{Z}$, are gauge identifications in the low energy theory.

The $U(1)_{A}$ axial symmetry that transforms all fermions by the same phase, however, is different. Suppose that $\epsilon_j=\epsilon$ is constant for $j=1,\ldots, F$. To keep the zero-mode monopole vertices invariant, one would like to shift the coordinates on those monopoles as,
\begin{equation}
    \vec{\alpha}_{n(j)} \cdot \vec{\sigma}\longmapsto \vec{\alpha}_{n(j)} \cdot \vec{\sigma}-\epsilon\qquad \forall j \in 1,\dots,F\,.
\end{equation}
For $N>F$, this shift does not satisfy the constraint $\sum_{j} \vec{\alpha}_{j} \cdot \vec{\sigma} =0$, unless one also shifts the  coordinates of a single monopoles without fermion zero modes by, 
\begin{equation}
    \vec{\alpha}_{b} \cdot \vec{\sigma}\longmapsto \vec{\alpha}_{b} \cdot \vec{\sigma} + F\epsilon\,.
    \label{eq:heavyetamodes}
\end{equation}
This shift does not leave that particular monopole vertex invariant and the  continuous $U(1)_A$ symmetry is not an exact symmetry of the theory. Note that there are $N-F$ different orthogonal choices for the $U(1)_{A}$ symmetry, each associated with a single heavy monopole coordinate $\vec{\alpha}_{b} \cdot \vec{\sigma}$ that shifts as in Eq.~\eqref{eq:heavyetamodes}. In this sense, we say the that the resulting massive modes of the infrared spectrum collectively act as an $\eta'$ meson. 

The fact that all $N-F$ different choices for the $U(1)_{A}$ symmetry are explicitly broken is the semiclassical realization of the four-dimensional axial anomaly. Indeed, the product of all $N$ monopole vertices has the quantum numbers of an ordinary four-dimensional instanton. Schematically, this product becomes the standard four-dimensional instanton vertex
\begin{equation}
    \prod_{k=1}^N\mathcal M_k\sim e^{-8\pi^2/g^2} \det_{i,j}\left(\bar\psi_{L,i}\psi_{R,j}\right).
    \label{eq:etainstanton}
\end{equation}

Thus, the full set of monopole constituents reproduces the usual four-dimensional breaking of the continuous $U(1)_A$ symmetry to its discrete remnant $U(1)_A\to \mathbb Z_{2F}$. The $\mathbb{Z}_{2F}$ symmetry  acts as a transformation with $\epsilon =2\pi/F$ on the monopoles with fermion zero modes, and a $-2\pi$ transformation on a single monopole coordinate without fermion zero modes. As shown in Eq.~\eqref{eq:heavyetamodes}, the latter transformation leaves the monopole vertex invariant. The fact that there are $N-F$ choices for this $\mathbb{Z}_{2F}$ symmetry will result in several different $\mathbb{Z}_{F}$ domain walls, as detailed in Section \ref{sec:deconf}. Indeed, after taking the quotient by the common fermion-parity or baryon-number element, the discrete axial symmetry acting faithfully is reduced from $\mathbb{Z}_{2F}$ to $\mathbb{Z}_F$.

\section{Chiral Lagrangian \& Deconfinement }
\label{sec:deconf}

In the following, we use the semiclassical description developed in the previous sections to construct the infrared spectrum of center-stabilized QCD with flavor-twisted boundary conditions.
The spectrum consists of $F-1$ massless pions, and $N-F$ heavier states, which we organize into $F$ versions of single-flavor QCD. Each version has a $\mathbb{Z}_{F}$ domain wall and we demonstrate that probe particles in representations of $N$-ality $N/\mathrm{gcd}(N,F)$ are deconfined on the wall whenever $N$ and $F$ are not coprime.

\subsection{Degree-of-freedom transmutation}\label{sec:condensate}
All fundamental quarks and the gauge bosons associated with broken generators acquire  masses of order $1/(NL)\sim m_W$ through both the color holonomy (Eq.~\eqref{eq:center symmetric holonomy}) and flavor boundary conditions (Eq.~\eqref{eq:generalflavor}). Hence, they are absent from the infrared spectrum, and the light perturbative theory below $m_W$ is described by the Cartan photons, or equivalently by the $N-1$ dual photons $\vec\sigma$. The fermion zero modes attached to the monopoles are not additional massless particles in this spectrum. They are normalizable zero-eigenfunctions of the Dirac operator in the monopole background and enter the corresponding saddle-point measure as Grassmann collective coordinates. 

Nevertheless, the microscopic gauge-invariant bilinear $\bar\psi_{L,j}\psi_{R,j}$ remains a well-defined local probe and its expectation value in the vacuum acts as an order parameter for chiral symmetry breaking. The order parameter can be calculated by performing a path integral in the semiclassical monopole background with an insertion of the operator $\bar\psi_{L,j}\psi_{R,j}$. This is equivalent to `closing up' the fermion zero mode legs in Fig.~\ref{fig:fermion-disttrib}. Evaluating the path integral, one finds that the semiclassical monopole gas induces the formation of a condensate in the fermion bilinear,
\begin{equation}
    \langle\bar\psi_j\psi_j\rangle \sim -2 \zeta \, \mathrm{Re}\, \langle e^{i \vec{\alpha}_{n(j)} \cdot \vec{\sigma}}\rangle_{\rm IR}\,,
    \label{eq:chiralcondensate}
\end{equation}
where $\zeta$ is defined in Eq.~\eqref{eq:potentialym}, and we are suppressing scheme-dependent regularization constants. In addition, we have indicated by $\langle \dots \rangle_\text{IR}$ the expectation value obtained by weighting the functional integral with the infrared action. Note that the fermion bilinear $\bar \psi_{L,j} \psi_{R,j}$ can be matched to $e^{i\vec\alpha_j \cdot \vec\sigma}$ at the operator level, beyond the equality of expectation values in Eq.~\eqref{eq:chiralcondensate}. This operator relationship is the precise sense in which the microscopic fermionic order parameter reappears as a magnetic variable in the infrared.

The formation of a fermionic condensate spontaneously breaks the $U(1)^{F-1}_{\mathcal{A}}$ axial symmetries (Eq.~\eqref{eq:u1s}) of the fermions. As a consequence, the resulting low energy spectrum contains $F-1$ Nambu--Goldstone bosons, i.e. the pions. The fermion condensate also spontaneously breaks the $\mathbb{Z}_{2F}$ unbroken axial subgroup of $U(1)_{A}$ down to $\mathbb{Z}_{2}$. Overall, the Chiral Lagrangian follows from the symmetry breaking pattern 
\begin{equation}
    U(1)^{F-1}_{\mathcal{V}}  \times U(1)^{F-1}_{\mathcal{A}} \times U(1)_{V} \times \mathbb{Z}_{2F}  \rightarrow U(1)^{F-1}_{\mathcal{V}} \times U(1)_{V} \times \mathbb{Z}_{2}   \,.
    \label{eq:symmetrybreaking}
\end{equation}
The absence of the full set of non-Abelian axial and vector symmetries in the symmetry breaking pattern of Eq.~\eqref{eq:flavorsym} is due to the flavor twisted boundary conditions. Consequently, the Chiral Lagrangian of twisted QCD does not lie in the same universality class as the four-dimensional Chiral Lagrangian consisting of $F^{2}-1$ pions. The absence of the off-diagonal flavor matrices also means that there is a less constrained mixing pattern among the pions' kinetic terms. 

Importantly, the connection between twisted QCD and ordinary QCD goes beyond the chiral symmetry breaking pattern. As shown in Appendix~\ref{app:cartan-cfc-anomalies}, for $F>1$ the four-dimensional mixed chiral-baryon-color-flavor anomaly remains nontrivial on backgrounds for the remaining faithful Cartan vector subgroup $U(1)^F/\mathbb Z_N \subset U(F)_{V}/\mathbb{Z}_N$, with the inherited color-flavor-quark-number identifications~\cite{Anber:2019nze}. This anomaly involves the $\mathbb{Z}_F$ chiral  symmetry relating vacua connected by the $\mathbb{Z}_{F}$ domain walls studied below. As discussed in the Introduction, this anomaly implies a potentially rich internal structure for $\mathbb{Z}_{F}$ domain walls. Its continued existence in twisted QCD provides a strong basis for the comparison with $\mathbb{Z}_{F}$ domain walls in QCD in a setting where semiclassical control makes their profiles and tensions accessible to explicit calculations.

\subsection{The Chiral Lagrangian of Twisted QCD}
\label{eq:lownenergychiralmixedgauge}
In this section, we aim to demonstrate that the infrared spectrum of twisted QCD is equivalent to the sum of $F$ versions of single-flavor QCD together with the aforementioned set of $F-1$ massless pions. 

Each of the $F$ versions of single-flavor QCD is equipped with a different color gauge group $SU(K_{i})$, where the integers $K_{i}$ are valued between 
$1\le K_{i} \le N-F+1$ and the sum satisfies $\sum_{i=1}^{F} K_{i} = N$. The exact sizes of each color gauge groups $K_{i}$ is set by the color-flavor twisted boundary conditions of Eq.~\eqref{eq:generalflavor}. We find that $K_{i}$ is determined by the distance between two consecutive fermion zero mode pairs $K_{i} = n(i+1) - n(i) \ \mathrm{mod} \ N$, where $n(i)$ indicates the location of the $i$-th fermion zero mode pair on the magnetic monopole lattice of Fig.~\ref{fig:fermion-disttrib}. 

In order to solve for this low energy spectrum, it is useful to translate the monopole vertex lattice of Fig.~\ref{fig:fermion-disttrib} into a dual photon potential. Each monopole without a fermion zero mode connecting the dual photons $\sigma_{i}$ and $\sigma_{i+1}$ contributes a factor proportional to $\cos{\left(\sigma_{i}-\sigma_{i+1} \right)}$ to the dual photon potential.  A monopole with fermion zero modes does not contribute to the dual photon potential. This results in a Lagrangian of the form
\begin{equation}
    \mathcal{L}= \frac{g^{2}}{8 \pi^{2}L} \sum_{i=1}^{N}(\partial\sigma_{i})^{2} -\sum_{b \in \mathcal{B}} \zeta \cos{\left(\sigma_{b} - \sigma_{b+1} \right)} \,, \qquad \sum_{i=1}^{N} \sigma_{i} =0\,.
    \label{eq:dualphotonpotzero}
\end{equation}

\begin{figure}
\centering
\begin{tikzpicture}[
    x=1.3cm,
    pt/.style={circle, fill=black, inner sep=1.6pt},
    bond/.style={thick}
]

\foreach \i in {1,...,9}{
    \node[pt] (a\i) at (\i,0) {};
    \node[below=5pt] at (\i,0) {$\sigma_{\i}$};
}
\draw[bond] (a1) -- (a2) -- (a3);
\draw[bond] (a4) -- (a5);
\draw[bond] (a7) -- (a8) -- (a9);

\foreach \i in {1,...,9}{
    \node[pt] (b\i) at (\i,-1.6) {};
    \node[below=5pt] at (\i,-1.6) {$\sigma_{\i}$};
}
\draw[bond] (b1) -- (b2) -- (b3);
\draw[bond] (b4) -- (b5) -- (b6);
\draw[bond] (b7) -- (b8) -- (b9);

\end{tikzpicture}
\caption{Dual-photon lattice as determined by the dual photon potential and the monopole lattice in Fig.~\ref{fig:fermion-disttrib} ($N=9$). The first row corresponds to zero modes on the pairs $(\sigma_{3},\sigma_{4}), (\sigma_{5},\sigma_{6}),(\sigma_{6},\sigma_{7}),(\sigma_{9},\sigma_{1})$. The second row is $\mathbb{Z}_{\mathrm{gcd}(N,F)}$-symmetric and has fermion zero modes on the pairs $(\sigma_{3},\sigma_{4}), (\sigma_{6},\sigma_{7}),(\sigma_{9},\sigma_{1})$.}
\label{fig:dualphotonpotlat}
\end{figure}
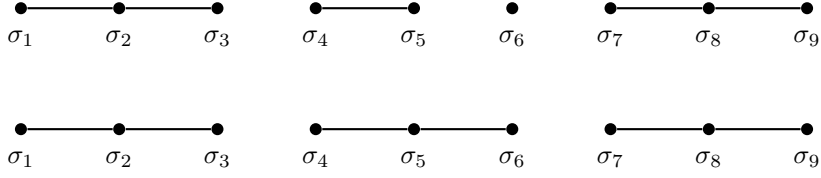

A useful visualization of this potential can be found  by dualizing the monopole vertex lattice of Fig.~\ref{fig:fermion-disttrib} into a dual photon lattice, where two dual photons $\sigma_{i}$ and $\sigma_{i+1}$ are connected if the monopole coordinate $\vec{\alpha}_{i} \cdot \vec{\sigma}$ does not have a fermion zero mode. The duals of Fig.~\ref{fig:fermion-disttrib} are drawn in Fig.~\ref{fig:dualphotonpotlat}.

The lattice of dual photons in Fig.~\ref{fig:dualphotonpotlat} consists of disconnected sub-lattices. If we denote by $n(j)$ the monopole coordinates of the $j$-th fermion zero mode pair, then a sublattice spans the dual photons $( \sigma_{n(j)+1},\dots, \sigma_{n(j+1)})$. Whilst the potential does not couple the different sublattices, they are still connected by the traceless sum constraint $\sum_{i=1}^{N} \sigma_{i} = 0$ on the dual photons. 

We first solve for the massless pion degrees of freedom of the infrared spectrum: on each separate sublattice, one can shift all the dual photons $(\sigma_{n(j)+1}, \sigma_{n(j)+2},\dots, \sigma_{n(j+1)})$ by the same amount, without changing the potential energy (Eq.~\eqref{eq:dualphotonpotzero}), as the potential only involves dual photon differences within each sublattice. The massless pions can therefore be identified as the excitations
\begin{equation}
    \pi_{j} = \sigma_{n(j)+1} + \dots +\sigma_{n(j+1)} \,, \qquad \sum_{j=1}^{F} \pi_{j} = 0\,.
    \label{eq:piondefinition}
\end{equation}

The fact that massless pions can be expressed as (linear combinations of) dual photons in the infrared is due the degree-of-freedom transmutation discussed in Section \ref{sec:condensate}. The sum constraint on the $F$ massless pions follows from the vanishing sum constraint on the dual photons $\sum_{i=1}^{N} \sigma_{i}=0$ and results in $F-1$ massless pion degrees of freedom. It can be seen from Eq.~\eqref{eq:pionshift} that the pions transform under the  $U(1)_{\mathcal{A}}^{F-1}$ symmetries as expected
\begin{equation}
    \pi_{j} \rightarrow \pi_{j} + \delta_{j} \,, \qquad \sum_{j=1}^{F} \delta_{j} =0\,.
    \label{eq:pionconstraiint}
\end{equation}
In general, the absence of the full $SU(F)_{A}$ axial symmetry group means that each such pion can have a distinct decay constant $f_{\pi}$, where $f_{\pi}^{2} \propto g^2/(4\pi^2L)$ is set by the kinetic term of the dual photons. A convenient, $F-1$ dimensional basis for the massless pions is provided by the Gell-Mann basis
\begin{equation}
    \tilde{\pi}_{1} = \pi_{1} - \pi_{2}\,, \qquad \tilde{\pi}_{2} = \pi_{1} + \pi_{2} - 2 \pi_{3}\,, \qquad \dots\,,
\end{equation}
although we will often work with the $F$ constrained pions of Eq.~\eqref{eq:piondefinition}.

The infrared spectrum also contains a set of massive modes. We now prove that the massive modes of each sublattice are completely decoupled from both the massless pions and the massive modes of other sublattices. This implies that each sublattice can be viewed as its own distinct unit. Of course, these statements are strictly only true when the quark masses are zero.

We define the size $K$ of a sublattice as the number of dual photons on that sublattice. According to Fig.~\ref{fig:dualphotonpotlat}, the size $K_{j}$ of the $j$-th sublattice is given by the distance $K_{j} = n(j+1)-n(j) \ \mathrm{mod} \ N$ between the  consecutive fermion zero mode pairs associated with monopole coordinates $n(j)$ and $n(j+1)$. 

Let us focus on a single sublattice of size $K_{i}$ in Fig.~\ref{fig:dualphotonpotlat}. This sublattice looks like a smaller version of the dual photon lattice of pure center-stabilized Yang--Mills (see Eq.~\eqref{eq:potentialym}), albeit missing the Kaluza--Klein monopole that connects the dual photons at the endpoints of the sublattice. One such sublattice is therefore equivalent to a version of center-stablized twisted QCD of gauge group $SU(K_{i})$ with one pair of fermion zero modes, corresponding to the missing Kaluza--Klein monopole. Hence, the massive spectrum of QCD with $F$ fundamental flavors breaks up into $F$ versions of single-flavor QCD with gauge group $SU(K_{i})$. 

We can solve for the massive modes of each such single-flavor QCD sublattice by again focusing on one sublattice $(\sigma_{n(j)+1}, \sigma_{n(j)+2},\dots, \sigma_{n(j+1)})$ of size $K$. The interactions between the dual photons constitute an open chain problem, with the missing cosine at the endpoints of the sublattice being equivalent to free (Neumann) boundary conditions. The normal modes, or equivalently the eigenvector directions of the open-chain discrete
Laplacian, are therefore of the form 
\begin{equation}
    (\sigma_{n(j)+1}, \sigma_{n(j)+2},\dots, \sigma_{n(j+1)}) = \chi(x)\sqrt{\frac{2}{K}}\left(\cos\frac{\pi k}{2K},\cos\frac{3\pi k}{2K},\ldots, \cos\frac{(2K-1)\pi k}{2K} \right)\, ,
    \label{eq:masssingleflavorqcd}
\end{equation}
where $\chi(x)$ is a solution to the corresponding three-dimensional plane wave equation. 

There are $k=1,\dots, K-1$ such orthogonal eigenmodes for each sublattice of size $K$. Each eigenmodes satisfies the local vanishing sum constraint $\sum_{\ell=n(j)+1}^{n(j+1)} \sigma_{\ell}=0$ on its own sublattice and can therefore be considered decoupled from all the massive eigenmodes on the other sublattices. The edge case of a sublattice of one dual photon of size $K_{i}=1$ only comprises a massless pion $\pi_i=\sigma_{n(i)+1}$, in accordance with Eq.~\eqref{eq:piondefinition}. 

We have thus shown that each sublattice contains $K$ modes: one massless excitation, the pion, and $K-1$ massive modes. The constraint in Eq.~\eqref{eq:pionconstraiint} requires that only $F-1$ of the $F$ massless excitations are genuine pions, as expected. Furthermore, since the sizes of the sublattices satisfy $\sum_{i=1}^{F}K_{i} = N$, we count $N-F$ massive modes in the spectrum. The mass of the massive eigenmodes of Eq.~\eqref{eq:masssingleflavorqcd} is set by twice the length $2K$ of the sublattice:
\begin{equation}
    m_{k}^{2} = m_{\phi}^{2} \sin^{2}\left(\frac{\pi k}{2K} \right) \,,  \qquad k=1,\dots, K-1 \qquad \mathrm{with} \qquad  m_{\phi}^2 = C\Lambda_{\mathrm{QCD}}^{2} \  \lambda^{-3} \ \eta^{\gamma}\, .
    \label{eq:massivespectrummass}
\end{equation}
Comparing Eq.~\eqref{eq:massmodes} with Eq.~\eqref{eq:massivespectrummass}, we understand that the difference between the massive spectrum of single-flavor QCD and that of pure Yang--Mills is the difference between Neumann and periodic boundary conditions. Here $\gamma$ should include the change in the running due to the quarks.

Whenever $N$ and $F$ are not coprime, the assignment of fermion zero modes over the magnetic monopole lattice can be taken to be color-flavor symmetric. In this case, the sublattices of dual photons repeat every $\mathrm{gcd}(N,F)$ steps (see bottom row of Fig.~\ref{fig:dualphotonpotlat}). The $\mathbb{Z}_{\mathrm{gcd}(N,F)}$ color-flavor center symmetry of the infrared spectrum then permutes the massive spectrum of the $j$-th sublattice to the massive spectrum of the $j+F/\mathrm{gcd}(N,F)$-th sublattice. Similarly, the massless pions are also permuted under the $\mathbb{Z}_{\mathrm{gcd}(N,F)}$ symmetry as
\begin{equation}
    \pi_{j} \rightarrow \pi_{j + F/\mathrm{gcd}(N,F)} \,.
\end{equation}

As previously mentioned, the color-flavor center symmetry of the infrared spectrum is the remnant of the larger $\mathbb{Z}_{N}$ permutation center symmetry of pure center-stabilized gluodynamics (Eq.~\eqref{eq:center symmetry action on Omega}). Having by now elucidated how this symmetry affects the distribution of the fermion zero modes over the $N$ monopoles of twisted QCD, we turn in what follows to its implications for probe-particle deconfinement on $\mathbb{Z}_F$ domain walls.

\subsection{Representations of Twisted QCD}
\label{sec:representationtheory}

In the previous section, we identified the infrared modes of color-flavor twisted QCD. The spectrum consists of $F$ versions of single-flavor QCD with gauge groups $SU(K_{i})$ and $F-1$ massless pions. In this section, we study how representations of the $SU(N)$ gauge theory decompose into representations of the $SU(K_{i})$ single-flavor QCDs and the massless pions. This will allow us to identify the $N$-ality of the representation of the deconfined particle on the $\mathbb{Z}_{F}$ domain wall in Section~\ref{sec:deconfnewzf}.  

The electric charges $\vec\omega$ of a probe particle under the $U(1)^{N-1}$ gauge groups of both center-stabilized pure gluodynamics and twisted QCD can be measured by a Gaussian surface that forms a $1$-dimensional closed line, say a circle $S^{1}$, around the charged particle:
\begin{equation}
    \frac{L}{g^{2}}\int_{S^{1}} \star \vec{F} = \vec{\omega} \,.
    \label{eq:gaussiansurface}
\end{equation}

The realizable electric charges $\vec{\omega}$ are determined by the ultraviolet breaking pattern $SU(N) \rightarrow U(1)^{N-1}$. Probe particles carrying color charge in the fundamental representation of $SU(N)$ break up into $N$ distinct probe particles with different weights $\vec \omega_{i}$. In terms of the $N$-dimensional basis of Eq.~\eqref{eq:ndimbasis}, these weights are given by $\vec\omega_i=\vec e_i-\frac{1}{N}\sum_{a=1}^{N}\vec e_a$, with $\sum_{i=1}^N\vec \omega_i=0$. 
The highest weight of the defining representation is  $\vec\omega_1=(1-1/N,-1/N,\ldots,-1/N)$, while the remaining weights can be reached by successive lowering with the simple roots $\vec\alpha_{i}$ as $\vec\omega_{i+1}=\vec\omega_i-\vec\alpha_i$. 

Any weight $\vec\omega$ corresponding to an ultraviolet representation of $N$-ality $p$ with respect to $SU(N)$ may be written as an integer linear combination of the weights of the fundamental representation,
\begin{equation}
    \vec \omega = \sum_{i=1}^{N} n_{i}\vec \omega_{i}\,, \qquad\sum_{i=1}^{N} n_{i} = p \ \mathrm{mod} \ N\,,\qquad n_i\in \mathds{Z}\, ,
    \label{eq:Nality}
\end{equation}
as can be verified by taking tensor products of the fundamental and anti-fundamental representations. 

From the definition of the dual photons in Eq.~\eqref{eq:dualphotons}, we see that the existence of an electrically charged probe particle inside the Gaussian circle of Eq.~\eqref{eq:gaussiansurface} is equivalent to a vortex (string) configuration in the dual photons described by a nontrivial winding around the $S^{1}$ circle, 
\begin{equation}
     \frac{1}{2\pi}\int_{S^{1}} d\vec{\sigma}  =  \vec{\omega} \,. 
    \label{eq:vortex}
\end{equation}

In pure center-stabilized Yang--Mills, all the dual photons obtain 
a mass from the monopole instantons. This implies that the dual photon profile around such a vortex string is confined into a domain wall.  The electrostatic potential between two charged probe particles is therefore set by the domain wall that stretches between the two vortex strings. The energy of this configuration is set 
by the distance between the string multiplied by the tension of the domain wall $T \sim \Lambda_{\rm QCD}^{2} \lambda^{-1/2} \eta^{\gamma/2 -1}$. This linear dependence of the potential on the distance between two charged particles is the hallmark of a confined phase. Below, we determine the dual photon profile around a vortex of the form~\eqref{eq:vortex} in twisted QCD. 

In the previous section, we learned that the low energy spectrum of twisted QCD is equivalent to $F$ versions of single flavor QCD with gauge groups $SU(K_{i})$ and $F-1$ massless pions. We now decompose the weights $\vec\omega$ of $SU(N)$ into weights $\vec\omega^{(K_{i})}$ of the $SU(K_{j})$ subgroups and weights $\vec\omega_{\pi}$ of the massless pions. The aim is to show the following decomposition holds: 
\begin{equation}
   \mathrm{Twisted}\ SU(N)\ \mathrm{with}\ F\ \mathrm{flavors}
   \quad\cong\quad
   \left[
   \bigoplus_{i=1}^{F}
   \left(SU(K_i)\ \mathrm{with}\ 1\ \mathrm{flavor}\right)
   \right]
   \mathbin{\scalebox{1.25}{$\times_{\mathrm{twist}}
 $}} \ 
   \mathrm{Massless\ Pions}\,.
   \label{eq:masterformulatwist2}
\end{equation}

That is, we aim to show that, in general, any single weight $\vec\omega$ of $SU(N)$ decomposes as a weight $\vec\omega^{(K_{i})}$ over a $SU(K_{i})$ subgroup and a weight $\vec \omega_{\pi}$ over the massless pions. The $\times_{\mathrm{twist}}$ twist in Eq.~\eqref{eq:masterformulatwist2} indicates that generically the weights of $SU(N)$ do not decompose into weights over just the $SU(K_{i})$ subgroups, but will generically also involve a pion weight $\vec\omega_{\pi}$.

A weight $\vec\omega^{(K_{i})}$ associated with a subgroup $SU(K_i)\subset SU(N)$ can be constructed by simply embedding $\vec\omega^{(K_{i})} \in \mathbb{R}^{K_i}$ in $\mathbb{R}^N$ in the trivial way. That is, the highest weight $\vec\omega^{(K)}_{1}$ of the fundamental representation of any $SU(K)$ sublattice reads
\begin{equation}
    \vec\omega^{(K)}_{1} = (\underbrace{0,\dots,0}_{\mathrm{Other \ Sublattices}}, \  \underbrace{1-1/K,-1/K,\dots,-1/K}_{K},\ \underbrace{0,\dots,0}_{\mathrm{Other \ Sublattices}})\,.
    \label{eq:weightofsuk}
\end{equation}
The non-zero indices are exactly those that span the $SU(K)$ subgroup inside $SU(N)$, or equivalently, those dual photons indices associated with the sublattice of size $K$ inside the dual photon lattice of size $N$. Notice the similarity with the weight of the defining representation $\vec\omega_{1}=(1-1/N, -1/N,\dots,-1/N)$ defined in the paragraph below Eq.~\eqref{eq:gaussiansurface} of the full $SU(N)$ gauge group. 

By construction, the weights of $SU(N)$ are  orthogonal to the constant sum $\sum_{a=1}^{N}\vec{e}_a=(1,1,\dots,1)\in\mathbb{R}^N$. The weights of a given subgroup $SU(K_i)\subset SU(N)$ satisfy the more constraining requirement that they are orthogonal to any constant vector $(1,1,\dots,1)\in\mathbb{R}^{K_j}$ associated with any one of the sublattices $K_{j}$ embedded in $\mathbb{R}^N$, not just the vector $(1,1,\dots,1)\in\mathbb{R}^{K}$ associated with its own sublattice. This is because the sum of the elements of  $\vec\omega^{(K)}_{1}$ over any given sublattice $K_{j}$ vanish.   

By Eq.~\eqref{eq:vortex}, any weight $\vec\omega$ corresponds to a probe particle with electric charge $\vec\omega$, or equivalently a vortex around which the dual photon winds as $\vec\sigma \rightarrow \vec\sigma + 2\pi \vec\omega$. The claim is that the pions do not wind when the dual photons wind by  $\vec\sigma \rightarrow \vec\sigma + 2 \pi \vec\omega^{(K)}_{1}$. To see this, note that according to Eq.~\eqref{eq:piondefinition}  the pions are linear combinations of dual photons with coefficients that are constant over each individual sublattice. In other words, the pions are linear combinations of the constant vectors $(1,1,\dots,1)\in \mathbb{R}^{K_{j}}$ as embedded in $\mathbb{R}^{N}$. As noted above, these these vectors are all orthogonal to the vector $\vec\omega^{(K)}_{1}$, and therefore the winding $\vec\sigma \rightarrow \vec\sigma + 2\pi \vec\omega^{(K)}_{1}$ does not include any pions. 

More in general, the vortices corresponding to the weights $\vec \omega^{(K)}$ only induce windings within the specific gauge subgroup $SU(K)$. As all the modes for this single-flavor QCD gauge group are massive, the dual photon profile around such a vortex is confined to a domain wall or flux tube, and the would-be probe particle is confined. This is similar to the confinement of probe particles in pure center-stabilized Yang--Mills discussed below Eq.~\eqref{eq:vortex}.

The weights $\vec \omega^{(K)}$ constructed in Eq.~\eqref{eq:weightofsuk} associated with a given $SU(K)$ sublattice  are not automatically weights in the original weight lattice of $SU(N)$. This implies that a probe particle with this charge could not have descended from a probe particle in a ultraviolet color representation of $SU(N)$. We thus restrict our attention to probe particles whose charges lie within the weight lattice of $SU(N)$.

We can decompose the charge of any such probe particle $\vec\omega$ of $SU(N)$ into weights $\vec \omega^{(K_{j})}_{1}$ under the $SU(K_{j})$ versions of single-flavor QCD and weights associated with the massless pions.  To appreciate this, consider again the highest weight $\vec\omega_{1} = (1-1/N, - 1/N,\dots,-1/N)$ of the fundamental representation of $SU(N)$ and suppose that the first dual photon, $\sigma_{1}$,  lies within a sublattice of size $K$. Then we can make the decomposition
\begin{equation}
    \vec\omega_{1} = \vec\omega^{(K)}_{1} + \vec\omega_{\pi} \,.
    \label{eq:decompositionintoreps}
\end{equation}
In this decomposition, $\vec\omega^{(K)}_{1}$ represents the highest weight of the fundamental representation of the first sublattice, which has size $K$ (see Eq~\eqref{eq:weightofsuk}):
\begin{equation}
    \vec\omega^{(K)}_{1} = (\underbrace{1-1/K,-1/K,\dots,-1/K}_{K},\  \underbrace{0,\dots,0}_{N-K})\,.
    \label{eq:highestweightexample}
\end{equation}
The second weight $\vec \omega_{\pi}$ is the difference between $\vec\omega_{1}$ and $\vec\omega^{(K)}_{1}$:
\begin{equation}
    \vec\omega_{\pi} = (\underbrace{1/K-1/N, \ 1/K-1/N,\dots,1/K-1/N}_{K},\  \underbrace{-1/N,\dots,-1/N}_{N-K})\,.
\end{equation}

The weight $\vec \omega_{\pi}$ is separately constant over the first sublattice $K$ and over all the remaining other sublattices. A weight that is constant over a sublattice does not activate any of the massive modes on that sublattice. This is because the massive mode vectors $\vec \sigma$ defined in Eq.~\eqref{eq:masssingleflavorqcd} are orthogonal to any constant vectors $(1,1,\dots,1)\in \mathbb{R}^{K_{j}}$ embedded in $\mathbb{R}^{N}$ by the roots of unity identity. Instead, a vector that is separately constant over each sublattice is associated with only pions winding. The weight $\vec \omega_{\pi}$ therefore corresponds to a massless pion string. The flux profile around such a string is uniform, instead of confined to a tubular region.

\begin{figure}
    \centering
\begin{tikzpicture}[scale=1.2]

\draw[very thick,red]
(0,0) -- (3,0);


\draw[->,very thick,red] (0,0) -- (1,0);

\draw[->,thick,black]
  (0.5,0) arc[start angle=0,end angle=320,radius=0.5];

\node[red] at (2,0.3) {Flux Tube};

\draw[dashed,gray] (0,0) -- (0,1.5);

 \foreach \radius in {1,1.5,2} { \foreach \angle in {30,60,90,120,150,210,240,270,300,330} { \draw[->,thin] ({\radius*cos(\angle)},{\radius*sin(\angle)}) -- ({(\radius+0.25)*cos(\angle)}, {(\radius+0.25)*sin(\angle)}); } }

\filldraw[fill=red, draw=black] (0,0) circle (4pt);

\node[below=4pt] at (0,-0.35) {Pions};

\end{tikzpicture}

    \caption{Flux profile in twisted QCD of a probe quark corresponding to the highest weight $\vec \omega_{1}$ of the fundamental representation of $SU(N)$. The flux profile separates into a confining flux tube (red) corresponding to a particle in the fundamental representation of a single-flavor QCD subgroup $SU(K_{i})$ with weight $\vec\omega^{(K)}_{1}$ and a uniform profile associated with the massless pion string weight $\vec \omega_{\pi}$, see Eq.~\eqref{eq:decompositionintoreps}. The radially outward going arrows around the pion string indicate the electric field associated with dualizing the pion into a gauge boson. The electric field associated with the particle's charge under the $SU(K)$ subgroup is confined to a tubular region.}
    \label{fig:fundamentalparticle}
\end{figure}
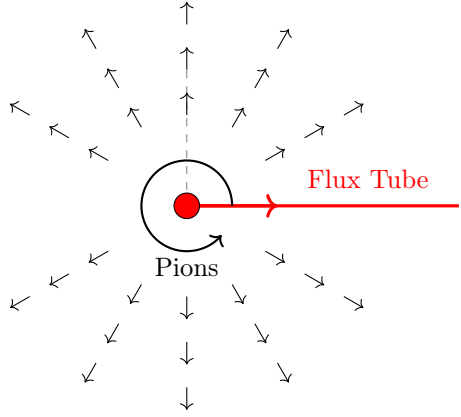

In Fig.~\ref{fig:fundamentalparticle}, we have drawn the dual photon profile in twisted QCD for a probe quark corresponding to the highest weight $\vec\omega_{1}$ of the fundamental representation of $SU(N)$. According to Eq.~\eqref{eq:decompositionintoreps}, this 
flux profile decomposes into both a confining flux tube associated with a weight $\vec\omega^{(K)}_{1}$ in the fundamental representation of $SU(K_{i})$ and a massless pion string profile associated with the weight $\vec\omega_{\pi}$.

Note that, in general, this procedure can be repeated for any weight of $SU(N)$ by first reducing that weight into linear combinations of the fundamental and anti-fundamental representation of $SU(N)$, and  then using the decomposition in Eq.~\eqref{eq:decompositionintoreps}.
 
We learn that probe charges in the fundamental representations of $SU(N)$ (i.e. a gauge vortex) are generally charged in the fundamental representation of one of the $SU(K_{i})$ subgroups, but also source a massless pion string according to Eq.~\eqref{eq:decompositionintoreps}. Bearing in mind the duality between a compact scalar field (a pion) and an Abelian gauge field in $d=2+1$ dimensions, we say that such a particle is (magnetically) charged under the massless pions. Dualizing the massless pions would reveal a uniform electric field pointing radially away from the vortex, as drawn in Fig.~\ref{fig:fundamentalparticle}. 

We briefly note that what holds for the representations also holds for the gauge identifications of the gauge theories. In fact, the vortex string in Eq.~\eqref{eq:vortex} is a gauge string that implements the gauge identification $\vec\sigma \equiv \vec\sigma+2\pi\vec\omega$. It can be checked that this is equivalent to the gauge identification on the magnetic monopole coordinates introduced earlier in  Eq.~\eqref{eq:gaugetransmagmon}. As we have seen, the only gauge identifications compatible with the ultraviolet gauge group being $SU(N)$ are those that act simultaneously on the massive modes of the $SU(K)$ sublattices and the set of massless pions. It is exactly this the sense in which the gauge identifications between the massive and massless modes are twisted in Eq.~\eqref{eq:masterformulatwist2}.

\subsection{Deconfinement on $\theta_{\mathrm{YM}}$ Domain Walls}
\label{sec:deconfzfdw}

In this section and the next, we study the deconfinement on the $\mathbb{Z}_{F}$ domain walls of massless flavor-twisted QCD. We start with a brief review (see ~\cite{Poppitz:2012nz,Chen:2020syd, Anber:2020xfk}) of deconfinement on CP domain walls in pure center-stabilized Yang--Mills at $\theta_{\mathrm{YM}} = \pi$, where our conventions for $\theta_{\mathrm{YM}}$ in the $4d$ action are
\begin{equation}
\mathcal{L} \supset \frac{f_{\theta_{\mathrm{YM}}}^{2}}{2} (\partial_\mu \theta_{\mathrm{YM}})^{2} + \frac{\theta_{\mathrm{YM}}}{16 \pi^2}\,\mathrm{Tr}\,G_{\mu\nu} \widetilde G^{\mu\nu}\,, \qquad \theta_{\mathrm{YM}} \equiv \theta_{\mathrm{YM}} + 2\pi \,\,.
\end{equation}
This is equivalent to studying axion domain walls. We start with this discussion to make contact with existing literature in $4d$ pure gluodynamics and ease the transition to the deconfinement on the $\mathbb{Z}_{F}$ domain wall. Crucial ingredients will be the $\mathbb{Z}_{N}$ center symmetry in pure gluodynamics and the $\mathbb{Z}_{\mathrm{gcd}(N,F)}$ color-flavor center symmetry in massless twisted QCD. 

\begin{figure}[ht]
    \centering
    \begin{tikzpicture}[>=Latex, thick]

\coordinate (Q) at (0,0);

\draw[->,blue] (-0.25,1.6) -- (-2.2,1.6) node[midway, below] {$E=-\frac{1}{2}$};
\draw[<-,blue] ( 0.25,1.6) -- ( 2.2,1.6) node[midway, below] {$E=-\frac{1}{2}$};

\draw[->,blue] (-0.25,0.8) -- (-2.2,0.8) node[midway, below] {$E=-\frac{1}{2}$};
\draw[->,red] ( 0.25,0.8) -- ( 2.2,0.8) node[midway, below] {$E=\frac{1}{2}$};

\draw (-3,0) -- (3,0);

\filldraw[fill=red, draw=black] (Q) circle (4pt);
\node[below=4pt] at (Q) {$+1$};

\end{tikzpicture}
    \caption{Deconfinement of a positive unit charge (red dot) at $\theta_{\mathrm{YM}} = \pi$ in $d=1+1$ QED. The background electric field is drawn as the top row of arrows. The total electric field is drawn as the bottom row of arrows. The bottom line corresponds to the one spatial dimension.}
    \label{fig:schwingerdecon}
\end{figure}
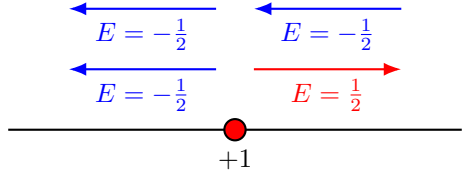

A simple toy model for deconfinement at $\theta_{\mathrm{YM}} =\pi$ exists in two dimensional QED. Two dimensional QED is naturally in a confined phase \cite{Schwinger:1962tp}, as electric flux cannot spread and electric fields are constant throughout the one spatial dimension.  An angle $\theta_{\mathrm{YM}} =\pi$ acts as a background electric field $E = -\frac{1}{2}$ (in appropriate units) \cite{Coleman:1976uz}. At the position of a  charged particle, the electric field jumps by one unit $\Delta E=1$ according to  Gauss' law. 
As the electric field to the left of the particle is $E =- \frac{1}{2}$ and to the right of the particle is $E = \frac{1}{2}$, the presence and position of the particle does not change the energy density $E^{2}$. The particle is therefore free to move and deconfined (see Fig.~\ref{fig:schwingerdecon}).

Deconfinement on $\theta_{\mathrm{YM}}$ domain walls is also expected to happen for particles with color charge  in pure four-dimensional Yang--Mills near the cuspy point $\theta_{\mathrm{YM}} = \pi$~\cite{Witten:1997ep}. It has been shown that this deconfinement is compatible with anomaly inflow arguments on the wall~\cite{Gaiotto:2017yup, Cordova:2019jnf}.

In center-stabilized pure $SU(N)$ Yang--Mills, the vacuum structure as a function of the $\theta_{\mathrm{YM}}$-angle is analytically tractable~\cite{Bhoonah:2014gpa,Aitken:2018mbb,Anber:2017rch,Aitken:2018kky} and the deconfinement on the axion domain wall can be explicitly studied. In full agreement with three-dimensional anomaly inflow arguments~\cite{Cox:2019aji,Anber:2018xek, Poppitz:2020tto},  probe color charges are deconfined on the domain wall $\theta_{\mathrm{YM}} \rightarrow \theta_{\mathrm{YM}} + 2\pi$ due to a rearrangement of the dual photons near the cusp  $\theta_{\mathrm{YM}} =\pi$~\cite{Poppitz:2012nz,Chen:2020syd, Anber:2020xfk}. This dual photon rearrangement is the CP domain wall at $\theta_{\mathrm{YM}} = \pi$. As we will require the same machinery to show deconfinement on the $\mathbb{Z}_{F}$ domain wall, we briefly review the deconfinement on CP domain walls in center-stabilized Yang--Mills here. 

The dual photon action including the $\theta_{\mathrm{YM}}$-angle in center-stabilized Yang--Mills can be obtained through the discussion in Section~\ref{sec:magsym}. In the absence of fermion zero modes, all monopoles are present in the potential. Since each magnetic monopole carries $1/N$ instanton number, the $\theta_{\mathrm{YM}}$-angle couples to each monopole as $e^{i\theta_{\mathrm{YM}}/N}$. Therefore, the dual photon Lagrangian is 
\begin{equation}
    \mathcal{L} =   \sum_{j=1}^{N}\frac{g^{2}}{8 \pi^{2}L}(\partial \sigma_{j})^{2} - \sum_{j=1}^{N} \zeta \cos\left(\sigma_{j}-\sigma_{j+1} + \frac{\theta_{\mathrm{YM}}}{N}\right)\, .
    \label{eq:YangMillsthetaaction}
\end{equation}
The coordinates of the $N$ distinct stationary points of the potential are independent of $\theta_{\mathrm{YM}}$. We label the stationary points by an index $k=1,\dots, N$, and they are given in monopole coordinates by
\begin{equation}
   \vec{\alpha}_{i \neq b_{\star}} \cdot \vec{\sigma} = -\frac{2\pi k}{N} \qquad \mathrm{and} \qquad \vec{\alpha}_{b_{\star}} \cdot \vec{\sigma} = 2\pi k -\frac{2\pi k}{N}\,, \qquad \mathrm{with} \qquad k = 1,\dots,N\, .
    \label{eq:axiondwfullset}
\end{equation}
For each of the $k$ stationary points, there are $N$ gauge-equivalent choices for the special index $b_\star$ in Eq.~\eqref{eq:axiondwfullset}. This index labels the magnetic monopole coordinate that undergoes an additional $2 \pi k$ shift, which is what guarantees that the mathematical identity $\sum_{i} \vec\alpha_{i} \cdot \vec\sigma =0$ is satisfied. 

Upon minimizing over the stationary points in Eq.~\eqref{eq:axiondwfullset}, the axion potential obtains $N$ branches:
\begin{equation}
V(\theta_{\mathrm{YM}}) =  \min_{k\in\mathbb{Z}_{N}}\left[-\zeta N\cos\left(\frac{\theta_{\mathrm{YM}}-2\pi k}{N}\right) \right]\,.
\label{eq:branchedaxionpot}
\end{equation}
The potential depends on the physical ratio $\theta_{\mathrm{YM}}/N$ and has cusps at $\theta_{\mathrm{YM}} =\pi, \ 3\pi, \ \mathrm{etc}$. Near the cusp, the heavy dual photons quickly rearrange between the stationary points. This happens at distance scales of order $\sim\sqrt{\lambda m_{W}/\zeta}$, much smaller than the length scales associated with axion domain walls $\sim \sqrt{Lf_{\theta_{\mathrm{YM}}}^{2}/\zeta}$. In order to solve for the Yang--Mills component of the $\theta_{\mathrm{YM}}$ domain wall, it is therefore enough to solve for the dual photon profile near the cusp $\theta_{\mathrm{YM}} = \pi$, e.g. the CP domain wall. 

At the cusp $\theta_{\mathrm{YM}} = \pi$, there are two stationary points with equal energy, $k=0$ and $k=1$. The two stationary points are CP conjugate, and the $\theta_{\mathrm{YM}}$ domain wall acquires an internal profile in the dual photons that interpolates between the CP conjugate vacua $k=0$ and $k=1$. 

There are $N$ choices for the $k=1$ vacuum associated with a choice for the gauge label $b_{\star}$ in Eq.~\eqref{eq:axiondwfullset}. Importantly, note that the $N$ choices of $b_\star$ are related by gauge transformations that act nontrivially on the boundary values of the field $\vec \sigma$. Such transformations are not redundancies of our description and each choice $b_{\star}$ yields a distinct domain wall. This is because applying a gauge transformation at only one end of a domain wall whilst keeping the other end of the domain wall fixed changes the relative gauge transformation and yields a physically distinct wall. This observation will play an important role shortly.

All choices for $b_{\star}$ result in CP domain walls with equal tensions, although they correspond to different internal profiles for the dual photons along the walls. This is not because of the gauge transformation, but because the dual photon boundary conditions specified by two different values $b_{\star, 1}$ and $b_{\star,2}$ 
are related by a permutation of the dual photons $\sigma_{i} \rightarrow \sigma_{i\pm\Delta b_{\star}}$, with $\Delta b_\star \equiv b_{\star,2}-b_{\star,1}$. In pure Yang--Mills, a permutation of dual photons corresponds to an element of the $\mathbb{Z}_{N}$ center symmetry group. The choice of boundary condition spontaneously breaks this symmetry, resulting in $N$ domain walls with equal tension.

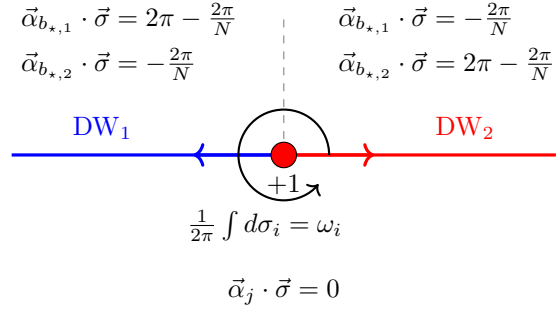
\begin{figure}
    \centering
\begin{tikzpicture}[scale=1.2]

\draw[very thick,blue]
(-3,0) -- (0,0);

\draw[very thick,red]
(0,0) -- (3,0);

\draw[->,very thick,blue] (0,0) -- (-1,0);
\draw[->,very thick,red] (0,0) -- (1,0);
\draw[->,thick,black]
  (0.5,0) arc[start angle=0,end angle=320,radius=0.5];
\node[blue] at (-2,0.3) {$\mathrm{DW}_1$};
\node[red] at (2,0.3) {$\mathrm{DW}_2$};

\draw[dashed,gray] (0,0) -- (0,1.5);
\filldraw[fill=red, draw=black] (Q) circle (4pt);
\node[below=4pt] at (Q) {$+1$};
\node[below left] at (0.75,-0.5) {$\frac{1}{2\pi}\int d\sigma_{i} = \omega_{i} $};

\node at (0,-1.5) 
{$\vec{\alpha}_{j}\cdot\vec{\sigma}=0$};


\node[anchor=west] at (-3,1.5) 
{$\vec{\alpha}_{ b_{\star,1}} \cdot \vec{\sigma} = 2\pi  -\frac{2\pi }{N}$};
\node[anchor=west] at (-3,1) 
{$\vec{\alpha}_{ b_{\star,2}} \cdot \vec{\sigma} =  -\frac{2\pi }{N}$};

\node[anchor=west] at (0.5,1.5) 
{$\vec{\alpha}_{ b_{\star,1}} \cdot \vec{\sigma} =  -\frac{2\pi }{N}$};
\node[anchor=west] at (0.5,1) 
{$\vec{\alpha}_{ b_{\star,2}} \cdot \vec{\sigma} = 2\pi  -\frac{2\pi }{N}$};

\end{tikzpicture}
    \caption{Deconfinement at $\theta_{\mathrm{YM}} =\pi$ in center-stabilized gluodynamics as described in Ref.~\cite{Anber:2020xfk}. Two domain walls of equal tension glued together. The string at the intersection is a probe color charge with fundamental weight $\omega_{i}$. As both domain walls pull equally, the charge is deconfined.}
    \label{fig:deconfonaxiodom}
\end{figure}

Two domain walls with distinct internal profiles at $\theta_{\mathrm{YM}} =\pi$ but gauge equivalent $k=1$ vacua at their endpoints can be glued together, without the need for a third domain wall connecting the vacua. We have drawn this configuration in Fig.~\ref{fig:deconfonaxiodom}. At the intersection of the two domain walls lives a vortex string, whose winding is set by the difference $\Delta b_{\star}$ in the boundary conditions. Whenever $\Delta b_{\star} =1$, the boundary conditions in Eq.~\eqref{eq:axiondwfullset} imply that the magnetic monopole coordinates wind around the vortex string upon performing the gauge transformation
\begin{equation}
\vec\alpha_{b_{\star}} \cdot \vec\sigma \rightarrow \vec\alpha_{b_{\star}} \cdot \vec\sigma + 2\pi \qquad \mathrm{and} \qquad \vec\alpha_{b_{\star}+1} \cdot \vec\sigma \rightarrow \vec\alpha_{b_{\star}+1} \cdot \vec\sigma - 2\pi \,.
\end{equation}

In dual photon coordinates, it can be checked that this corresponds to a winding of the vortex at the intersection set by a weight of the fundamental $\vec\omega_{i}$ of $SU(N)$ (see Section \ref{sec:representationtheory} and Eq.~\eqref{eq:vortex}),
\begin{equation}
   \frac{1}{2\pi} \int_{S^1} d\vec{\sigma} = \vec{\omega}_{i} \,.
   \label{eq:vortexstringaxionspec}
\end{equation}
As the two $\mathbb{Z}_{N}$-related domain walls (flux tubes) have equal tensions, the net force on the probe particle vanishes and the particle is deconfined ~\cite{Cox:2019aji, Poppitz:2012nz,Chen:2020syd, Anber:2020xfk}. The deconfinement of the probe particle in Fig.~\ref{fig:deconfonaxiodom} should be compared to the two-dimensional picture of deconfinement at $\theta_{\mathrm{YM}} =\pi$ in Fig.~\ref{fig:schwingerdecon}. 

Importantly, note that we could have run this process in reverse and started by studying the flux tube profile around a probe particle corresponding to a vortex string with charge in Eq.~\eqref{eq:vortexstringaxionspec}. We would have found that at $\theta_{\rm YM} =0$, the flux of the particle confines into a single flux tube. At $\theta_{\rm YM} = \pi$, the flux of the particle is separated over two flux tubes of equal tension, and the particle is deconfined. 

Formally, the exact representation that the deconfined probe color charge is in is unobservable at long distances, and only the $N$-ality is well-defined. This is because center-stabilized Yang--Mills contains massive off-diagonal color charged bosons at the scale $m_{W}= 2\pi/(NL)$. The massive charged bosons can be pair produced and screen the color charge of the probe particle. Any such pair production will not alter the $N$-ality of the representation, which for the fundamental representation is $1$. These statements are equivalent to the fact that the dual photons have an emergent  
$U(1)^{N-1}_{\mathrm{elec}}$ electric $1$-form symmetry, that can be used to measure the charges of particles. The massive charged bosons explicitly break this down to the original $\mathbb{Z}_{N}$ $1$-form center symmetry of $SU(N)$.

\subsection{Deconfinement on $\mathbb{Z}_{F}$ Domain Walls}
\label{sec:deconfnewzf}

Standard lore suggests that there is a link between $\theta_{\mathrm{YM}}$ domain walls in pure gluodynamics and $\eta'$ domain walls in massless QCD, hinting at a similar deconfinement phenomenon on $\mathbb{Z}_{F}$ domain walls in QCD. In twisted QCD, probe color particles in representations of $N$-ality $N/\mathrm{gcd}(N,F)$ are deconfined on $\mathbb{Z}_{F}$ domain walls whenever $N$ and $F$ are not coprime, as we argue below. 

In massless twisted QCD, because of the pairing between the fermion zero modes and the monopole coordinates (see Fig.~\ref{fig:fermion-disttrib} and the discussion below Eq.~\eqref{eq:etainstanton}), $\mathbb{Z}_{F}$ domain walls correspond to transformations of all monopole coordinates associated with fermion zero modes by a common phase,
\begin{equation}
    \vec{\alpha}_{n(j)} \cdot \vec\sigma \rightarrow  \vec{\alpha}_{n(j)} \cdot \vec\sigma + \frac{2\pi}{F} \ \qquad \mathrm{and} \qquad \vec{\alpha}_{b_{\star}} \cdot \vec{\sigma} \rightarrow \vec{\alpha}_{b_{\star}} \cdot \vec{\sigma} - 2\pi \,.
\end{equation}
Here $n(j)$ indicates all monopole coordinates with fermion zero modes. The shift of the single monopole coordinate with index $b_{\star}$ occurs on a monopole without fermion zero modes and guarantees that $\sum_{i} \vec{\alpha}_{i} \cdot \vec{\sigma}=0$. Different choices for $b_{\star}$ are related by a gauge transformation, similar to the index $b_{\star}$ for the axion domain wall in Eq.~\eqref{eq:axiondwfullset}. 

A $\mathbb{Z}_{F}$ domain wall therefore interpolates between the two sets of QCD vacua:
\begin{equation}
   \vec\alpha_{i}\cdot  \vec{\sigma}(-\infty) = 0 \qquad \rightarrow \qquad \vec{\alpha}_{n(j)} \cdot \vec\sigma(+\infty) = \frac{2\pi}{F} \,, \qquad \vec{\alpha}_{b_{\star}} \cdot \vec{\sigma}(+\infty) = - 2\pi \,.
    \label{eq:boundaryconditionszfwalls}
\end{equation}

In complete analogy with the $\theta_{\mathrm{YM}}$ domain wall, there are $N-F$ such choices for the domain wall corresponding to the index $b_{\star}$. In general, these $N-F$ different choices have different internal dual photon profiles and tensions. 

The $\mathbb{Z}_{F}$ domain wall will contain both a massless pion profile and a well-defined core of size $\sim m_{\phi}^{-1}$ associated with the massive modes of a single-flavor QCD sublattice. The fact that the $\mathbb{Z}_{F}$ domain wall has both a pion and massive profile is because of the twisted nature of the compactified theory, as discussed at the end of Section~\ref{sec:representationtheory}: any deconfined particle sitting at the intersection of two $\mathbb{Z}_{F}$ domain walls generically carries a pion profile. This can already be seen in Fig.~\ref{fig:fundamentalparticle}, where a probe quark in the fundamental of $SU(N)$ sources both a massive flux tube and a massless pion string. In light of this, in order to study $\mathbb{Z}_{F}$ domain walls in twisted QCD, one would have to regulate the massless pion profile by considering the theory at finite volume only. 

Similarly to the axion domain wall, the $\mathbb{Z}_{F}$ domain wall can host a deconfined particle. This happens whenever $N$ and $F$ are not coprime: the infrared spectrum has a $\mathbb{Z}_{\mathrm{gcd}(N,F)}$ CFC symmetry that permutes the dual photon index $i \rightarrow i+ N/\mathrm{gcd}(N,F)$ and leaves the boundary conditions for the monopole coordinates with fermion zero modes in Eq.~\eqref{eq:boundaryconditionszfwalls} invariant. However, the symmetry also permutes the index $b_{\star} \rightarrow b_{\star} + N/\mathrm{gcd}(N,F)$, making two $\mathbb{Z}_{F}$ domain walls with $\Delta b_{\star} = b_{\star,1} - b_{\star,2} = N/\mathrm{gcd}(N,F)$ have equal tension while still featuring different internal profiles.

In complete analogy with the different choices for the axion domain wall, two $\mathbb{Z}_{F}$ domain walls with different $b_{\star}$ indices can be glued together without the need for a third domain wall. At the intersection, there sits a vortex string and thus a probe color charge. If the two walls have equal tension (i.e. if $\Delta b_{\star} = N/\mathrm{gcd}(N,F)$), the color charge of the particle at the intersection is, 
\begin{equation}
    \frac{1}{2\pi}\int_{S^1} d\vec{\sigma} =  \vec\omega_{b_{\star}+1} + \dots + \vec\omega_{b_{\star}+N/\mathrm{gcd}(N,F)} \,.
    \label{eq:weightsforzf}
\end{equation}

This weight corresponds to that of a probe color particle in a representation of $N$-ality $N/\mathrm{gcd}(N,F)$. In full analogy with the $\theta_{\mathrm{YM}}$ domain wall, there is no net force on the probe particle at the intersection between the two domain walls (flux tubes) and it is deconfined. In this context, the $\theta_{\mathrm{YM}}$ domain wall in pure gluodynamics appears as the special  $\mathrm{gcd}(N,0)=N$ case.

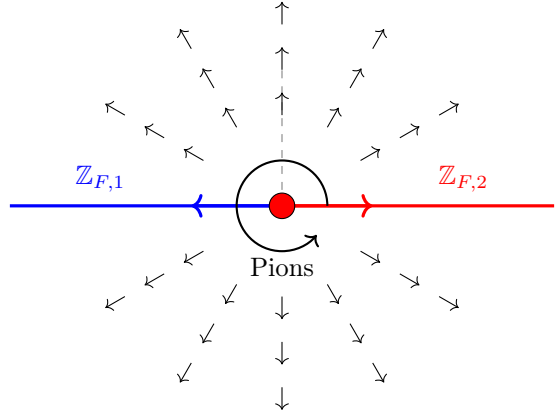
\begin{figure}
    \centering
\begin{tikzpicture}[scale=1.2]

\draw[very thick,blue]
(-3,0) -- (0,0);

\draw[very thick,red]
(0,0) -- (3,0);

\draw[->,very thick,blue] (0,0) -- (-1,0);
\draw[->,very thick,red] (0,0) -- (1,0);

\draw[->,thick,black]
  (0.5,0) arc[start angle=0,end angle=320,radius=0.5];

\node[blue] at (-2,0.3) {$\mathbb{Z}_{F,1}$};
\node[red] at (2,0.3) {$\mathbb{Z}_{F,2}$};

\draw[dashed,gray] (0,0) -- (0,1.5);

 \foreach \radius in {1,1.5,2} { \foreach \angle in {30,60,90,120,150,210,240,270,300,330} { \draw[->,thin] ({\radius*cos(\angle)},{\radius*sin(\angle)}) -- ({(\radius+0.25)*cos(\angle)}, {(\radius+0.25)*sin(\angle)}); } }

\filldraw[fill=red, draw=black] (0,0) circle (4pt);

\node[below=4pt] at (0,-0.35) {Pions};

\end{tikzpicture}

    \caption{Deconfinement on $\mathbb{Z}_{F}$ domain walls in twisted QCD. Two $\mathbb{Z}_{F}$ domain walls of equal tension glued together. The string at the intersection is a probe color charge in a representation of $N$-ality $N/\mathrm{gcd}(N,F)$. The charge is pulled equally by the two flux tubes associated with the massive modes of the different $SU(K_{i})$ subgroups. The probe charge also emits a uniform `electric' field associated with the massless pion modes.}
    \label{fig:deconfonzffig}
\end{figure}

We can obtain the flux profile drawn in Fig.~\ref{fig:deconfonzffig} from assigning the weights in Eq.~\eqref{eq:weightsforzf} to appropriate $SU(K_{i})$ sublattices. We will assume that $\vec{\omega}_{b_{\star}}$ is the first weight of the fundamental of $SU(N)$, with $\sigma_{b_\star}$ lying within a sublattice of size $K$. The sum of the weights of the gauge string sitting at the intersection of the two $\mathbb{Z}_{F}$ domain walls of Eq.~\eqref{eq:weightsforzf} can be divided over the sublattices according to the following decomposition 
\begin{equation}
\underbrace{
\vec\omega_{b_{\star}+1} + \dots + \vec\omega_{b_{\star}+K-1}
}_{%
\begin{tikzpicture}[baseline=(current bounding box.north),scale=0.85]

\draw[very thick,blue]
(0,0) -- (-3,0);


\draw[->,very thick,blue] (0,0) -- (-1,0);

\draw[->,thick,black]
  (0.5,0) arc[start angle=0,end angle=-320,radius=0.5];

\node[blue] at (-2,0.3) {Flux Tube};

\draw[dashed,gray] (0,0) -- (0,1.5);

 \foreach \radius in {1,1.5,2} { \foreach \angle in {30,60,90,120,150,210,240,270,300,330} { \draw[->,thin] ({\radius*cos(\angle)},{\radius*sin(\angle)}) -- ({(\radius+0.25)*cos(\angle)}, {(\radius+0.25)*sin(\angle)}); } }

\filldraw[fill=red, draw=black] (0,0) circle (4pt);

\node[below=4pt] at (0,-0.25) {Pions};

\end{tikzpicture}
}
\quad+\quad
\underbrace{\vec\omega_{b_{\star}+K}+\dots+
\vec\omega_{b_{\star}+N/\mathrm{gcd}(N,F)-1}}_{%
\begin{tikzpicture}[baseline=(current bounding box.north),scale=0.85]

\draw[->,thick,black]
  (0.5,0) arc[start angle=0,end angle=-320,radius=0.5];

\draw[dashed,gray] (0,0) -- (0,1.5);

 \foreach \radius in {1,1.5,2} { \foreach \angle in {30,60,90,120,150,210,240,270,300,330} { \draw[->,thin] ({\radius*cos(\angle)},{\radius*sin(\angle)}) -- ({(\radius+0.25)*cos(\angle)}, {(\radius+0.25)*sin(\angle)}); } }

\filldraw[fill=red, draw=black] (0,0) circle (4pt);

\node[below=4pt] at (0,-0.25) {Pions};

\end{tikzpicture}
}
\quad+\quad
\underbrace{\vec\omega_{b_{\star}+N/\mathrm{gcd}(N,F)}}_{%
\begin{tikzpicture}[baseline=(current bounding box.north),scale=0.85]

\draw[very thick,red]
(0,0) -- (3,0);


\draw[->,very thick,red] (0,0) -- (1,0);

\draw[->,thick,black]
  (0.5,0) arc[start angle=0,end angle=320,radius=0.5];

\node[red] at (2,0.3) {Flux Tube};

\draw[dashed,gray] (0,0) -- (0,1.5);

 \foreach \radius in {1,1.5,2} { \foreach \angle in {30,60,90,120,150,210,240,270,300,330} { \draw[->,thin] ({\radius*cos(\angle)},{\radius*sin(\angle)}) -- ({(\radius+0.25)*cos(\angle)}, {(\radius+0.25)*sin(\angle)}); } }

\filldraw[fill=red, draw=black] (0,0) circle (4pt);

\node[below=4pt] at (0,-0.25) {Pions};

\end{tikzpicture}
} \,.
\label{eq:weightdecomppionpict}
\end{equation}

The first set of weights in Eq.~\eqref{eq:weightdecomppionpict}, i.e. $\vec\omega_{b_{\star}+1},\dots,\vec{\omega}_{b_\star+K-1}$, consists of $K-1$ of the fundamental $SU(N)$ weights associated with the chosen sublattice of size $K$. We can rewrite this weight decomposition in terms of the original weight $\vec\omega_{b_\star}$. Indeed, according to Eq.~\eqref{eq:decompositionintoreps}, each of these weights decomposes into an $SU(K)$ component and a common pion component
\begin{equation}
    \vec\omega_{b_\star+i}= \vec\omega^{(K)}_{i+1}+\vec\omega_\pi\,,\qquad i=0,\dots,K-1\,.
\end{equation}

The $SU(K)$ components are the $K$ weights of the $SU(K)$ fundamental representation and therefore satisfy $\sum_{i=0}^{K-1}\vec\omega^{(K)}_i=0$. Consequently, the sum of the first set of $K-1$ weights is
\begin{equation}
    \vec\omega_{b_\star+1}+\dots+\vec\omega_{b_\star+K-1}=-\vec\omega_1^{(K)}+(K-1)\vec\omega_\pi
    \label{eq:first-set-weight-dec}
\end{equation}

Thus, the first set of weights in Eq.~\eqref{eq:weightdecomppionpict} sources an anti-fundamental $SU(K)$ flux profile together with a massless pion profile. This is illustrated on the LHS of Eq.~\eqref{eq:weightdecomppionpict}. The $N-$ality of this set of weights is $K-1$, as $K-1$ weights of the fundamental of $SU(N)$ are involved. 

The middle set of weights in Eq.~\eqref{eq:weightdecomppionpict}, i.e. $\vec\omega_{b_\star+K},\dots,\vec\omega_{b_\star+N/\mathrm{gcd}(N,F)-1}$, consists  of only $SU(N)$ weights that involve exactly all the indices of the sublattices they span, provided that $b_\star$ labels the first site of the chosen sublattice of size $K$, as we assume here. Within each sublattice, the $SU(K_i)$ components sum to zero, leaving only the sum of the pion components. The middle set therefore sources a pion-only profile. Since it contains $N/\mathrm{gcd}(N,F)-K$ fundamental $SU(N)$ weights, its total $N$-ality is $N/\mathrm{gcd}(N,F)-K$. 

The final weight in Eq.~\eqref{eq:weightdecomppionpict} corresponds to the first fundamental weight of that sublattice, by virtue of the CFC symmetry. It corresponds to a flux tube and massless pion profile of the form in Fig.~\ref{fig:fundamentalparticle} and a particle of $N$-ality $1$. Together, the sum of all the three contributions in Eq.~\eqref{eq:weightdecomppionpict} constitutes the deconfined particle in a representation of $N$-ality $N/\mathrm{gcd}(N,F)$ in Fig.~\ref{fig:deconfonzffig} at the intersection of two $\mathbb{Z}_{F}$ domain walls. We now briefly comment on several aspects of the $\mathbb{Z}_{F}$ domain wall composites of Fig.~\ref{fig:deconfonzffig} that are distinct from the axion domain walls.

Firstly, in contrast to center-stabilized gluodynamics, twisted QCD does not have a $\mathbb{Z}_{N}$ $1$-form symmetry in the ultraviolet to measure the $N$-ality of the probe particle. This implies that the charge and $N$-ality of the deconfined probe particle on any $\mathbb{Z}_{F}$ domain wall can be completely screened at long distances by particle-anti-particle pair production, in contrast to the aforementioned $N$-ality $1$ deconfined probe particle on axion domain walls in pure gluodynamics. 

Secondly and curiously, as the domain walls of Fig.~\ref{fig:deconfonzffig} on both sides of the probe particle are associated with different $SU(K_{i})$ single-flavor QCDs, they are completely decoupled at this order in perturbation theory and there is no force between the two flux tubes and they are free to move around.

Finally, the individual flux tubes in Fig.~\ref{fig:deconfonzffig} could decay by ending on a gauge string of the form in Fig.~\ref{fig:fundamentalparticle}. The decay rate along the flux tube depends on the string tension and the domain wall tension. In four-dimensional QCD, the tension of the string and the domain wall are both expected to scale as $\sim N$, giving a suppressed decay rate per unit area $\Gamma \sim e^{-N}$ at large $N$ in $4d$ QCD \cite{Komargodski:2018odf}. 

In twisted QCD at finite $N$, the tension of the $\mathbb{Z}_{F}$ domain wall can be estimated by the mass of the dual photons (see Eq.~\eqref{eq:massivespectrummass}) and their kinetic term as $T_{\mathrm{3d}} \sim \lambda^{-1/2} \eta^{\gamma/2-1}\Lambda_{\mathrm{QCD}}^{2}$, whilst the string tension can be estimated as $T_{\mathrm{s}} \sim \lambda m_{W}$. This gives a decay rate per unit length equal to
\begin{equation}
    \Gamma_{\mathrm{3d}} \sim \mathrm{exp}\left(- \frac{T_{\mathrm{s}}^{2}}{T_{\mathrm{3d}}} \right)\sim \mathrm{exp}\left(-A \lambda^{5/2} \eta^{-1-\gamma/2} \right)\,,
    \label{eq:metastable}
\end{equation}
where the coefficient $A$ is an unknown $\mathcal{O}(1)$ number. For small enough $\eta$ and positive $\gamma$, the $\mathbb{Z}_{F}$ domain walls are therefore also metastable. The difference in the exponent between the $N$-suppressed decay rate in four-dimensional QCD and the $N$-independent exponent in twisted QCD is due to the $N$-dependent size of the compact dimension $L \ll (N\Lambda_{\mathrm{QCD}})^{-1}$, see Eq.~\eqref{eq:etasmall}. We stress that Eq.~\eqref{eq:metastable} is an estimate at finite $N$. We will solve for the $N$-scaling of $T_{\mathrm{3d}}$ at large $N$ in the next Section.

\section{Domain Walls \& Emergent Dimensions}\label{sec:emerge}

In the previous sections, we discussed the rich internal structure of the $\theta_{\mathrm{YM}}$ domain wall and of the $\mathbb{Z}_{F}$ domain wall. In this section, we aim to study the tension of the $\theta_{\mathrm{YM}}$ domain wall and of the $\mathbb{Z}_{F}$ domain wall at large $N$ in center-stabilized compactifications\footnote{The tension of domain walls in center-stabilized pure Yang--Mills have been extensively studied~\cite{Unsal:2008ch,Anber:2018xek,Anber:2015kea,Poppitz:2017ivi,Poppitz:2012nz,Chen:2020syd, Anber:2020xfk} at finite $N$ due to the connection to the tension of confining  flux tubes in pure $4d$ gluodynamics. In this regard, several conjectures considering color flux tubes and the $N$-ality dependence can be tested analytically. For example, Ref.~\cite{Poppitz:2017ivi} established a preference for an $N$-ality scaling similar to the square root of the Casimir scaling as would be expected from the MIT bag model~\cite{Johnson:1975sg}, in favor of the supersymmetrically motivated Sine law or the linear Casimir scaling law.}. The results obtained in this section are easily generalised to any domain wall at large $N$ in center-stabilized compactifications.

In contrast to the singular (see Introduction) $\eta'$ domain walls in  the large-$N$ planar limit of QCD \cite{Witten:1980sp, DiVecchia:1980yfw} or the softly-broken limit of supersymmetric QCD \cite{Csaki:2023yas}, the tension of the $\mathbb{Z}_{F}$ domain wall in center-stabilized compactified theories is analytically tractable as the theory posses no singularity or cusp. In four-dimensional supersymmetric Yang--Mills and QCD, both 
$D$-brane  \cite{Witten:1997ep,Witten:1998uka}  and supersymmetric estimates \cite{Dvali:1996xe,Kovner:1997ca,Kogan:1997dt}  suggest a tension scaling as $T _{\mathrm{4d}}\sim N\Lambda_{\mathrm{YM}/\mathrm{QCD}}^{3}$ for respectively the $\theta_{\mathrm{YM}}$ in pure gluodynamics and the $\mathbb{Z}_{F}$ domain wall in QCD. We find that in center-stabilized compactification, the $\mathbb{Z}_{F}$ domain wall tension scales as $T_{\mathrm{4d}} \sim N\Lambda_{\mathrm{QCD}}^{3}$, whilst the $\theta_{\mathrm{YM}}$ domain wall tension scales as $T_{\mathrm{4d}} \sim N \ln(N)\Lambda_{\mathrm{YM}}^{3}$. The difference between the $\mathbb{Z}_{F}$ domain wall tension in twisted QCD and the $\theta_{\mathrm{YM}}$ domain wall tension in pure Yang--Mills does highlight limits to the common lore that such domain walls should be similar. The disparity between the two tension is caused by the massless pion direction in twisted QCD, which allow the tension of the $\mathbb{Z}_{F}$ domain wall to relax compared to the tension of the $\theta_{\mathrm{YM}}$ domain wall.  

We work in the double-scaling regime where we take the limit $N\rightarrow \infty$ while keeping the product $NL$ fixed. This guarantees that both the masses of the off-diagonal gauge bosons of $SU(N)$,  $\sim m_{W} =2\pi/(NL)$, remain fixed and that the gauge coupling stops running at a fixed scale, see Eq.~\eqref{eq:etasmall}. In this large-$N$ double-scaling limit, it was demonstrated in Ref.~\cite{Cherman:2016jtu} that the $N-1$ dual photons span an emergent dimension, and that fermion zero modes act as stationary, localized branes in this extra-dimension. In this section, we establish that this tower of dual photons is equivalent (dual) to an emergent Abelian gauge field and that the localized fermionic branes act as perfect conductors in the higher dimension. We utilize these facts to establish an equivalence between domain walls in center-stabilized Yang--Mills/QCD and electrostatic charge distributions in the emergent dimension. In this way, we can analytically solve both for the axion domain wall profile and for the $\mathbb{Z}_{F}$ domain wall profile at large $N$. The tension of the domain wall becomes equivalent to the electrostatic energy stored in the charge distribution, allowing us to solve for the tension scaling as $N$ grows large.  

Let us briefly mention that in the large-$N$ planar limit \cite{tHooft:1973alw} of four-dimensional gluodynamics and QCD, there are $\sim N^2$ gluonic degrees of freedom and the four-dimensional vacuum energy density is expected to scale as $E_{\mathrm{4d}} \sim N^{2}$. In contrast, in the double-scaling limit of twisted QCD, the vacuum contains only $\sim N$ dual photons and the three-dimensional vacuum energy density scales as $E_{\mathrm{3d}} \sim N$. These results are consistent since the compactified direction has size $L \sim (N\Lambda_{\mathrm{QCD}})^{-1}$ and therefore any $3d$ energy density $E_{\mathrm{3d}}$ is related to a $4d$ energy density by $E_{\mathrm{4d}} \sim N \Lambda_{\mathrm{QCD}} E_{\mathrm{3d}}$ by a factor of $N$. Importantly, this implies that three-dimensional tensions $T_{\mathrm{3d}}$ calculated hereafter differ from their four-dimensional counterparts by a factor $L \sim ( N\Lambda_{\mathrm{QCD}})^{-1}$ as $T_{\mathrm{4d}} \sim N \Lambda_{\mathrm{QCD}} T_{\mathrm{3d}}$ and we restore this factor at the end of each section.

\subsection{Large-$N$ CP Domain Wall at $\theta_{\mathrm{YM}} =\pi$}
\label{sec:LargeNCPWallsTension}

We start by analytically solving for the CP domain wall at $\theta_{\mathrm{YM}} =\pi$ in pure center-stabilized Yang--Mills ($F=0$). This domain wall offers a simple pedagogical example to introduce our electrostatic interpretation of the emergent dimension. As $N$ grows large, we demonstrate that the tension of the CP domain wall scales as $\hat{T}_{\mathrm{3d}}\sim \pi \ln{N}$, where we denote by $\hat{T}_\text{3d}$ the 3$d$ tension of the wall stripped of all dimensionful factors. Solving for the dual photon profile on the CP domain wall is equivalent to solving for the dual photon profile on the axion domain wall $\theta_{\mathrm{YM}} \rightarrow \theta_{\mathrm{YM}} + 2\pi$, as the dual photons are so heavy that they are only excited near the cuspy point $\theta_{\mathrm{YM}} =\pi$. The tension obtained in this section could thus also be the dominant contribution  to the axion domain wall tension at large $N$. 

From Eq.~\eqref{eq:YangMillsthetaaction}, the dual photon action including the $\theta_{\mathrm{YM}}$-angle in center-stabilized Yang--Mills is
\begin{equation}
    S = \int d^3x\ \left[\sum_{j=1}^{N}\frac{g^{2}}{8 \pi^{2}L}(\partial \sigma_{j})^{2} -\sum_{j=1}^{N}  \zeta \cos{\left( \sigma_{j+1}-\sigma_{j} - \frac{\pi}{N}\right)} +N\zeta \cos{\left(\frac{\pi}{N} \right)}\right]\, ,
    \label{eq:unsupressed-action}
\end{equation}
where we have added an additional constant $N\zeta\cos{\left(\pi/N\right)}$ offset that sets the minimum of the potential to zero. As explained in Section \ref{sec:deconfzfdw}, the dual photon potential in Eq.~\eqref{eq:unsupressed-action} has two gauge-inequivalent 
CP-conjugate vacua at $\theta_{\mathrm{YM}} = \pi$ with the same energy:
\begin{equation}
    \vec{\alpha}_j\cdot\vec{\sigma} = 0 \qquad \mathrm{or} \qquad \vec{\alpha}_{j} \cdot \vec\sigma = - \frac{2 \pi}{N} \quad \mathrm{with} \qquad \vec\alpha_{N} \cdot \vec\sigma = 2\pi - \frac{2\pi}{N}\,,
    \label{boundaryconditoinsCPwall}
\end{equation}
where we have chosen the $b_{\star}$ gauge in which the additional $2\pi$ winding occurs for the $N$-th monopole coordinate. The boundary conditions in Eq.~\eqref{boundaryconditoinsCPwall} indicate that the differences between adjacent fields $\sigma_{j}$ and $\sigma_{j+1}$ become small as $N$ grows large, with one crucial exception: in the $N\to\infty$ limit, $\sigma_N-\sigma_1$ goes from $0$ to $2\pi$ across the domain wall. 

From this point on, let $\sigma_j\equiv\sigma_j(z)$ be the domain-wall profile of the $j$-th dual photon, with $z$ the coordinate transverse to the wall (the profile is constant along the remaining directions). Evaluating the action in Eq.~\eqref{eq:unsupressed-action} on the full set of dual photon $\{\sigma_i\}_{i=1,\dots,N}$, we obtain
\begin{equation}
    T_\text{3d}=\dfrac{S}{V_\parallel} = \int dz\ \left[\sum_{j=1}^{N}\frac{g^{2}}{8 \pi^{2}L}(\partial_{z} \sigma_{j})^{2} -\sum_{j=1}^{N}  \zeta \cos{\left( \sigma_{j+1}-\sigma_{j} - \frac{\pi}{N}\right)} +N\zeta \cos{\left(\frac{\pi}{N} \right)}\right]\, ,
\end{equation}
where $T_\text{3d}$ is the 3$d$ tension of the CP domain wall, while $V_\parallel$ is its worldvolume. We find it convenient to introduce dimensionless spatial variables: 
\begin{equation}
    \hat{z} = z \sqrt{\frac{4 \pi^{2}L\zeta}{g^{2}} } = z \sqrt{\frac{8 \pi^{3}\zeta}{\lambda m_{W}}} \,.
\label{eq:rescaling-z}
\end{equation}

This convention is independent of $N$ and greatly simplifies our formulae. We can define a rescaled tension accordingly,
\begin{equation}
    \hat{T}_\text{3d} \equiv T_\text{3d} \sqrt{\frac{8\pi^{3}}{\lambda m_{W} \zeta}} = \int d \hat{z} \ \left[\sum_{j=1}^{N}\frac{1}{2}(\partial_{\hat{z}} \sigma_{j})^{2} -\sum_{j=1}^{N}  \cos{\left( \sigma_{j+1}-\sigma_{j} - \frac{\pi}{N}\right)} +N\cos{\left(\frac{\pi}{N} \right)}\right]\, .
    \label{eq:thetalagpot}
\end{equation}

We will arrange the periodic dual photons $\sigma_{j} \equiv \sigma_{j+N}$ on the sites of a periodic lattice of length $2\pi$. The position of the $j$-th lattice site can be specified by a coordinate $\varphi_j$, and the spacing between adjacent sites is $\varphi_{j+1}-\varphi_{j}=2\pi/N$. In the limit $N\to \infty$, the dual photons span an emergent dimension, as the set $\{\varphi_j\}_{j=1,\dots,N}$ becomes dense in the interval $(-\pi,\pi]$. We are free to parametrize the emergent dimension with a continuous periodic coordinate $\varphi\in(-\pi,\pi]$. Furthermore, the collection $\{\sigma_1 (\hat z),\dots,\sigma_N(\hat z)\}$ 
can be thought of as a single dual photon field $\sigma(\hat{z},\varphi)$, using the identification
\begin{equation} 
\sigma_j(\hat{z}) \simeq \sigma(\hat{z},\varphi_j)\,,\qquad\text{as $N\to\infty$}\,.
    \label{eq:dual-photon-field}
\end{equation}

Relying on the dual photon field description, we can readily translate the discrete differences between neighboring dual photons into a derivative by considering a suitable incremental ratio:
\begin{equation}
  \frac{\sigma_{j+1}(\hat{z})-\sigma_{j}(\hat{z})}{2\pi/N} \simeq \partial_{\varphi}\sigma(\hat{z},\varphi_j)\,,\qquad \text{as $N\to\infty$}\,.
   \label{eq:finitedifop}
\end{equation}

If we substitute Eqs.~\eqref{eq:dual-photon-field}, \eqref{eq:finitedifop} into Eq.~\eqref{eq:thetalagpot} and assume that the cosines in Eq.~\eqref{eq:thetalagpot} can be well approximated by their quadratic Taylor expansion around zero, the three-dimensional tension becomes
\begin{align}
   \hat{T}_\text{3d} &\simeq\int d\hat{z} \, d\varphi\ \left[   \frac{N}{4\pi}(\partial_{\hat{z}} \sigma)^{2} + \frac{1}{2} \frac{2\pi}{N}\left(\partial_{\varphi}\sigma  - \frac{1}{2}\right)^{2} -\frac{\pi}{4N}\right]\,,\qquad\text{as $N\to\infty$}\,.
    \label{eq:emergdim}
\end{align}

We stress that Eq.~\eqref{eq:emergdim} does not exactly correspond to the large-$N$ limit of Eq.~\eqref{eq:thetalagpot}. This is because we assumed Eq.~\eqref{eq:finitedifop} to hold for any $j=1,\dots,N$. However, the limit in Eq.~\eqref{eq:finitedifop} is not well defined when considering the last difference $\sigma_N-\sigma_1$. Indeed, this difference is not small everywhere: for $N\to\infty$, the difference $\sigma_N-\sigma_1 \rightarrow 2\pi$ on one of the domain wall boundaries, according to Eq.~\eqref{boundaryconditoinsCPwall}. Moreover, as the difference $\sigma_{N}-\sigma_{1}$ is $\mathcal{O}(1)$ across the domain wall, the cosine corresponding to this single difference in  Eq.~\eqref{eq:thetalagpot} is poorly approximated by the quadratic Taylor polynomial around zero. 

Nonetheless,  
we solve for the analytical domain-wall profiles of the dual photons using Eq.~\eqref{eq:emergdim}. 
This approximation  
introduces an error, but we show that it does not affect our conclusions. Later in this section, we also briefly comment on how this procedure could, in principle, be improved.

\begin{figure}[ht]
\centering

\begin{tikzpicture}[scale=2]

    \def\N{8}

    \def\R{1.5}

     \def\ArcR{0.45}

    \draw[->] (-1.8,0) -- (1.8,0) node[right] {};
    \draw[->] (0,-1.8) -- (0,1.8) node[above] {};

\foreach \k [evaluate=\k as \i using int(\k+1)] in {0,...,\numexpr\N-1\relax}{
    \pgfmathsetmacro{\ang}{180/\N + 360*\k/\N}

    \draw[thick,brown]
        (0,0) --
        ({\R*cos(\ang)},{\R*sin(\ang)});

}

\node at
    (-{1.12*\R*cos(180/\N)},
     -{1.12*\R*sin(180/\N)})
    {$\sigma_1$};

\node at
    (-{1.12*\R*cos(180/\N + 360/\N)},
     -{1.12*\R*sin(180/\N + 360/\N)})
    {$\sigma_2$};

\node at
    (-{1.12*\R*cos(180/\N + 360*(\N-2)/\N)},
     -{1.12*\R*sin(180/\N + 360*(\N-2)/\N)})
    {$\sigma_{N-1}$};

\node at
    (-{1.12*\R*cos(180/\N + 360*(\N-1)/\N)},
     -{1.12*\R*sin(180/\N + 360*(\N-1)/\N)})
    {$\sigma_N$};

\pgfmathsetmacro{\angA}{180/\N + 2*360/\N}
\node at
(-{1.15*\R*cos(\angA)},
 -{1.15*\R*sin(\angA)})
{$\cdots$};

\pgfmathsetmacro{\angB}{180/\N + (\N-3)*360/\N}
\node at
(-{1.15*\R*cos(\angB)},
 -{1.15*\R*sin(\angB)})
{$\cdots$};

    \draw[red,thick]
        ({\ArcR*cos(180/\N)},{\ArcR*sin(180/\N)})
        arc[
            start angle={180/\N},
            end angle={180/\N + 360/\N},
            radius=\ArcR
        ];

    \node[red] at
        ({1.25*\ArcR*cos(180/\N + 180/\N)},
         {1.25*\ArcR*sin(180/\N + 180/\N)})
        {$\ \ \frac{2\pi}{N}$};
   \draw[red, dashed, thick]
    (-1,0.05) to (-2,0.05);
   \node[
    draw=red,
    text=red,
    fill = white,
    circle,
    minimum size=5mm,
    inner sep=3pt] (A) at (-1,0) {$\bm{+1}$};

      \node[
    draw=blue,
    text=blue,
    fill = white,
    circle,
    minimum size=1mm,
    inner sep=1pt] (A) at (0,0) {$\bm{-\frac{1}{2}}$};


\end{tikzpicture}

 \caption{Vortex distribution of $\sigma(\omega)$ in the emergent dimension for pure Yang--Mills ($F=0$) and general $N$ at $\theta_{\mathrm{YM}} = \pi$. The $-\frac{1}{2}$ charge (blue) at the origin provides a background field not included in $\sigma(\omega)$. The unit positive charge (red), provides the winding at $r=\infty$. The radial lines (brown) corresponds to the dual photons profiles $\sigma_{j}(\hat{z})$. The value of the field $\sigma_{j}$ at a given radius $r$ is the angle that point makes with respect to the location of the positive unit charge. A branch cut emanates from the positive charge to the left (red dashed line). }
    \label{fig:F0thetapiemerg}
    \end{figure}
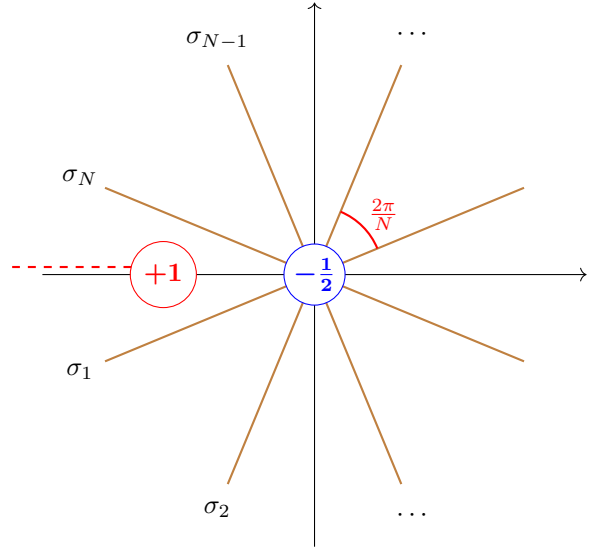

For our purposes, a more convenient form of Eq.~\eqref{eq:emergdim} can be found by conformally mapping the two-dimensional cylinder $(\hat{z},\varphi) \in (-\infty,\infty) \times (-\pi,\pi]$ spanned by the coordinates $\hat{z}$ and $\varphi$ to the $2d$ plane with radial coordinate $r = \mathrm{exp}(2\pi \hat{z}/N)$ and angular coordinate $\varphi$. Moving radially away from the center of this plane corresponds to increasing $\hat{z}$, whilst angles are still associated with the coordinate $\varphi$. We have drawn this plane in Fig.~\ref{fig:F0thetapiemerg}.

By applying the above change of coordinates to Eq.~\eqref{eq:emergdim}, we obtain the action of a two-dimensional compact scalar with a shifted angular derivative:
\begin{equation}
    \hat{T}_\text{3d} \simeq \frac{1}{2}\int dr \, d\varphi \left[  r (\partial_{r}\sigma)^{2} + \frac{1}{r} \left(\partial_{\varphi} \sigma - \frac{1}{2}\right)^{2} - \frac{1}{4r} \right] \, .
    \label{eq:largeNtension}
\end{equation}

We may now draw an instructive analogy with electromagnetism by reintroducing the suppressed time dimension.\footnote{ Remarkably, the analogy holds even though the 2$d$ plane parametrized by $(r,\varphi)$ is not purely a spacetime plane: while $r$ is the image of the spacetime coordinate $\hat{z}$ under the conformal map, the angle $\varphi$ emerges from the continuum limit of an auxiliary discrete coordinate $\varphi_j$ used to label the dual photons, and therefore lives in an internal space.} In this picture, we can regard the dual photon field as a \textit{static} compact scalar field in $d=2+1$ dimensions and Eq.~\eqref{eq:largeNtension} as its action per unit time. Through the duality with an Abelian gauge field ($\partial_{i}\sigma_a \propto \epsilon_{ijk}  \partial_{j}A^{a}_{k}$), Eq.~\eqref{eq:largeNtension} can then be read as the electrostatic energy of the corresponding gauge field configuration. 

Electrically charged particles under this gauge field correspond to vortices in the $\sigma$ dual frame and their charge is determined by a closed integral measuring the winding $\frac{1}{2\pi}\int d\sigma$ around the object, similar to what is  detailed in Section~\ref{sec:representationtheory}. The constant $\theta_{\mathrm{YM}} = \pi$ shift of the angular derivative in Eq.~\eqref{eq:largeNtension} corresponds to a $-\frac{1}{2}$ winding and thus a background fractional charge $-\frac{1}{2}$ sitting at $r=0$. 

In two spatial dimensions, the electrostatic energy of a particle is logarithmically divergent, as electric fields in two spatial dimensions decay like $\frac{1}{r}$. The final term in Eq.~\eqref{eq:largeNtension} subtracts the divergent self-energy of the fractional background charge and corresponds to the final term in Eq.~\eqref{eq:thetalagpot} that sets the minimum of the potential to zero. We have drawn the negative background charge in Fig.~\ref{fig:F0thetapiemerg} 
for visualization purposes. 

We now solve for the domain-wall profile $\sigma(\hat{z},\varphi)$ in the presence of this charge $-\frac{1}{2}$ background field. Under the aforementioned small-difference assumption, the limit in Eq.~\eqref{eq:finitedifop} exists and is finite for $\hat{z}\to\pm\infty$ and for all $\varphi\in(-\pi,\pi]$ , allowing us to recast the boundary conditions for the CP domain wall profile given by Eq.~\eqref{boundaryconditoinsCPwall} in the continuum as follows:
\begin{equation}
    \partial_{\varphi}\sigma(\hat{z}=-\infty,\varphi) = 0\,,  \qquad \partial_{\varphi}\sigma(\hat{z}=+\infty,\varphi) = 1\, .
    \label{eq:diffopbound}
\end{equation}
Since these imply a total winding of $2\pi$ across $(-\pi,\pi]$ at $r=\hat z=+\infty$, we must also introduce a branch cut along the negative real axis, to accommodate the $2\pi$ jump of the dual photon field:
\begin{equation}
    \sigma(+\infty,\pi)-\sigma(+\infty,-\pi)=\int_{-\pi}^{\pi} d\varphi \ \partial_\varphi \sigma(+\infty,\varphi) = 2\pi\,.
    \label{eq:branch-cut}
\end{equation}

The boundary conditions of Eq.~\eqref{eq:diffopbound} are such that at $r=0$ there is no winding in the solution, whilst at $r\to\infty$, the solution has a rotational winding equal to $1$. This implies that there must be a unit charge sitting somewhere at finite $r>0$. In terms of the dual photon field $\sigma$, this corresponds to a  vortex that has been displaced away from the center of the coordinate system, see Fig.~\ref{fig:F0thetapiemerg}. The solution is therefore of the form
\begin{equation}
    \sigma(\omega) =\arg{\left( \omega - c\right)} -\mathrm{arg}(-c)\,, \qquad \mathrm{with} \qquad \omega = r e^{ i \varphi} \,.
    \label{eq:largeNthetapiprof}
\end{equation}

The constant $c \neq 0$ in Eq.~\eqref{eq:largeNthetapiprof} determines the position of the unit charge. In order to find it, consider the discrete picture drawn in Fig.~\ref{fig:F0thetapiemerg}. We choose to rotate the coordinate system such that the emergent angles $\varphi_{j}$ satisfy $\varphi_{j} = -\pi\frac{N+1-2j}{N}$, for $j=1,\dots,N$. In this way, the branch cut is placed between $\sigma_N$ and $\sigma_1$, consistently with the boundary conditions in Eq.~\eqref{boundaryconditoinsCPwall}.

We can partition the 2$d$ conformal plane into wedges $\Delta_{j,j+1}$ subtended by the angles $\phi_{j}$ and $\phi_{j+1}$, for $j=1,\dots,N$. In light of the analogy with electromagnetism we drew above, we can then find the position of the unit charge by measuring the electrostatic flux though each wedge. The total charge contained in the wedge $\Delta_{j,j+1}$ is given by
\begin{equation}
    Q_{j,j+1} = \frac{1}{2\pi}\oint_{\Delta_{j,j+1}} d\sigma \simeq \frac{1}{2\pi} \big[\sigma(+\infty,\varphi_{j}) - \sigma(+\infty,\varphi_{j+1}) \big]\,,\qquad\text{as $N\to\infty$}\,.
    \label{eq:chargeinwedge}
\end{equation}

According to the large-$N$ limit of the boundary conditions in Eq.~\eqref{boundaryconditoinsCPwall} which reproduces Eq.~\eqref{eq:branch-cut}, the unit charge resides in the wedge $\Delta_{N,1}$, as drawn in Fig.~\ref{fig:F0thetapiemerg}. The charge sits on the negative real axis, and by the $r\rightarrow \lambda r$ scaling symmetry of Eq.~\eqref{eq:largeNtension}, we are free to set $c=-1$.\footnote{The symmetry $r \rightarrow \lambda r$ is ordinarily not a symmetry of Eq.~\eqref{eq:largeNtension} due to the need to introduce a scale-dependent regulator, such as a finite core size, for the logarithmically divergent energy in $(2+1)$d electromagnetism. In our case, the regulator is the scale-independent finite angular resolution $\Delta \varphi = \frac{2\pi}{N}$ in Fig.~\ref{fig:F0thetapiemerg}, and this rescaling is a symmetry. } 

\begin{figure}[ht]
        \centering
        \includegraphics[width=0.75\textwidth]{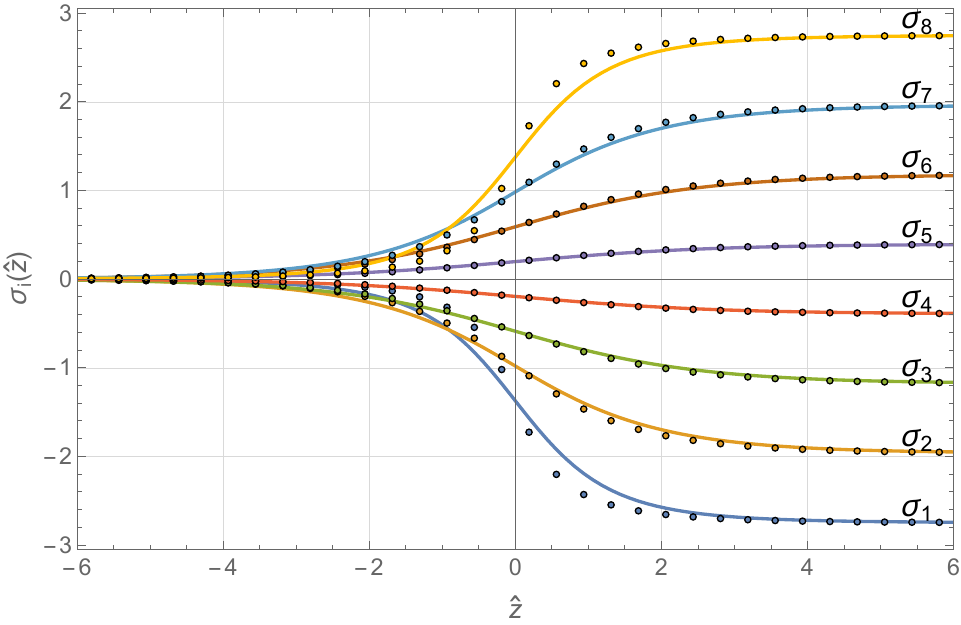}

\hspace{1cm}

    \caption{Profile of $\sigma(\omega)$ in the emergent dimension for $F=0$ and $N =8$ at $\theta_{\mathrm{YM}} = \pi$. The numerical profiles (lines) are found by minimizing the action~\eqref{eq:thetalagpot} with boundary conditions in Eq.~\eqref{boundaryconditoinsCPwall}. The analytical profiles (circles) correspond to the profiles in Eq.~\eqref{eq:largeNthetapiprof}. Numerical and analytical profiles show good agreement at large $N$, apart from  $\sigma_{1}$ and $\sigma_{8}$, because their difference at the boundary does not decrease as $N\to \infty$.}
    \label{figF0thetapiprofile}
\end{figure}

The full charge configuration is drawn in Fig.~\ref{fig:F0thetapiemerg}. In Fig.~\ref{figF0thetapiprofile}, we compare the analytical profile (circular dots) of Eq.~\eqref{eq:largeNthetapiprof} at $N=8$ to the numerical profile (continuous lines). The numerical profiles were found by minimizing the action~\eqref{eq:thetalagpot} with boundary conditions in Eq.~\eqref{boundaryconditoinsCPwall}. The analytical domain-wall profile $\sigma_j(\hat{z})$ of the $j$-th dual photon was found by placing the dual photon field solution \eqref{eq:largeNthetapiprof} into Eq.~\eqref{eq:dual-photon-field}, yielding:
\begin{equation}
    \sigma_j(\hat{z})\simeq\arg(\omega_j+1),\qquad \omega_j=\exp\left(\frac{2\pi\hat{z}}{N}+i\varphi_j\right)\,.
    \label{eq:analytical-profile}
\end{equation}
Note that, given our choice of the lattice points $\varphi_j=-\pi\frac{N+1-2j}{N}$, which are symmetric under $\varphi_j\to-\varphi_{N+1-j}$ and avoid the branch cut, these approximate dual photon profiles satisfy the usual constraint traceless sum constraint $\sum_j \sigma_j=0$.

By looking at Fig.~\ref{figF0thetapiprofile}, we observe a general good agreement between analytical and numerical profiles for all dual photons, except for the outermost ones, $\sigma_{1}$ and $\sigma_{8}$. As we remarked earlier, this discrepancy can be traced back to their difference at the boundary not decreasing as $N\to \infty$. The small-difference approximation allowed us to bypass this issue entirely. Had we not made it, we would have had to deal with two obstacles we mentioned before:
\begin{enumerate}
    \item The derivative in Eq.~\eqref{eq:finitedifop} is ill-defined for $j=N$.  
    \item The cosine for the last difference $\sigma_{N}-\sigma_{1}$ in Eq.~\eqref{eq:thetalagpot} can no longer be approximated by the second-order Taylor polynomial around zero  and anharmonicities will contribute. 
\end{enumerate}

Furthermore, because of the cosine terms in the action \eqref{eq:unsupressed-action} paring each $\sigma_j$ to its neighboring $\sigma_{j+1}$, the error cascades all the way down the the dual photon lattice, propagating first to $\sigma_{N-1}$ and $\sigma_2$, then to a lesser extent to $\sigma_{N-2}$ and $\sigma_3$ through $\sigma_{N-1}$ and $\sigma_2$, then to an even lesser extent to $\sigma_{N-3}$ and $\sigma_4$ through $\sigma_{N-2}$ and $\sigma_3$, and so on. This explains why the the agreement between analytical and numerical domain-wall profiles increases as we move inward closer to the $\hat{z}$-axis in Fig.~\ref{figF0thetapiprofile}, and confirms that Eq.~\eqref{eq:dual-photon-field} is not exact over the entire domain of $\hat{z}$, as expected. 

We briefly note that one could potentially improve upon the approximations made in this section by substituting the analytical profiles $\sigma_j(\hat{z})$ for $1<j<N$ back into Eq.~\eqref{eq:thetalagpot}, and then solving for $\sigma_{1}-\sigma_{N}$ separately. We will nevertheless refrain from pursuing this here, since our approximations suffice to estimate the leading-order contribution to the CP domain-wall tension at $\theta_{\rm YM}=\pi$ at large-$N$, as we show in the following.

The 3$d$ tension cannot be computed from Eq.~\eqref{eq:largeNtension} by direct substitution of the analytical profile in Eq.~\eqref{eq:largeNthetapiprof} due to logarithmic singularities at the string core: if we use the identities
\begin{equation}
    r\partial_{r}\sigma(\omega) = \mathrm{Im} \left( \frac{\omega}{\omega - c} \right)\,, \qquad 
    \partial_{\varphi}\sigma(\omega) =  \mathrm{Re} \left( \frac{\omega}{\omega - c} \right)  
    \label{eq:emdif}
\end{equation}
and compute the tension from Eq.~\eqref{eq:largeNthetapiprof},
\begin{align}
   \hat{T}_\text{3d}&=\frac{1}{2}\int \frac{dr}{r}\,d\varphi \ \left\{\mathrm{Im}\left( \frac{\omega}{\omega+1}\right)^{2} + \left[ \mathrm{Re}\left( \frac{\omega}{\omega+1}\right) - \frac{1}{2}\right]^{2} - \frac{1}{4} \right\} = -\frac{1}{2} \int_{-\pi}^{\pi} d\varphi \ \varphi  \cot{\left(\varphi \right)}\, ,
   \label{eq:tensioncpwall-continuous}
\end{align}
we get a logarithmically divergent result. This is expected on the grounds of the analogy with electrostatics we delineated above: the self-energy of a point charge in $d=2+1$ dimensions, or equivalently the tension of a string in the same number of dimensions (obtained by letting the charge extend along the time direction), diverges logarithmically as one approaches the core of the string ($\varphi\to\pm\pi$).\footnote{ The other logarithmic divergence that would usually  contribute to the energies of strings far away from their core at large radii is canceled by the additional  offset in Eq.~\eqref{eq:emergdim} and the energy density decays exponentially quickly at large radii. }

In fact, it is perhaps surprising that the electrostatic energy $\hat{T}_\mathrm{3d}$ of the charge distribution in Fig.~\ref{fig:F0thetapiemerg} would not depend on the distance between the charges and that the configuration is static at all. In $(2+1)$d electromagnetism, the energy between two unit point charges at a distance $r$ is $\sim \ln{\frac{r}{a}}$ where $a$ is a scale-dependent regulator, usually taken to be the core size of the charges. In the large-$N$ limit under consideration, the divergence of at the core of the charges is instead regulated by the angular lattice spacing of the emergent dimension, the angular resolution $\Delta \varphi = \frac{2\pi}{N}$. This scale-independent regulator implies that the action in Eq.~\eqref{eq:largeNtension} has a symmetry $r\rightarrow r\lambda$, and any finite energy configuration cannot depend on the distance $r$.  

To explicitly calculate the tension with this finite angular resolution, we therefore step back from the continuum description and regulate the integral $\int d\varphi$ in Eq.~\eqref{eq:tensioncpwall-continuous} with a discrete sum $\frac{2\pi}{N}\sum_{j=1}^{N}$ on the $N$ dual-photon lattice sites. By virtue of Eq.~\eqref{eq:dual-photon-field}, we can repeat the same steps that led from Eq.~\eqref{eq:largeNtension} to Eq.~\eqref{eq:tensioncpwall-continuous} and arrive at the regulated, large-$N$ limit expression for the 3$d$ tension:
\begin{equation}
    \hat{T}_\text{3d} \simeq -\frac{\pi} {N}\sum_{j=1}^{N} \varphi_j  \cot{\left(\varphi_j\right)}\, .
   \label{eq:tensioncpwall}
\end{equation}

The angles $\varphi_{j}$ are given by $ \varphi_{j}= -\pi\frac{N+1-2j}{N}$ for $j=1,\dots,N$. We approach the core of the string only up to an angular distance $\sim \frac{1}{N}$ (see Fig.~\ref{fig:F0thetapiemerg}) and there are $\mathcal{O}(N)$ points close enough to the core divergence at $\varphi_j \rightarrow \pm\pi$ for the sum in Eq.~\eqref{eq:tensioncpwall} to behave like a harmonic series. The other $\mathcal{O}(N)$ lattice points each contribute a roughly constant $\mathcal{O}(1/N)$ tension due to the prefactor in Eq.~\eqref{eq:tensioncpwall}, and therefore collectively only add an $\mathcal{O}(1)$ overall contribution. Including both the harmonic contribution from the near core points and the overall $\mathcal{O}(1)$ yields a tension that scales as
\begin{equation}
    \hat{T}_\text{3d} \simeq \sum_{j=1}^{\mathcal{O}(N)} \frac{\pi}{j} + \mathcal{O}(1) \simeq \pi \ln{N} + \mathcal{O}(1) \,.
    \label{eq:tensionexp}
\end{equation}

Restoring units and working with a four-dimensional tension $T_{\mathrm{4d}} \sim N \Lambda_{\mathrm{YM}} T_{\mathrm{3d}}$, we estimate a tension for the CP or axion domain wall,
\begin{equation}
    T_{\mathrm{4d}} \sim \pi N \ln{(N)} \Lambda_{\mathrm{YM}}^{3} + \mathcal{O}(N (\ln N)^0)\,,
\end{equation}
up to polynomial prefactors in the ’t Hooft coupling $\lambda$ and in $\eta$. This is the main result of this section on CP/axion domain walls.

\begin{figure}[ht]
    \centering
    \includegraphics[width=0.75\linewidth]{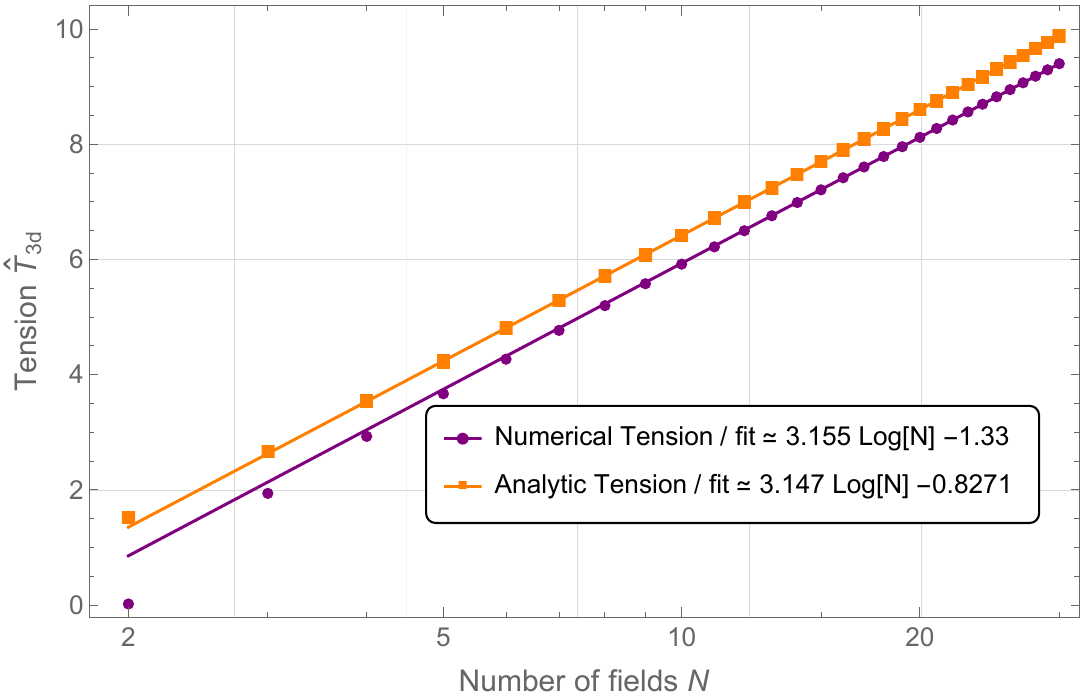}
    \caption{Scaling of the tension $\hat{T}_\text{3d}$ of the CP domain wall in pure Yang--Mills at $\theta_{\mathrm{YM}} = \pi$ on a linear axis versus the number of dual photons $N \in \left( 2,30\right)$ on a logarithmic axis. The orange square markers correspond to the tension of the analytical profile~\eqref{eq:largeNthetapiprof}, whilst the purple round markers correspond to the tension of the numerical profile found by minimizing~\eqref{eq:thetalagpot}. Best fit values for the curves $\hat{T}_\text{3d} = a \ln{N} + b$ for $N>15$ are shown. The fits agree well with a logarithmically scaling tension at large $N$.  }
    \label{fig:tensionscalingthetapiF0}
\end{figure}

In Fig.~\ref{fig:tensionscalingthetapiF0}, we have plotted both the tension of the analytical profile~\eqref{eq:analytical-profile} and the numerical profile found by minimizing the action in Eq.~\eqref{eq:largeNtension}. The tension of both profiles is calculated using the full Lagrangian~\eqref{eq:thetalagpot} and shows a clear logarithmic scaling with $N$ with a slope in near perfect agreement with $ \pi$, as expected from Eq.~\eqref{eq:tensionexp}. The two tensions differ, however, by an $\mathcal{O}(1)$ number ($\Delta T \simeq 0.5$) at large $N$. We conclude that the small-difference approximation was sufficient to reproduce the leading-order contribution to the domain-wall tension, but failed to capture the sub-leading $\mathcal{O}(1)$ number predicted by Eq.~\eqref{eq:tensionexp}. Indeed, this number is incalculable, since we assumed that the difference $\sigma_{N}-\sigma_{1}$ is everywhere small, which is not true. 

It is one of the beauties of doing physics that, by taking the large-$N$ limit of center-stabilized Yang--Mills and by considering the resulting emergent dimension spanned by its infrared spectrum, we arrive at the $(2+1)$-dimensional analogue of the $(1+1)$-dimensional Schwinger model~\cite{Schwinger:1962tp, Coleman:1976uz}, which we depicted in Fig.~\ref{fig:schwingerdecon} and which should be compared to charges in Fig.~\ref{fig:F0thetapiemerg}. In both models, individual unit charges are (log-)confined. In both models the angle $\theta_{\mathrm{YM}} =\pi$ is equivalent to adding a constant electric background field associated with a fractional charge. In both models, a unit charge in this background is deconfined and free to move. In the $(2+1)$-dimensional model at hand, this deconfinement, the fact that the energy density did not depend on the position between the charges, is due to the scale-independent, finite angular resolution $\Delta \varphi = 2\pi/N$. The ability to change the position of the unit charge along the horizontal axis, $\mathrm{Re}(c) \rightarrow \mathrm{Re}(c) + \mathrm{constant}$, corresponds to the translational zero mode of the domain wall. The ability to rotate the unit charge in the plane by an angle $\mathrm{arg}(c) \rightarrow \mathrm{arg}(c) +2\pi/N$ corresponds to the $\mathbb{Z}_{N}$ permutation symmetry of the dual photons. This rotation changes the boundary conditions in Eq.~\eqref{boundaryconditoinsCPwall} and results in a distinct domain wall profile: the $N$ distinct CP domain walls with equal tension. This concludes our study of the CP domain wall at $\theta_{\mathrm{YM}} = \pi$  in center-stabilized Yang--Mills at large $N$. 

\subsection{Large-$N$ $\mathbb{Z}_{F}$ Domain Wall}
\label{sec:LargeNZFWallsTension}

In this section, we study the domain wall profiles and tension scaling of the $\mathbb{Z}_{F}$ domain walls of Section~\ref{sec:deconfnewzf} for the case of massless twisted QCD. In the large-$N$ double-scaling limit, the domain wall profiles can once again be associated with charge distributions in the emergent dimension. We find that the three-dimensional, dimensionless tension of the $\mathbb{Z}_{F}$ domain wall scales as $\hat{T}_{\text{3d}} \sim  \mathcal{O}(1)$, in sharp contrast with the logarithmic tension scaling of the $\theta_{\mathrm{YM}}$ domain wall in the previous section. 

In the dimensionless units introduced in Section~\ref{sec:LargeNCPWallsTension}, the Lagrangian in the dual photon basis reads
\begin{equation}
    \hat{\mathcal{L}} = \sum_{i=1}^{N}\frac{1}{2}(\partial_{\hat{z}} \sigma_{i})^{2} -\sum_{b\in \mathcal{B}}  \cos{\left( \sigma_{b}-\sigma_{b+1}\right)} + (N-F) \,.
    \label{eq:laglargeN}
\end{equation}
This time, the sum only includes the cosines without fermion zero modes and the last term is added to set the potential at its minimum to zero. The difference between successive $\sigma_{i}$'s will be small at large $N$, apart from a finite number of differences. In the following, we once again employ the small-difference approximation of Section~\ref{sec:LargeNCPWallsTension} and rely on the continuum description therein to solve for the analytical domain-wall profiles of the dual photons.

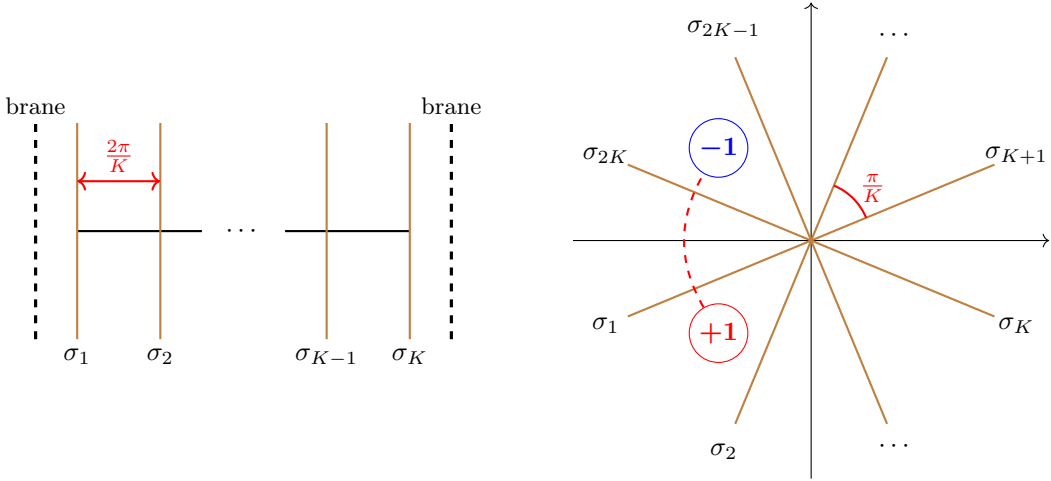
\begin{figure}[ht]
\centering
\begin{minipage}{0.48\textwidth}
\centering
\begin{tikzpicture}[scale=1.1]

    \def\H{1.3}

    \draw[thick] (1,0) -- (2.5,0);
    \draw[thick] (3.5,0) -- (5,0);
    \node at (3,0) {$\cdots$};

    \foreach \x in {0.5,5.5}{
        \draw[very thick, dashed] (\x,-\H) -- (\x,\H);
        \node[above, font=\small] at (\x,\H) {brane};
    }

    \foreach \x/\lab in {1/\sigma_1, 2/\sigma_2, 4/\sigma_{K-1}, 5/\sigma_K}{
        \draw[thick, brown] (\x,-\H) -- (\x,\H);
        \node[below] at (\x,-\H) {$\lab$};
    }

    \draw[<->, red, thick] (1,0.6) -- (2,0.6);
    \node[red, above] at (1.5,0.6) {$\frac{2\pi}{K}$};

\end{tikzpicture}
\end{minipage}
\hfill
\begin{minipage}{0.48\textwidth}
    \begin{tikzpicture}[scale=1.75]

    \def\N{8}

    \def\R{1.5}

     \def\ArcR{0.45}

    \draw[->] (-1.8,0) -- (1.8,0) node[right] {};
    \draw[->] (0,-1.8) -- (0,1.8) node[above] {};

\foreach \k [evaluate=\k as \i using int(\k+1)] in {0,...,\numexpr\N-1\relax}{
    \pgfmathsetmacro{\ang}{180/\N + 360*\k/\N}

    \draw[thick,brown]
        (0,0) --
        ({\R*cos(\ang)},{\R*sin(\ang)});

}

\node at
    ({1.12*\R*cos(180/\N)},
     {1.12*\R*sin(180/\N)})
    {$\sigma_{K+1}$};

\node at
    ({1.12*\R*cos(180/\N + 360/\N)},
     {1.12*\R*sin(180/\N + 360/\N)})
    {$\cdots$};

\node at
    ({1.12*\R*cos(180/\N + 360*3/\N)},
     {1.12*\R*sin(180/\N + 360*3/\N)})
    {$\sigma_{2K}$};

\node at
    ({1.12*\R*cos(180/\N + 360*4/\N)},
     {1.12*\R*sin(180/\N + 360*4/\N)})
    {$\sigma_{1}$};

\node at
    ({1.12*\R*cos(180/\N + 360*(\N-2)/\N)},
     {1.12*\R*sin(180/\N + 360*(\N-2)/\N)})
    {$\cdots$};

\node at
    ({1.12*\R*cos(180/\N + 360*(\N-1)/\N)},
     {1.12*\R*sin(180/\N + 360*(\N-1)/\N)})
    {$\sigma_{K}$};

\pgfmathsetmacro{\angA}{180/\N + 2*360/\N}
\node at
({1.15*\R*cos(\angA)},
 {1.15*\R*sin(\angA)})
{$\sigma_{2K-1}$};

\pgfmathsetmacro{\angB}{180/\N + (\N-3)*360/\N}
\node at
({1.15*\R*cos(\angB)},
 {1.15*\R*sin(\angB)})
{$\sigma_{2}$};

    \draw[red,thick]
        ({\ArcR*cos(180/\N)},{\ArcR*sin(180/\N)})
        arc[
            start angle={180/\N},
            end angle={180/\N + 360/\N},
            radius=\ArcR
        ];

    \node[red] at
        ({1.25*\ArcR*cos(180/\N + 180/\N)},
         {1.25*\ArcR*sin(180/\N + 180/\N)})
        {$\ \ \frac{\pi}{K}$};

   \node[
    draw=red,
    text=red,
    fill = white,
    circle,
    minimum size=5mm,
    inner sep=2pt] (A) at (-0.7,-0.7) {$\bm{+1}$};

     \node[
        draw=blue,
        text=blue,
     fill=white,
        circle,
        minimum size=5mm,
        inner sep=2pt
    ] (B) at (-0.7,0.7) {$\bm{-1}$};


\draw[red, dashed, thick]
    (A) to[bend left=30] (B);

\end{tikzpicture}
\end{minipage}
\caption{Dual photon fields $\sigma(\omega)$ in the emergent dimension for a sublattice of size $K$. The leftmost diagram displays the open interval $\sigma_{1} - \dots -\sigma_{K}$ with the fermion branes (dashed lines) before $\sigma_{1}$ and after $\sigma_{K}$. The rightmost diagram displays the equivalent orbifold  periodic chain $\sigma_{1}-\dots-\sigma_{2K}$ with orbifold parity $\sigma_{j}=\sigma_{2K+1-j}$. In the emergent orbifold, the charge distribution  consists of a positive unit charge (red) at $r=1$ and angle $\frac{2\pi}{2K}$ together with a negative unit charge (blue) at $r=1$ and angle $-\frac{2\pi}{2K}$. The branch cut lying between them is drawn as a dashed red line.}
    \label{fig:F1emerge}
    \end{figure}

The key difference between the $\theta_{\mathrm{YM}}$/CP domain wall and the $\mathbb{Z}_{F}$ domain wall are the fermion zero modes and the missing cosines in Eq.~\eqref{eq:laglargeN}. In Section~\ref{sec:deconfnewzf}, we observed that twisted QCD consists of $F$ sublattices or open chains of size $K_{i}$, stretched between dual photons associated with each two consecutive fermion zero mode pairs. Fermions therefore cut the full angular circle of the $2d$ conformal plane into disconnected wedge segments or open intervals (see LHS of Fig.~\ref{fig:F1emerge}). Equivalently, they act as localized branes/boundaries in the emergent dimension, as was established in Ref.~\cite{Cherman:2016jtu}. This behavior  is reminiscent of their diagrammatic role in the planar large-$N$ limit~\cite{tHooft:1973alw}.

The dual photons $\sigma_{i}$ at the endpoints of the open chains are not fixed. In the continuum, this corresponds to Neumann boundary conditions $\partial_{\varphi}\sigma =0$ on both sides of a fermion brane, that is a vanishing tangential electric field. The fermion branes in the emergent dimension act as perfect conductors and the different open chains are completely independent. 

In Section \ref{sec:deconfnewzf}, we argued that the boundary conditions of the $\mathbb{Z}_{F}$ domain wall are of the form
\begin{equation}
    \vec{\alpha}_{n(j)} \cdot \vec\sigma \rightarrow  \vec{\alpha}_{n(j)} \cdot \vec\sigma + \frac{2\pi}{F} \,, \qquad \vec{\alpha}_{b_{\star}} \cdot \vec{\sigma} \rightarrow \vec{\alpha}_{b_{\star}} \cdot \vec{\sigma} - 2\pi \,.
    \label{eq:boundaryzfdomainwalldual}
\end{equation}
Here $n(j)$ indicated all monopole coordinates with fermion zero modes and the index $b_{\star}$ occurs on a single monopole without fermion zero modes. Shifting the index $b_{\star}$ corresponded to a gauge transformation, as in the case of the axion domain wall, and its presence guarantees that $\sum_{i} \vec{\alpha}_{i} \cdot \vec{\sigma}=0$. 

In Section \ref{sec:deconfnewzf}, we also argued that the $\mathbb{Z}_{F}$ domain wall generally contains a massive flux tube profile and a massless pion domain wall (see Fig.~\ref{fig:fundamentalparticle}). The tension of the massless pion domain wall scales inversely proportional with the length of the domain wall and is negligible at large volumes. In this section, we therefore study the tension of just the massive flux tube. The index $b_{\star}$ in the boundary conditions in Eq.~\eqref{eq:boundaryzfdomainwalldual} tell us which sublattice $K$ this massive flux tube is associated with. In this section, we simply take this to be the sublattice $(\sigma_{1},\dots,\sigma_{K})$ without a loss of generality. 

The boundary conditions for the massive flux tube part of the $\mathbb{Z}_{F}$ domain wall were given by the weight $\vec \omega^{(K)}_1$  in Eq.~\eqref{eq:highestweightexample} and the domain wall interpolates between the two boundaries,
\begin{equation}
    (\sigma_{1},\dots,\sigma_{K}) : \qquad (0,0,\dots,0) \qquad \rightarrow \qquad  2\pi \left(1-\frac{1}{K},-\frac{1}{K},\dots,-\frac{1}{K}\right)
    \label{eq:boundarycondzf}
\end{equation}
and all other dual photons fixed to zero. 

We assume that the flavor twisted boundary conditions of the fermions around the compact circle are such that the size of the subgroup $K$ scales linearly with $N$, such that we can use the small difference approximation of the previous section as $N$ grows large. For this to happen, it is sufficient to hold $F$ constant. In this case, regardless of the choice of fermion boundary conditions around the compact circle, there is always at least one sublattice whose size is scaling with the increasing number of total lattice size, $N$.
 
We solve for the domain wall profile with boundary conditions~\eqref{eq:boundarycondzf} using the standard method of images in electrostatics. In the continuum limit, where correspondingly $K\sim N$ are taken to be large, the extra-dimensional open interval of length $K$ with Neumann boundary conditions at both endpoints is equivalent to a full periodic circle of length $2K$ with a $\mathbb{Z}_{2}$ parity `orbifold' identification, under which the lower and upper halves of the circle are identified. The discrete analogue of this equivalence consists in doubling an open chain $\sigma_{j}$ with $j=1,\dots,K$ to a periodic closed chain $\sigma_{j}$ with $j=1,\dots,2K$ with the $\mathbb{Z}_{2}$ orbifold identification $\sigma_{j} = \sigma_{2K+1-j}$. We have drawn the equivalent open and closed chain in Fig.~\ref{fig:F1emerge}. The conductor boundary conditions induced by the fermion brane correspond to a vanishing radial electric field on the horizontal axis of the plot. The extra orbifold $\sigma_{i >N}$ fields inherit the boundary conditions of their orbifold partners in Eq.~\eqref{eq:boundarycondzf}. In what follows, we solve for the charge distribution on the $\mathbb{Z}_{2}$ orbifold with fields $\sigma_{j}$, $j=1,\dots,2K$.

All dual photons take on the same boundary values, apart from $\sigma_{1}$ (and its orbifold partner $\sigma_{2K}$). Employing the wedge-charge formula~\eqref{eq:chargeinwedge}, we learn that there must exist a positive unit charge in the wedge between $\sigma_{1}$ and $\sigma_{2}$ and a negative unit charge sitting somewhere between $\sigma_{1}$ and $\sigma_{K}$: a region that encompasses the whole upper half plane. We can fix the placement of the negative unit charge by the $\mathbb{Z}_{2}$ orbifold parity, as shown in Fig.~\ref{fig:F1emerge}.

The full discrete setup consists of a unit charge-anti-charge cluster stretched across $\sigma_{1}$ and $\sigma_{2K}$. Between the two charges lies a branch cut, drawn as a dashed line. Dual photons that cross the branch cut pick up a shift of $2\pi$ at $r=\infty$, compared to the other dual photons. Which dual photons do so is fixed by the boundary conditions. For the case studied here, Eq.~\eqref{eq:boundarycondzf} singles out the first dual photon $\sigma_{1}$ and its  orbifold partner $\sigma_{2K}$ to be the ones that cross the cut.

\begin{figure}[ht]
           \centering \includegraphics[width=0.75\textwidth]{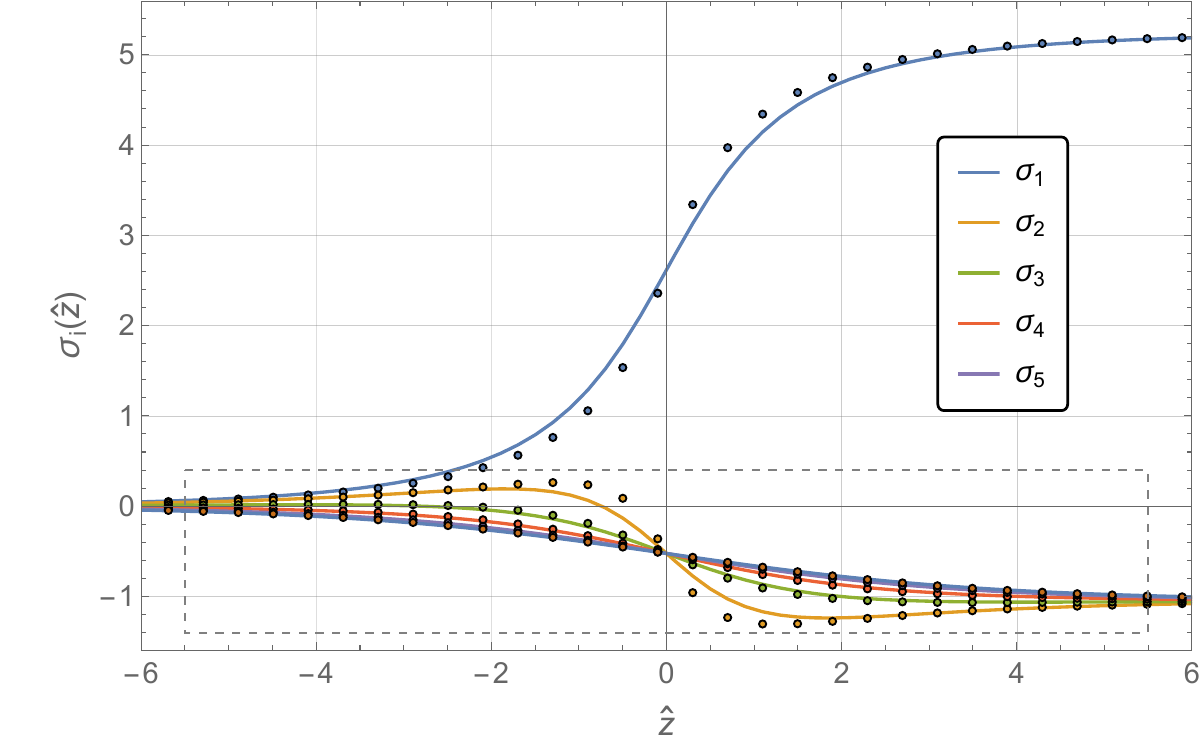}
    \caption{Profiles of the dual photons fields $\sigma_{i}(\hat{z})$ for $K=5$ on $\mathbb{Z}_{F}$ domain walls with boundary conditions~\eqref{eq:boundarycondzf}. The numerical solutions corresponding to minimizing the Lagrangian~\eqref{eq:laglargeN} are drawn as continuous lines. The analytical profiles (circular dots) corresponding to the dual photon field profiles~\eqref{eq:F1string} agree well with the numerical solutions with deviations for $\sigma_{1}$ and $\sigma_{2}$ explained by their finite boundary difference at large $N$. A zoom of the grey boxed area with $\hat{z}\in(-5.5,5.5)$ is displayed in Fig.~\ref{fig:F1zoomprofile}.}
    \label{fig:F1profile}
\end{figure}

\begin{figure}[ht]
           \centering \includegraphics[width=0.75\textwidth]{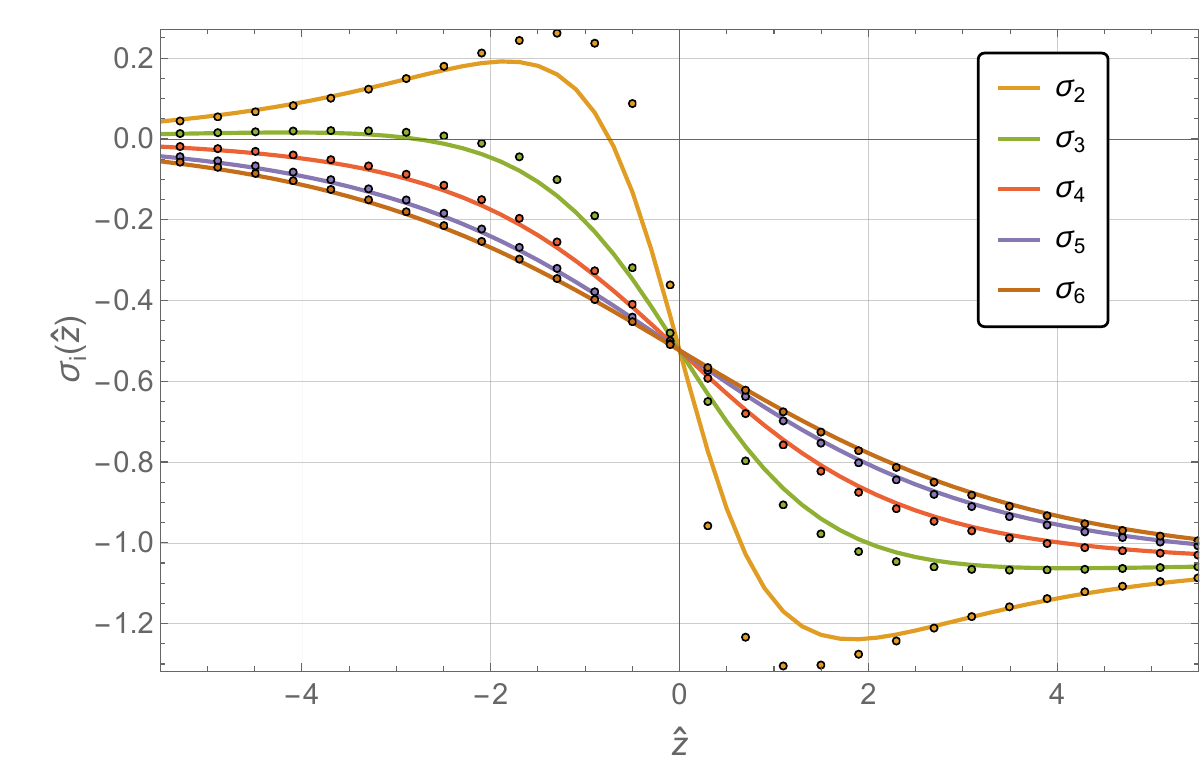}
    \caption{Profiles of the dual photons fields $\sigma_{i>2}(\hat{z})$ for $K=5$ on $\mathbb{Z}_{F}$ domain walls with boundary conditions~\eqref{eq:boundarycondzf} zoomed in on the domain $\hat{z} \in (-5.5,5.5)$ corresponding to the grey box in Fig.~\ref{fig:F1profile}. The numerical and analytic profiles agree well with deviations for the $\sigma_{2}$ profile explained in Fig.~\ref{fig:F1profile}.}
    \label{fig:F1zoomprofile}
\end{figure}

Solving for dual photon field profile on the emergent orbifold dimension in the continuum, we find 
\begin{equation}
    \sigma(\omega) =    \mathrm{arg}{\left(\omega -e^{i\pi+i \pi/K} \right)}-\mathrm{arg}{\left(\omega  -e^{i\pi -i\pi/K}\right)}-\frac{2\pi}{K}   \qquad \mathrm{with} \qquad \omega_{j}= \sqrt{r}\,e^{i \varphi_{j}/2}\, .
    \label{eq:F1string}
\end{equation}
The lattice angles on the orbifold are given by $\varphi_{j}/2 = -\pi \frac{2K+1-2j}{2K}$. In Fig.~\ref{fig:F1profile}, we have plotted the analytical dual photon profiles (obtained upon substitution of Eq.~\eqref{eq:F1string} into Eq.~\eqref{eq:dual-photon-field}) for $K=5$ (circular dots). The true numerical field profiles found by minimizing the Lagrangian in Eq.~\eqref{eq:laglargeN} with the specific boundary conditions in Eq.~\eqref{eq:boundarycondzf} are drawn as continuous lines. Overall, the analytic and numerical profiles agree well, with the exception of $\sigma_1$ and $\sigma_2$, for which the analytic result deviates from the numerics. Once more, this is due to the $\mathcal{O}(1)$ difference between $\sigma_{1}$ and $\sigma_{2}$ at large $N$ on across the wall, which renders the derivative in Eq.~\eqref{eq:finitedifop} ill-defined and the $\mathcal{O}(1)$ difference implies anharmonic contributions from the cosine $\cos{\left(\sigma_{1}-\sigma_{2} \right)}$. Moreover, in complete analogy with Section~\ref{sec:LargeNCPWallsTension}, the cosine terms in Eq.~\eqref{eq:laglargeN} paring neighboring $\sigma_j$'s propagate the error along the dual photon lattice, with each successive dual photon inheriting a smaller error than the previous one; see Figure~\ref{fig:F1zoomprofile}.

We now calculate the tension of the flux tube by regularizing its divergence on the discrete dual-photon lattice, as done in Section~\ref{sec:LargeNCPWallsTension}. The domain wall system on the orbifold consists of a single charge-anti-charge pair separated by an angular distance $2\pi/K$ with  vanishing total charge $Q=0$. In this case, the (regularized) divergence near the cores of both charges contributes a harmonic series to the tension similar to the single charge in Eq.~\eqref{eq:tensionexp}:
\begin{equation}
     \hat{T}_\text{3d} = \sum_{j=1}^{\mathcal{O}(N)} \frac{\pi}{j} - \sum_{j=1}^{\mathcal{O}(N)} \frac{\pi}{j+\ell} + \mathcal{O}(1) = \mathcal{O}(1)\,,
     \label{eq:clustercharge}
\end{equation}
where $\ell =2$ defines the angular separation between the charge. In contrast to the tension of a single positive charge in Eq.~\eqref{eq:tensionexp}, the two individually divergent contributions from the core of the strings combine into a telescoping series. The latter is bounded above by an $\mathcal{O}(1)$ number, resulting in a tension that scales as $\hat{T}_{\mathrm{3d}} \sim \mathcal{O}(1)$. This reflects the fact that the latticized dimension cannot see the core of any individual charge before it is screened by the divergence of the other charge. The total tension of the domain wall is therefore an incalculable constant at large-$N$, yielding a $4d$ tension $T_{\mathrm{4d}} =c N \Lambda_{\mathrm{QCD}}^{3}$, where $c$ is an undetermined constant. The $4d$ tension we have just derived is compatible with the D-brane and SUSY estimates~\cite{Witten:1997ep, Dvali:1996xe, Kovner:1997ca, Kogan:1997dt}. 

\begin{figure}[ht]
    \centering
    \includegraphics[width=0.75\linewidth]{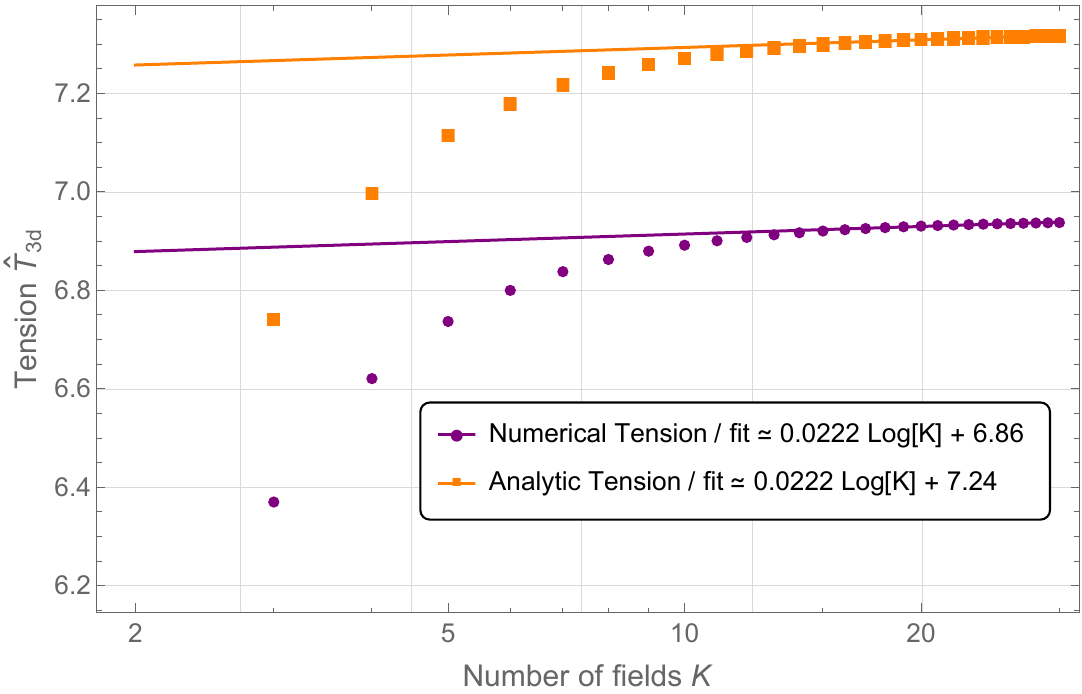}
    \caption{Scaling of the  tension $\hat{T}_{\mathrm{3d}}$ of the $\mathbb{Z}_{F}$ domain wall on a linear axis versus the number of dual photons $K\in \left( 3,30\right)$ on a logarithmic axis. The orange square markers correspond the tension of the analytical profile~\eqref{eq:F1string}, whilst the purple round markers correspond to the tension of the numerical profile found by minimizing Eq.~\eqref{eq:laglargeN}. Best fit values for the curves $\hat{T}_{\mathrm{3d}} = a \ln{K} + b$ for data points $K>15$ are shown. The fits agree well with a saturating tension at large $K\sim N$. }
    \label{fig:tensionscalingF1}
\end{figure}

We have plotted the (three-dimensional, rescaled) analytical and numerical tensions versus the number of fields $K$ (on a logarithmic scale) in Fig.~\ref{fig:tensionscalingF1}. The growth of the tension saturates at large $K\sim N$, and the difference between the saturation points of the analytical and numerical tensions is roughly $\Delta T \simeq 0.4$.

The large $N$ results of this section also apply to the flux tubes in center-stabilized Yang--Mills. The flux tubes in center-stabilized Yang--Mills are associated with domain walls with boundary conditions set by the highest weight $\vec\omega_1=(1-1/N,-1/N,\ldots,-1/N)$ of the fundamental representation. These domain wall boundary conditions for the dual photons are similar to those considered in this section in Eq.~\eqref{eq:boundarycondzf} when $K \sim N$. The tension $\hat{T}_{\mathrm{3d}}$ of these flux tubes will therefore also be a $\mathcal{O}(1)$ incalculable number. This $\mathcal{O}(1)$ tension will differ from the tension of the flux tubes calculated in this section as center-stabilized Yang--Mills has one more Kaluza--Klein-monopole that connect $\sigma_{1}$ and $\sigma_{N}$ compared to the $SU(K)$ sublattices.   

The remarkable parallels between domain wall profiles and electrostatic charge distributions, and the tensions of said domain wall and the energy densities of the charge distributions in the emergent dimension motivates us to briefly generalize the results of Sections \ref{sec:LargeNCPWallsTension} and \ref{sec:LargeNZFWallsTension} to any domain wall in center-stabilized compactifications at large $N$. In general, the results of the previous sections indicate that the tension of flux tubes or domain walls in the large-$N$ double-scaling limit depends on the  way in which the  charges in the emergent dimension are distributed, as set by the boundary conditions of the domain wall. We define a cluster of charges as a set of charges whose angular separation goes to zero as $N$ grows large. Examples include the charge-anti-charge cluster in Fig.~\ref{fig:F1emerge}. Within a cluster, core divergences add together and the latticized dimension cannot probe the core of any charge before it is screened by the other charges. Therefore, a cluster with total charge $Q_\text{cluster}$ gives a logarithmically divergent contribution to the tension proportional to the square of the charge $Q_\text{cluster}^{2}$, as expected for an energy density:
\begin{equation}
    T_{\text{4d}} = \pi Q^{2}_{\mathrm{cluster}} N\ln{(N)}\Lambda_\text{QCD}^3 + \mathcal{O}\left(N (\ln N)^0\right) \,.
\end{equation} 

If charges group into clusters and the angular separation between the individual clusters remains of order $\mathcal{O}(1)$ at large $N$, then the latticized dimension has enough resolution to probe each cluster individually. As a consequence, the total tension of the resulting charge distribution approaches the sum of the different cluster tensions: 
\begin{equation}
   T_{\mathrm{4d}} \sim \pi  \sum_{\mathrm{clusters}} Q_{\mathrm{cluster}}^{2} N\ln{(N)} \Lambda_{\mathrm{QCD}}^{3} + \mathcal{O}\left(N (\ln N)^0\right) \,,
\label{eq:largeNtotaltensionpred}
\end{equation}
where we have ignored polynomial factors in the ’t Hooft coupling $\lambda$ and in $\eta$. This equation yields a general master formula for domain wall tensions in center-stabilized compactifications based solely on the boundary conditions of domain walls.  

\section{The 4$d$ Chiral Lagrangian and the Structure of the $\eta'$ Potential}\label{sec:real}
Before concluding, it is useful to discuss explicitly the relation between the main results of the paper, well-known facts about the ordinary $4d$ Chiral Lagrangian and aspects of the $\eta'$ potential that are currently object of speculation.

At the classical level, when all quark masses vanish, $4d$ QCD with $F$ quarks in the fundamental representation has a global $(U(F)_L \times U(F)_R)/\mathbb{Z}_N$ 0-form symmetry, but quantum mechanically the $U(1)_A$ factor is
explicitly broken. At low energy, below the confinement scale, only the diagonal
subgroup $U(F)_{L+R}$ is linearly realized, resulting in $F^2-1$ massless Nambu--Golstone bosons (the pions). Identical quark masses explicitly break $U(F)_L \times U(F)_R$ down to the linearly realized subgroup $U(F)_{L+R}$, giving masses to the pions. In this work we set quark masses to zero and focused mostly on the states of our $3d$ model closest to the ordinary $\eta'$, studying the part of their potential that stays non-zero when quark masses vanish.

The $\eta'$ is the would-be Nambu--Goldstone boson of the $U(1)_A$ symmetry. It receives a mass at the QCD scale from the chiral anomaly, 
\be
\partial_\mu j^\mu_A =  \frac{F}{8\pi^2} \mathrm{Tr}\,G_{\mu \nu}\widetilde G^{\mu \nu}\, ,
\ee
that explicitly breaks the $U(1)_A$ symmetry. Below the confinement scale, it is easiest to write the most general two-derivative Lagrangian consistent with all the symmetries if we introduce the pion matrix
\be
U = e^{i \eta'/F} e^{i \pi^a T^a}\, ,
\ee
where $T^a$ are the generators of $SU(F)_A$. As anticipated in the Introduction, we normalize scalar mesons to their decay constants. Thus, for us the symbols $\eta'$ and $\pi^a$ denote dimensionless fields. Under $SU(F)_L\times SU(F)_R\times U(1)_A$ the pion matrix transforms as
\be
U \to e^{2i \alpha} U_L U U_R^\dagger\, .
\ee
It is also convenient to introduce the quark mass matrix
\be
M= m_q \mathds{1}_{F\times F}\, .
\ee
At this point, we can write the lowest order in the derivative expansion of the Chiral Lagrangian
\be
\mathcal{L}=\frac{f_\pi^2}{4}{\rm Tr}[(\partial_\mu U^\dagger)(\partial^\mu U)]- a \Lambda_{\rm QCD}f_\pi^2\left( {\rm Tr}[M U]+{\rm h.c.}\right)- b \Lambda_{\rm QCD}^2 f_\pi^2 H(\det U)\, ,
\ee
where $a, b$ are unknown $\mathcal{O}(1)$ coefficients and $H$ is an unknown function. The last term in the above expression reproduces the effect of the chiral anomaly, giving mass to the $\eta'$.

The results of this work can tell us something about $H$. Its analytic structure is unknown in QCD, but it is known in certain models that are connected to QCD in specific limits. At large-$N$~\cite{Witten:1980sp, DiVecchia:1980yfw}
\be
\lim_{N \to \infty} H(\det U) \sim \frac{(\log \det U)^2}{N} \, ,
\ee
while in supersymmetric  QCD (with supersymmetry breaking from anomaly mediation) we have for $N\neq F$~\cite{Csaki:2023yas}
\be
H(\det U) \propto (\det U)^{\frac{1}{|N-F|}}\, . \label{eq:SUSY}
\ee
Both results imply a branched structure for the $\eta'$ potential that remains $2\pi$ periodic (i.e. $H(\eta')=H(\eta'+2\pi)$) because the lowest energy branch of the log (or of the fractional power in Eq.~\eqref{eq:SUSY}) changes with every $2\pi/|N-F|$ displacement of the field~\cite{Witten:1980sp, Csaki:2023yas}. In formulae,
\be
V(\eta') \propto \min_{k}\left[-\cos\left(\frac{\eta'+2\pi k}{|N-F|}\right)\right]\, , \label{eq:toyeta}
\ee
where we set to zero the QCD $\theta$-angle for simplicity, but we will return to its connection with the $\eta'$ below.

The two results are consistent with each other and incompatible with the naive expectation from a dilute instanton gas approximation (i.e. $H\propto\det U$),  which is often used in the literature. 

Some care must be taken when trying to infer the form of $H$ in the real world, where $N=F$ or $N=F+1$. According to Eq.~\eqref{eq:SUSY}, we would expect a single-valued function with no branches, consistent with the calculation from instantons. However, the calculation in Ref.~\cite{Csaki:2023yas} is performed with the SUSY breaking scale $\widetilde M_S$ smaller than the confinement scale $\Lambda_{\rm QCD}$ and the limit from this configuration to the real world, i.e. $\widetilde M_S \gg \Lambda_{\rm QCD}$ is not smooth. In particular, for $N=F$ the result is obtained by starting from $N=F+1$ and integrating out a quark supersymmetrically and expecting no branches is inconsistent with the symmetry arguments in Ref.~\cite{Kitano:2025ezk}. The argument in Ref.~\cite{Kitano:2025ezk} uses two ingredients: 1) the mixed anomaly between $U(F)/\mathbb{Z}_N$ and the shift $\bar\theta \to \bar\theta + 2\pi$ whenever gcd($N,F$)$>1$~\cite{Cordova:2019jnf, Cordova:2019uob}, and 2) the spurionic transformation of $\bar\theta$ under the $U(1)_A$ to promote $V(\bar\theta)$ to a function of the $\eta'$. These two ingredients allow to conclude that the $\eta'$ potential must inherit the structure of the $\bar\theta$ potential and have a number of branches proportional to gcd($N,F$).

This result is trivially consistent with the large-$N$ results presented above (since gcd($N,0$)$=N$) and also with the SUSY calculation for $N<F$ and $N>F$. However, we should be careful when trying to deduce properties of the $\eta'$ potential from the vacuum energy as a function of $\theta$. We can think of the latter as the axion potential that is left after integrating out all heavier fields. When a heavy field is integrated out, its branched structure leaves no trace in the low energy theory. For example, when we integrate out a $\eta'$ with the potential in Eq.~\eqref{eq:toyeta}, we are going to be in a minimum of the potential with a well-defined value of $k=k_*$. In the EFT obtained integrating out the $\eta'$, we cannot access energies that allow us to move over the potential barriers between branches, since this can only be done at energies $E \gtrsim m_{\eta'}$.

Therefore we can have a $\eta'$ potential with $|N-F|$ branches and, in the same theory, an axion with gcd($N,F$) branches. This is indeed the result that was found in supersymmetric QCD in Ref.~\cite{Csaki:2023yas}. Summarizing, all results currently present in the literature are (to the best of our knowledge)  consistent with each other whenever $N>F$ or $N<F$ since $|N-F|$ and gcd($N, F$) are proportional in these cases. In this work, we are unable to make definitive statements when $N=F$, as the infrared spectrum of twisted QCD does not contain any massive modes when $N=F$. Instead, we used a calculable model to deepen our understanding of the $\eta'$ potential in the following ways:
\begin{enumerate}
    \item We write a consistent effective theory that includes the leading effects of fields heavier than the $\eta'$ and show how one can move between branches of the $\eta'$ potential along the heavier directions in field space, i.e. the $\eta'$ potential is single valued before integrating out the heavy fields.
    \item In a specific large-$N$ limit of our $3d$ QCD model, we find a duality between the low energy description of $3d$ QCD and a $4d$ theory with an emergent gauge field. This duality is extremely useful to calculate analytically the tension of $\eta'$ and axion domain walls. We find that the two tensions are not always the same, challenging the lore that one can deduce properties of the $\eta'$ from $\theta$. 
\end{enumerate}

In this section, we want to better understand the first point above (the second being already discussed in detail in Section~\ref{sec:emerge}). The first thing to do is to be more explicit about which degrees of freedom play the role of the $\eta'$ in compactified QCD. The simplest way to get an idea is to trace the effect of the anomaly from the ultraviolet to the infrared and see what states get a mass from the explicit breaking of the $U(1)_A$ that the anomaly induces. After compactification, the usual $\theta$-term of $4d$ QCD survives inside the dual photon potential of Eq.~\eqref{eq:dualphotonpotzero} as 
\be
 \mathcal{L} \supset -\sum_{b \in \mathcal{B}} \zeta \cos{\left(\sigma_{b} - \sigma_{b+1} + \frac{\theta}{N-F}\right)}\, .
 \label{eq:ChSimple}
\ee
From the discussion in the previous sections, we know that $N-F$ states get a mass from this part of the Chiral Lagrangian and they can be parametrically heavier than the remaining $F-1$ pions that only get masses proportional to the quark masses, when they are non-zero. In Eq.~\eqref{eq:ChSimple} we have rotated the chiral fermions in such a way that $\theta$ only appears in the potential of the heavy modes.

As each single-flavor QCD $SU(K_{i})$ sublattice forms a completely decoupled sector of the infrared spectrum, one can in principle identify a single version of $\eta'$ per sublattice, that we can call $\eta'_{(K_{i})}$. In this sense, we say that the $\eta'$ generically fractionalizes into $F$ $\eta'_{(K_{i})}$'s in twisted QCD. We note that the $\eta'_{(K_i)}$'s are not mass eigenstates of the model, and that the dynamics of twisted QCD can be correctly described only by taking into account the complete set of its $N-F$ massive states, as we shall see momentarily.

For a given sublattice of size $K$, a possible choice is to define  $\eta'_{(K)}$ as the common phase of the cosines in Eq.~\eqref{eq:ChSimple} across a given sublattice,\footnote{This definition differs from other definitions of $\eta'$ in the literature, see Ref.~\cite{Cherman:2017dwt}.}
\begin{equation}
    \eta'_{(K)} = \sum_{b \in (K-1)} \left(\sigma_{b} - \sigma_{b+1} \right) 
\end{equation}
where $b \in (K-1)$ denotes the first $K-1$ dual photon indices in the sublattice $K$ such that the sum only contains dual photons of that sublattice. This is motivated by the position of the $\theta$-angle spurion in Eq.~\eqref{eq:ChSimple}, which sits as a common phase across the cosines of each sublattice. For a given sublattice, this state has a large overlap with its lightest massive eigenstate. Moreover, under the $\mathbb{Z}_{F}$ domain wall associated with the sublattice $SU(K)$, $\eta'_{(K)}$ correctly interpolates between the values $\eta'_{(K)}$ and $\eta'_{(K)} \rightarrow \eta'_{(K)} + 2\pi$ and the others $\eta'_{(K_{i}\neq K)}$ remain trivially unchanged. 

After integrating out the heavier states of the sublattice, one obtains the following potential for this $\eta'_{(K)}$ degree of freedom,
\begin{equation}
    V_{(K)} \supset \min_{k}\left[-\zeta\  (K-1)\cos{\left( \frac{\eta'_{(K)} - 2\pi k }{K-1} + \frac{\theta}{N-F} \right)}\right]+\mathcal{O}(m_q \Lambda_{\rm QCD}^2)
    \label{eq:etaindiv}
\end{equation}
where $k$ runs from $1$ to $K-1$ and denotes the different minima associated with the heavier states. 

The potential in Eq.~\eqref{eq:etaindiv} for the individual $\eta'_{(K)}$ is of the same form as the $\eta'$ potential in large-$N$ QCD \cite{Witten:1980sp} and softly-broken sQCD \cite{Csaki:2023yas}. Importantly however, the potential in Eq.~\eqref{eq:etaindiv} was recovered by integrating out the heavier states of the sublattice. We stress that the masses of these heavier states are approximately integer multiples of the $\eta'_{(K)}$ mass (see Eq.~\eqref{eq:massivespectrummass}), and the branched potential in Eq.~\eqref{eq:etaindiv} is therefore a poor approximation to the dynamics of the sublattice at these energy scales. Instead, an appropriate way to capture the dynamics is found by including all the sublattice modes and using the smooth,  unbranched potential in Eq.~\eqref{eq:ChSimple}. 

In conclusion we find that in compactified QCD the number of branches of the $\eta'$ (or $\eta'_{(K)}$) potential is a dynamical quantity that cannot be determined using only symmetry arguments, since the values of $K$ depend on the choice of fermion boundary conditions and there are several different choices compatible with the same symmetries. Additionally, and perhaps more importantly, we confirm the intuition that the cusps (and thus the branches) in the $\eta'$ potential arise only when integrating out states parametrically heavier than the $\eta'$. This situation occurs in approximations of QCD where the $\eta'$ degree of freedom is parametrically lighter than the QCD resonances, such as large-$N$ QCD or softly-broken sQCD, but might not be a realistic representation of the effective theory in ordinary $4d$ QCD near the $\eta'$ mass. 

In ordinary $4d$ QCD, the mass of the $\eta'$ meson is close to that of other mesons. This is similar to how the mass of the $\eta'_{(K)}$ degrees of freedom in twisted QCD are close to the heavier states on the same sublattice. It is entirely possible that the situation realized in Nature is that of a smooth potential in terms of the massive mesons, with no cusps, no branches and multiple accessible directions in field space, as this is the situation for the sublattice modes in Eq.~\eqref{eq:ChSimple}. In the previous sections, we have shown that such a potential is capable of matching all the anomalies of $3d$ twisted QCD without the need for branches in the effective infrared theory.

\section{Discussion}\label{sec:conclusion}
In this paper, we investigated the dependence of the confining vacuum of quantum chromodynamics (QCD) on the number of colors $N$ and fundamental flavors $F<N$ of the theory. We accomplished this by compactifying QCD with $N$ colors on a small circle of size $L$ with a center-stabilizing potential for the Polyakov loop around the compact direction. The $F$ fundamental fermions were given flavor-twisted boundary conditions on the small circle. This set-up allowed us to study the vacuum and $\mathbb{Z}_{F}$ domain walls interpolating between the vacua in a QCD-like theory without singularities. In this way, both the tension and rich internal structure of $\mathbb{Z}_{F}$ domain walls could be analytically studied. 

In Section~\ref{sec:deconf}, we established that whenever $N$ and $F$ are not coprime, probe particles in representations of $N$-ality $N/\mathrm{gcd}(N,F)$ become deconfined on $\mathbb{Z}_{F}$ domain walls. When $\mathrm{gcd}(N,F)>1$, the infrared spectrum has an additional $\mathbb{Z}_{\mathrm{gcd}(N,F)}$ color-flavor-center symmetry. This symmetry results in $\mathrm{gcd}(N,F)$ domain walls with equal tension. The $\mathbb{Z}_{F}$ domain walls can be glued together and the intersection between two domain walls has the quantum numbers of a probe particle in a representation of $N$-ality  $N/\mathrm{gcd}(N,F)$. As the two domain walls have equal tension, the intersection is free to move and the probe particle is deconfined. This deconfinement whenever $N$ and $F$ are not coprime shows remarkable similarity with the deconfinement of probe particles of unit $N-$ality on axion domain walls in pure Yang--Mills.

In the last section, Section~\ref{sec:emerge}, we studied the profile and tension of the axion and $\mathbb{Z}_{F}$ domain walls in a large-$N$ limit with $NL$ held fixed. In this limit, the $N$ degrees of freedom in the infrared spectrum span an emergent dimension. Their dynamics is equivalent (dual) to an emergent Abelian gauge field and the domain wall is dual to a charge distribution in this dimension. The energy of this charge distribution (tension of the domain wall) depends on the boundary conditions of the fermions around the compact dimension. Thanks to this duality, we established a tension scaling $T_{\mathrm{4d}} \sim \pi N \Lambda_{\mathrm{QCD}}^{3}$ at large $N$. This agrees with the estimated tension scaling of the $\mathbb{Z}_{F}$ domain wall in the four-dimensional large-$N$ planar limit. Moreover, we established that the tension of the gluonic core of the axion domain wall scales as $T_{\mathrm{4d}} \sim \pi N \ln{(N)} \Lambda_{\mathrm{YM}}^{3}$ in this double-scaling limit. Therefore $\mathbb{Z}_{F}$ domain walls in QCD and axion domain walls in pure gluodynamics do not have the same properties in center-stabilized compactifications. 

In Section~\ref{sec:real} we tried to make contact between our results and the $\eta'$ potential in the real world. In our $3d$ model, the would-be $\eta'$ mesons have masses close to the other QCD resonances. Their overall dynamics is governed by a smooth, single-valued potential. One can see very explicitly that cusps and branches in the potential only arise when integrating out the slightly heavier states. The situation in QCD is likely to be similar to that in our compactified world, where the $\eta'$ lives close in mass to many other resonances and their collective potential is smooth and single-valued.

The large-$N$ double-scaling methods developed in this paper offer exciting avenues for future research. At large $N$, it is expected that the $\Delta^{++}$ baryon can be described by a bounded $\eta'\rightarrow \eta'+ 2\pi F$ domain wall. It would be interesting to study the properties of this baryon in twisted QCD and compare its properties to the known properties of the baryon such as its size and mass. This would allow one to test the role of vector mesons \cite{Kitano:2020evx} on the boundary of the wall in an analytically controlled scenario. Moreover, it would be equally interesting to demonstrate the deconfinement of probe quarks on the baryon worldsheet in the case $\mathrm{gcd}(N,F)=1$ when the $\mathbb{Z}_{\mathrm{gcd}(N,F)}$ symmetry is absent, as even in the case that $F=1$, the four-dimensional $\eta'$ domain walls are expected to host a $SU(N)_{1}$ Chern--Simons theory \cite{ Kitano:2025ezk,Komargodski:2018odf} on which quarks deconfine.

In general, it could be interesting to use the analytic large-$N$ methods developed for extended objects in this paper to reanalyze results obtained either analytically or only numerically at finite $N$ in the literature on YM/QCD compactifications. This could include a study of the the $N$-ality scaling of gluonic flux tubes tensions in center-stabilized Yang--Mills in the large-$N$ double-scaling limit in order to test the Casimir/Sine scaling law predictions~\cite{Poppitz:2017ivi}. On a similar note, one might be able to extend the large-$N$ methods developed here to study the tension scaling of the (collapsed) double-gluonic strings in compactified SYM or QCD(adj) at large $N$ as studied in  Refs.~\cite{Anber:2015kea,Bub:2020mff} at finite $N$. This would allow comparisons with ordinary planar large-$N$ expectations.

\section*{Acknowledgments}
AP would like to acknowledge helpful conversations with Kai Bartnick. AP acknowledges the support of the Klarman Fellowship. PB, PS and RTD acknowledge ANR grant ANR-23-CE31-0024 EUHiggs for partial support. RTD acknowledges support from the Institute for Advanced Study and an anonymous donor. PS acknowledges financial support from the Spanish Ministry of Science and Innovation (MICINN) through the Spanish State Research Agency, under Severo Ochoa Centres of Excellence Programme 2025-2029 (CEX2024001442-S). This work is also part of the R\&D\&i project PID2023-146686NB-C31, funded by MICIU/AEI/10.13039/501100011033/ and by ERDF/EU.
PB is grateful for the hospitality of the Enrico Fermi Institute and the Leinweber Institute for Theoretical Physics at the University of Chicago, where part of the revision of this work was completed, and acknowledges partial support from the France And Chicago Collaborating in The Sciences (FACCTS) program.
The authors would like to thank Munich Institute for Astro-, Particle and BioPhysics (MIAPbP) for support and hospitality funded by the Deutsche Forschungsge- meinschaft (DFG) Excellence Strategy – EXC-2094 – 390783311 and the organisers of the workshop Navigating New Horizons. The authors would also like to thank ICTP-SAIFR for hosting the \textit{Horizons Beyond the Standard Model: from Colliders to Dark Matter and Axions} workshop.

\newpage
\appendix

\section{Anomalies of the residual flavor symmetry}
\label{app:cartan-cfc-anomalies}

The flavor twist removes the off-diagonal flavor transformations, but this reduction of symmetry need not remove the mixed anomalies of QCD. In this appendix we show that the four-dimensional mixed chiral-baryon-color-flavor anomaly remains nontrivial when the vector flavor symmetry is restricted to its faithful Cartan subgroup. We describe the allowed backgrounds and exhibit one for which a discrete chiral transformation gives a nontrivial phase.

We consider $F\geq2$ massless Dirac quarks in the fundamental representation of $SU(N)$ and evaluate the anomaly on a closed, oriented spin four-manifold $X_4$.

\subsection{The faithful vector symmetry and the CFC action}
\label{app:anomaly-symmetries}

We normalize the quark number by $Q(\psi)=1$ and baryon number by $B=Q/N$. An $N$-quark baryon therefore has $B=1$. The faithful vector symmetry is 
\begin{equation}
    G_V=\frac{U(F)_V}{\mathbb Z_N}\cong\frac{SU(F)_V\times U(1)_Q}{\mathbb Z_F\times \mathbb Z_N}\, .
\end{equation}
The two quotient generators can be represented on color, flavor, and quark-number transformations as
\begin{equation}
    (z_N\mathbf 1_N,\mathbf 1_F,z_N^{-1})\,,\qquad (\mathbf 1_N,z_F \mathbf 1_F,z_F^{-1})\,, \qquad z_N=e^{2\pi i /N}\,,\qquad z_F=e^{2\pi i/F}\, .
\end{equation}
Each acts trivially on a quark. The first identifies a common quark-number rotation with a color gauge transformation. The second is the usual identification $U(F)\cong [SU(F)\times U(1)]/\mathbb Z_F$. This is the quotient structure used to introduce BCF backgrounds in Refs.~\cite{Anber:2019nze,Cordova:2019jnf,Cordova:2019uob}.

For a flavor twist with distinct eigenvalues, the surviving vector subgroup is
\begin{equation}
    H_V=\frac{U(1)^F}{\mathbb Z_N}\cong \frac{T_F\times U(1)_Q}{\mathbb Z_F\times \mathbb Z_N}\,,\qquad T_F=S\left(U(1)^F \right)\, .
    \label{eq:surviving-vector-subgroup}
\end{equation}
The finite quotients remain present after restricting to the Cartan. Although $H_V$ is abstractly an $F$-dimensional torus, the quotient fixes its charge normalization and its allowed background fluxes in the microscopic quark fields. Replacing $H_V$ by $F$ independently normalized $U(1)$ symmetries and forgetting the quotients changes the allowed backgrounds and can erase the anomaly one is trying to detect. For example, the local diagonal connections $A_j$ can have fluxes
\begin{equation}
    \frac{1}{2\pi}\int_{\Sigma}(dA_1,\ldots, dA_F)=(m_1,\ldots , m_F)+\frac{r}{N}(1,\ldots, 1)\,,\qquad m_j,r\in \mathbb{Z}\, .
    \label{eq:anomaly-Cartan-fluxes}
\end{equation}
Thus the differences have integral flux but the individual diagonal entries need not. The common fractional flux in Eq.~\eqref{eq:anomaly-Cartan-fluxes} is compensated by the color bundle.

Consider now the flavor-center-symmetric twist
\begin{equation}
    U_{\rm F}=e^{i\theta_V}e^{-\pi i(F-1)/F}\mathrm{diag}\,(1,\nu,\nu^2,\ldots, \nu^{F-1})\,,\qquad \nu=e^{2\pi i/F}\, .
\end{equation}
Set $h=F/d$ with $d=\gcd(N,F)$. We define the unitary flavor-permutation matrix $\mathsf{P}\in U(F)_V$ by $\mathsf{P}e_j=e_{j+h}$, with flavor labels understood modulo $F$. It consists of $h$ disjoint cycles of length $d$, so $\mathsf P^d=\mathbf1_F$ and $\det\mathsf P=(-1)^{h(d-1)}$. Its action on the twist satisfies
\begin{equation}
    \mathsf{P}U_{\rm F}\mathsf{P}^{-1}=z_{d}^{-1}U_{\rm F}\,,\qquad z_d=e^{2\pi i/d}\in \mathbb{Z}_N\, .
\end{equation}
A color transformation $g(x_4)$ with $g(L)=z_d g(0)$ changes the quark boundary condition by $z_d$. Its combination with $\mathsf{P}$ therefore preserves $U_{\rm F}$. This defines the CFC generator $\mathsf{C}$. Its $d$-th power is a periodic color gauge transformation and acts trivially on gauge-invariant observables. The existence of this symmetry for a nontrivial common divisor was established in Ref.~\cite{Cherman:2017tey}.

The action of $\mathsf{C}$ on a diagonal vector transformation $V=\mathrm{diag}\,(v_1,\ldots, v_F)$ is
\begin{equation}
    \mathsf{C}[V]\mathsf{C}^{-1}=[\mathsf{P}V\mathsf{P}^{-1}]=[\mathrm{diag}\,(v_{1-h},\ldots, v_{F-h})]\, .
    \label{eq:anomaly-CFC-on-Cartan}
\end{equation}
Here, square brackets denote the equivalence class modulo the common $\mathbb{Z}_N$. Conjugation by $\mathsf{P}$ preserves this identification, but permutes the entries of a general diagonal matrix $V$.

We write a general transformation in the ordered form $[V]\mathsf C^k$ and represent it by the pair $([V],k)\equiv [V]\mathsf C^k$, with $k\in\mathbb Z_d$. Using Eq.~\eqref{eq:anomaly-CFC-on-Cartan}, the product of two such transformations is
\begin{equation}
    ([V],k)([W],\ell)=[V\mathsf P^kW\mathsf P^{-k}]
    \mathsf C^{k+\ell}\equiv ([V\mathsf{P}^{k}W\mathsf{P}^{-k}],k+\ell)\, ,
\end{equation}
where addition in the second entry is understood modulo $d$.  Thus the subgroup generated by vector rotations and CFC is
\begin{equation}
    G_{V,\mathrm{CFC}}=H_V \rtimes_{\rho} \mathbb{Z}_d^{\rm CFC}\,,\qquad \rho([V])=[\mathsf{P}V\mathsf{P}^{-1}]\, .
\end{equation}
For $N=F$, this vector-CFC subgroup was studied in Ref.~\cite{Nishimura:2019umw}, together with a discrete chiral factor in the analysis of high-temperature twisted-QCD walls. The general-$N,F$ expression above follows from the permutation by $F/d$ in Eq.~\eqref{eq:anomaly-CFC-on-Cartan}.

\subsection{Cartan backgrounds and the discrete chiral anomaly}
\label{app:anomaly-cartan-backgrounds}

The quotient data can be described using transition functions~\cite{Anber:2019nze}. Choose a good open cover $M_4=\bigcup_i U_i$. For an ordinary $SU(N)$ gauge configuration, the transition function $g_{ij}:U_{ij}\to SU(N)$ specifies how the local gauge description on $U_j$ is related to that on $U_i$ over their overlap. Let $g_{ij}^{c}$, $g_{ij}^{f}$, and $g_{ij}^V$ be local lifts of the color, flavor, and quark-number transition functions. On a triple overlap $U_i \cap U_j \cap U_k$ they may obey
\begin{equation}
    g_{ij}^c g_{jk}^c g_{ki}^c=e^{2\pi i b_{ijk}^c/N}\mathbf{1}_N\,,\qquad g_{ij}^f g_{jk}^f g_{ki}^f=e^{2\pi i b_{ijk}^f/F}\mathbf{1}_F\,,\qquad g_{ij}^V g_{jk}^V g_{ki}^V=e^{-2\pi i (b_{ijk}^c/N+b_{ijk}^f/F)}\, .
    \label{eq:anomaly-cocycles}
\end{equation}
The integers $b^c_{ijk}$ and $b^f_{ijk}$ are defined modulo $N$ and $F$, respectively. Locally, the quarks are patched according to $\psi_i=(g_{ij}^c\otimes g_{ij}^f\otimes g_{ij}^V)\psi_j$. The three phases in Eq.~\eqref{eq:anomaly-cocycles} cancel in the transition function of the quark on a triple overlap. Consequently the complete quark bundle is well defined, even when the separate color and flavor lifts are projective. These are the BCF compatibility conditions of Ref.~\cite{Anber:2019nze}.

Restricting $g_{ij}^f$ to diagonal matrices does not set $b^f$ to zero: the flavor center is itself diagonal. We retain the backgrounds satisfying these compatibility conditions that admit diagonal flavor lifts.

To analyze the discrete chiral anomaly we consider the order-$F$ chiral transformation
\begin{equation}
    \chi:\qquad \psi_L\longmapsto e^{2\pi i/F}\psi_L\,,\qquad \psi_R\longmapsto\psi_R\,.
    \label{eq:anomaly-chiral-generator}
\end{equation}
It differs from a generator of the axial $\mathbb{Z}_{2F}$ by a vector rotation. Under such transformation the path integral measure transforms by
\begin{equation}
    \mathcal{J}_{\chi}=\exp\left(\frac{2\pi i}{F}\, \mathcal{I}\right)\,,\qquad \mathcal I=\operatorname{Ind}\slashed D_{\mathscr E}\in\mathbb Z\,.
    \label{eq:anomaly-index-Jacobian}
\end{equation}
In a trivial vector background, $\mathcal{I}$ contains $F$ times the color instanton number, so Eq.~\eqref{eq:anomaly-index-Jacobian} is unity.

Let $G_c$ and $F_f$ denote the local traceless color and flavor curvatures, and let $f_V$ be the local quark-number curvature. The total curvature on the quark bundle is~\cite{Anber:2019nze}
\begin{equation}
    \mathcal{F}=G_c\otimes \mathbf{1}_F+\mathbf{1}_N\otimes F_f+f_V\mathbf{1}_{NF}\,.
\end{equation}
The Atiyah--Singer index theorem gives~\cite{Atiyah:1968mp,Harvey:2005it,Bilal:2008qx}
\begin{equation}
    \mathcal{I}=\frac{1}{8\pi^2}\int_{X_4}\mathrm{Tr}_{c,f}(\mathcal{F}\wedge \mathcal{F})-\frac{NF}{24}\int_{X_4}p_1(TX_4)\, ,
    \label{eq:anomaly-index}
\end{equation}
where $p_1$ is the first Pontryagin class representing the gravitational counterpart of an instanton-density class, and $TX_4$ denotes the collection of the tangent spaces over all points in $X_4$, together with their smooth patching.
Since $\mathrm{Tr}_c\,G_c=\mathrm{Tr}_f\, F_f=0$, expanding the first term gives
\begin{equation}
    \mathrm{Tr}_{c,f}(\mathcal{F}\wedge \mathcal{F})=F\,\mathrm{Tr}_c\, G_c\wedge G_c+N\, \mathrm{Tr}_f\,F_f\wedge F_f+NF\, f_V\wedge f_V\, .
\end{equation}
The individual terms can have fractional integrals, it is their sum in Eq.~\eqref{eq:anomaly-index} that is the integer index. For a diagonal flavor curvature $F_f=\mathrm{diag}\,(F_1,\ldots, F_F)$ with $\sum_j F_j=0$, its contribution is simply $\mathrm{Tr}_f F_f\wedge F_f=\sum_j F_j\wedge F_j$. Commutativity of the diagonal matrices does not make this expression vanish.
\newline

A nontrivial anomaly can already be detected with diagonal flavor curvatures. To see this, first take $X_4=S^2\times S^2$. Let $u$ and $v$ be normalized two-forms on the first and second spheres, respectively, pulled back to $X_4$. These are smooth closed two-forms representing the two integral generators of $H^2(X_4,\mathbb Z)$. They satisfy
\begin{equation}
    \int_{S^2_1}u=1\,,\qquad \int_{S^2_2}v=1\,,\qquad u\wedge u=v\wedge v=0\,,\qquad \int_{X_4}u\wedge v=1\,.
\end{equation}
The self-products vanish because each form is pulled back from a two-dimensional factor, while their mixed product has support along all four directions.
Define the traceless diagonal matrices
\begin{equation}
    H_N=\mathrm{diag}\,\left(1-\frac{1}{N},-\frac{1}{N},\ldots,-\frac{1}{N}\right)\,,\qquad H_F=\mathrm{diag}\,\left(1-\frac{1}{F},-\frac{1}{F},\ldots,-\frac{1}{F}\right)\, .
\end{equation}
Following the Cartan-flux construction of Ref.~\cite{Anber:2019nze}, consider the curvatures
\begin{equation}
    \frac{G_c}{2\pi}=H_N u\,,\qquad \frac{F_f}{2\pi}=H_Fv\,,\qquad \frac{f_V}{2\pi}=\frac{u}{N}+\frac{v}{F}\,.
    \label{eq:anomaly-S2S2-background}
\end{equation}
For these backgrounds, the full diagonal vector curvatures $dA_j=F_j+f_V$ obey
\begin{equation}
    \frac{1}{2\pi}(dA_1,\ldots,dA_F)=v(1,0,\ldots,0)+\frac{u}{N}(1,\ldots,1)\,.
\end{equation}
On the first sphere, the fluxes have the allowed common fractional part $1/N$; on the second, only the first flavor has one unit of flux. Thus the background belongs to the faithful Cartan subgroup specified in Eq.~\eqref{eq:anomaly-Cartan-fluxes}.

The separate color and flavor fluxes are fractional. Their combination on a quark is nevertheless integral. Indeed, the component of color $a$ and flavor $j$ sees
\begin{equation}
    \frac{\mathcal F_{aj}}{2\pi}=\left(\delta_{a1}-\frac1N\right)u+\left(\delta_{j1}-\frac1F\right)v+\frac{u}{N}+\frac{v}{F}=\delta_{a1}u+\delta_{j1}v\,.
\end{equation}
Thus every color--flavor component has the flux of an ordinary line bundle, and the complete quark bundle is well defined.

The tangent bundle of $S^2\times S^2$ is the direct sum of the tangent bundles pulled back from the two spheres. Since $p_1$ has degree four and is additive under this direct sum, $p_1(TX_4)=0$. The index is therefore
\begin{equation}
    \mathcal I=\frac{1}{2}\sum_{a=1}^N \sum_{j=1}^F\int_{X_4}\left(\delta_{a1}u+\delta_{j1}v\right)^2=\sum_{a,j}\delta_{a1}\delta_{j1}\int_{X_4}u\wedge v=1\, .
\end{equation}
Consequently,
\begin{equation}
    \mathcal J_\chi=e^{2\pi i/F}\,,
\end{equation}
which is nontrivial for $F>1$.

This calculation establishes that the four-dimensional mixed chiral anomaly remains nontrivial after restricting the vector flavor symmetry to its faithful Cartan subgroup. The calculation does not by itself determine which anomaly components are detected by backgrounds of the three-dimensional theory at a fixed compactification holonomy, or the spectrum and tension of its domain walls. The result needed here is that removing the off-diagonal flavor transformations does not eliminate the four-dimensional anomaly component exhibited above.

\bibliography{ref}
\bibliographystyle{JHEP}

\end{document}